\documentclass[natbib]{svjour3}                     % onecolumn (standard format)
\smartqed  % flush right qed marks, e.g. at end of proof
\usepackage{graphicx}
 \usepackage{aps-bibstyle}  % use this style if you don't use BibTeX.
\begin{document}

\title{The Three-Dimensional Shapes of Galaxy Clusters 
%The Need for a Multiprobe Approach
}
%\subtitle{Combining Lensing, X-ray and Sunyaev Zel'dovich Observations}
\subtitle{}


\author{Marceau Limousin, Andrea Morandi, 
	Mauro Sereno, Massimo Meneghetti, Stefano Ettori,
	Matthias Bartelmann \& Tomas Verdugo
}

\authorrunning{Limousin et~al.}

\institute{Marceau Limousin \at
Aix Marseille Universit\'e, CNRS, LAM (Laboratoire d'Astrophysique de Marseille) UMR 7326, 13388, Marseille, France \\
              \email{marceau.limousin@oamp.fr} \\
	       \& \\
		Dark Cosmology Centre, Niels Bohr Institute, University
of Copenhagen\\	
		Juliane Maries Vej 30,
		DK-2100 Copenhagen, Denmark \\
              \email{marceau.limousin@oamp.fr} \\
           \and
	Andrea Morandi \at
	Raymond and Beverly Sackler School of Physics and Astronomy,
Tel Aviv University, Tel Aviv, 69978, Israel
	\and
	Mauro Sereno \at 
		Dipartimento di Fisica, Politecnico di Torino, corso Duca degli Abruzzi 24, 10129 Torino, Italy \& \\
		INFN, Sezione di Torino, via Pietro Giuria 1, 10125, Torino, Italy \\
	\and
	Massimo Meneghetti \at
		INAF - Osservatorio Astronomico di Bologna, via Ranzani 1, 40127, Bologna, Italy \& \\
		INFN, Sezione di Bologna, Viale Berti Pichat 6/2, 40127, Bologna, Italy \\
	\and
	Stefano Ettori  \at
		INAF - Osservatorio Astronomico di Bologna, via Ranzani 1, 40127, Bologna, Italy \& \\
		INFN, Sezione di Bologna, Viale Berti Pichat 6/2, 40127, Bologna, Italy \\
	\and
	Matthias Bartelmann \at 
Zentrum f\"ur Astronomie, Institut f\"ur Theoretische Astrophysik, Albert-\"Uberle Str. 2, 69120, Heidelberg, Germany
	\and
	Tomas Verdugo  \at 
	Centro de Investigaciones de Astronom\'ia (CIDA), Apartado Postal 264, M\'erida 5101-A, Venezuela
}

\date{Received: date / Accepted: date}

\maketitle

\begin{abstract}
While clusters of galaxies are considered one of the most important cosmological
probes, the standard spherical modelling of the dark matter and the intracluster
medium is only a rough approximation. Indeed, it is well established both 
theoretically and observationally that galaxy clusters are much better approximated as triaxial objects.
However, investigating the asphericity of galaxy clusters is still in its infancy.
We review here this topic which is currently gathering a growing interest from the cluster
community.
We begin by introducing the triaxial geometry.
Then we discuss the topic of deprojection and demonstrate the need for combining
different probes of the cluster's potential.
We discuss the different works that have been addressing these issues.
We present a general parametric framework intended to simultaneously fit complementary
data sets (X-ray, Sunyaev Zel'dovich and lensing data).
We discuss in details the case of Abell 1689 to show how different models/data sets 
lead to different haloe parameters.
We present the results obtained from fitting a 3D NFW model to X-ray, SZ, and lensing
data for 4 strong lensing clusters.
We argue that a triaxial model generally allows 
to lower the inferred value of
the concentration parameter compared to a spherical analysis.
This may alleviate tensions regarding, \emph{e.g.} the over-concentration problem.
However, we stress that predictions from numerical simulations rely on a
\emph{spherical analysis} of triaxial halos.
Given that triaxial analysis will have a growing importance in the observational
side, we advocate the need for simulations to be analysed in the very same way,
allowing reliable and meaningful comparisons.
Besides, methods intended to derive the three dimensional shape of galaxy clusters should be
extensively tested on simulated multi-wavelength observations.
% Finally, we propose some directions of improvement of the current modeling
% techniques to be tested on multi-wavelength simulations.

\end{abstract}

\section{Three-Dimensional Shape of Galaxy Clusters}
\label{intro}

Spectroscopic galaxy redshift surveys and numerical N-body simulations
have revealed a large scale distribution of matter in the Universe featuring
a complex network of interconnected filamentary galaxy associations.
Vertices, i.e. intersections among the filaments, correspond to the very 
dense compact nodes within this
\emph{cosmic web} where one can find massive galaxy clusters.

In this review, we concentrate on the shape of galaxy clusters.

\subsection{Galaxy Clusters are \emph{not} Spherical}
\label{not_spherical}

There is much observational evidence for clusters not being spherical
objects, from the non-circular projection of various probes: 
optical, from the density maps of cluster galaxies \citep{carter80,bingelli82};
X-ray, from the surface brightness maps \citep{fabricant84,buote92,buote96,kawahara,lau2012};
Sunyaev Zel'dovich pressure maps \citep{sayers2011a};
strong gravitational lensing \citep{soucail87}, and
weak gravitational lensing \citep{evans09,oguri10,oguri11}.
Recently, the azimuthal variation of galaxy kinematics has been detected for the first time
in a stacked sample
of 1\,743 galaxy clusters from the SDSS \citep{skielboe}. They find that the line of sight
velocity dispersion of galaxies lying along the major axis of the central galaxy is larger
than those that lie along the minor axis.
This detection provides further evidence for the asphericity of galaxy clusters.

On the numerical side, haloes forming in cosmological simulations have been found
to be triaxial in shape, with a preference for prolateness over oblateness \citep{frenk88,dubinski91,warren92,cole96,jing2002,hopkins05,bailin05,kasunevrard05,paz06,allgood,bett07,munozcuartas,phoenix}.
These simulations also predict an evolution of the shape with mass and redshift:
low mass haloes appear more spherical than high mass haloes
\citep[see, however, an opposite conclusion by][]{rossi11}, and for a given
mass, lower redshift haloes are more spherical than high redshift haloes.

However, more than forty years after the first observational evidence of the
asphericity of galaxy clusters,
the majority of cluster studies use the spherical assumption.
Historically, this was due to the fact that the quality of the data may not
allow a triaxial model to be constrained.
Besides, studies often rely on the analyses of a single data set
(\emph{e.g.} lensing, X-ray, SZ or dynamics of cluster members only), while to
recover triaxiality one needs to combine data sets, unless some priors
are used (see Section~2).

Besides the expectation for triaxial collapse coming from first principles
(see Section~\ref{matthias}),
another hint regarding the need for non-spherical models is the mass 
discrepancy found
between, \emph{e.g.} lensing and X-ray data when spherical symmetry is assumed
\citep{roco03,clowe04,gavazzi05,corless07}.

\subsection{Triaxiality: a Consequence of Gravitational Collapse for an Initial
Gaussian Random Field of Density Fluctuations}
\label{matthias}

Interestingly, triaxial collapse is a straightforward prediction of structure growth driven by self-gravity of 
Gaussian density fluctuations. Well into the mildly non-linear regime, gravitational structure formation can be 
described by the Zel'dovich approximation, which models the motion of dark-matter particles as inertial motion in a 
suitably adapted time coordinate. This assumption defines a map between initial, Lagrangian, and final, Eulerian 
coordinates of any particle. The Jacobian of this map is called the Zel'dovich deformation tensor, here abbreviated 
as $F$. The matter density is then simply given by the inverse determinant of $F$, times
the initial density.

Under the (reasonable) assumption of an initially irrotational flow, a velocity potential $\psi$ exists in Lagrangian 
space whose gradient is the Lagrangian velocity field. The Zel'dovich deformation tensor can be represented by the 
unit matrix plus the Hessian (curvature) matrix of the velocity potential,
\begin{equation}
  F_{ij} = \delta_{ij}+\partial_i\partial_j\psi\;.
\end{equation}
It is thus symmetric and has three real eigenvalues $(1+\lambda_i)$, $i = 1, 2, 3$.

Since determinants are invariant under orthogonal transforms, the inverse determinant of the Zel'dovich tensor is
\begin{equation}
  (\det F)^{-1} = \left[\prod_{i=1}^3(1+\lambda_i)\right]^{-1}\;.
\end{equation}
Collapse will set in according to the Zel'dovich approximation whenever any of the eigenvalues $\lambda_i$ approaches $-1$. If all eigenvalues $\lambda_i$ were the same, collapse was spherical; if they were all different, collapse was triaxial.

It was shown already by \citet{doroshkevich} that the probability distribution for the $\lambda_i$ in a Gaussian random 
field is proportional to the product of absolute differences
\begin{equation}
  |\lambda_i-\lambda_j|
\end{equation}
for any pair $i, j$ with $j \ne i$. This means that equal eigenvalues are not allowed in a Gaussian random field, thus excluding 
spherical and even spheroidal collapse: in any realistic triple of eigenvalues, no pair of them can be equal. Collapse 
will proceed first along the principal axis belonging to the largest eigenvalue, forming the sheet-like structures 
called `pancakes' by Zel'dovich. Those will then contract along the principal axis of the second-largest eigenvalue, 
forming essentially one-dimensional bridges of matter that finally shrink to triaxial haloes along their remaining axis. 
This inherent and necessary triaxiality of the mildly non-linear collapse is retained through the non-linear collapse 
of the haloes which is stopped by virialization.

\subsection{Triaxiality: an Outgrowth of the Large Scale Structure Formation Scenario}
The standard cosmological framework ($\Lambda$CDM), which consists of a cosmological 
constant and cold dark matter (DM) with Gaussian initial conditions, envisages structure 
formation as a hierarchical merging process. With this perspective, gravity is 
constantly pulling lumps of matter together to form increasingly larger structures. 
The structures we see in the Universe today (galaxies, clusters, filaments, sheets 
and voids) are predicted to have formed in this way, with galaxy clusters sitting 
atop this hierarchy and being the largest virialized structures that have had time to 
collapse under the influence of their own gravity.
In particular, being a tracer of the primordial density perturbations, this large 
scale structure scenario leads to a picture where the matter is distributed as a 
network of gigantic dense (filaments) and empty (voids) regions, creating a vast 
foam-like structure called the "cosmic web", with the densest regions of the dark 
matter cosmic web hosting massive clusters of galaxies.
Numerical simulations, in agreement with analytic predictions, indicate that the infall of material into the most 
massive  dark matter haloes is not spherical but is expected to be preferentially 
funnelled through the
filaments where the haloes are embedded. The cluster mass haloes would indeed acquire 
most of their mass from major mergers along the filaments, hence leading to an 
alignment between the major axis of the host haloe and the large-scale filament
\citep{bailin05,altay06,patiri06,aragon07,brunino07}.
Clusters would then relax from this chaotic initial state to a 
quasi-equilibrium via violent relaxation. This process leads to equilibrium
which is plausibly related to bounded triaxial DM distribution with a
'universal' density structure, regardless of haloe mass, of cosmological
parameters, and of the initial fluctuation spectrum.
Therefore, triaxiality appears to be a direct outgrowth of the large scale structure
formation scenario, providing a record of the initial conditions in the Universe and of the
topology of the cosmic structures.

\section{From 2D Observables to a 3D Mass Model: Deprojection}

\subsection{An Under-Constrained Problem}

The problem of deprojection is typically under-constrained: we have access to 2D projected
informations (X-ray surface brightness maps, or lensing mass density maps)
and aim to derive the 3D properties of the triaxial structure.

As an example, let's consider gravitational lensing observations.
Fitting triaxial models with lensing data only is an intrinsically
under-constrained problem, since lensing can never give full information about the
3D triaxial structure: it is constraining only the 2D projected mass
density.
This translates into much larger error bars on the parameters of a triaxial
haloe constrained using lensing observations alone. These error bars fairly represent
the true extent of our limited
knowledge of the structure of galaxy cluster lenses, and make clear the
importance of combining constraints from other theoretical or other
observational data in order to narrow down the parameter space of a more
realistic triaxial model.

The first attempts to determine three dimensional morphologies were based on 
statistical approaches consisting in the inversion of the distribution of apparent 
shapes. \citet{hub26} first determined the relative frequencies with which galaxies
of a given intrinsic ellipticity, oriented at random, are observed as having 
various apparent projected ellipticities. Several following studies have then 
applied similar methods to different classes of astronomical 
objects~\citep{noe79,bin80,bi+de81,fa+vi91,det+al95,moh+al95,bas+al00,coo00,th+ch01,al+ry02,ryd96,pli+al04,paz+al06}.
With the exception of disc galaxies, prolate-like shapes appear to dominate all 
cosmic structure on a large scale.

\subsection{Combining Data Sets}

We will consider three different types of data sets: gravitational lensing
(both weak and strong), X-ray emission, and the thermal SZ effect.
For details on these different data sets, we refer the reader to the other reviews 
published in this volume.

When combining X-ray and SZ data sets, the idea is that one can infer the 3D
properties of a cluster by taking advantage of the different dependences
of the X-ray and SZ signals on the gas density and temperature:
the SZ effect is proportional to the electron pressure integrated along the line of sight, 
whereas the X-ray surface brightness is proportional to the integral along the line of sight
of the square of the electron density.
% ADD EQUATIONS HERE
Besides, gravitational lensing provides a direct probe of the two dimensional
mass distribution projected along the line of sight.

Combining complementary data sets to reconstruct the three dimensional 
properties of galaxy clusters is not a new idea, and different
authors have proposed different approaches.

\subsubsection{Theoretical Studies}
We review here the works that have developed methods which have been tested
on simulated data sets only.

\citet{zaroubi98} proposed a non parametric deprojection method specifically 
designed for the deprojection of X-ray, SZ and lensing maps of galaxy clusters,
under the assumption of \emph{axial symmetry} of the cluster. This method was first applied
to a simple analytic model for cluster dark matter and gas distributions,
then it was tested on cosmological hydrodynamical simulations of galaxy 
clusters \citep{zaroubi2001}. The authors found a good agreement between
the actual (simulated) and reconstructed three-dimensional properties.

Some studies proposed a non parametric deprojection method based on Abel's
integrals. This inversion is applied on
SZ and X-ray data in order to infer temperature and density profiles of the
ICM.
\citet{silk_white} were the first to propose to apply the Abel inversion to the
X-ray and SZ profiles, but their approach was limited to the estimation of the
central values of gas density and temperature aiming at the determination
of cosmological parameters.
More recently, \citet{yoshikawa} used this technique for a non parametric
reconstruction of radial density and temperature profiles using analytical and
simulated cluster models.
This method assumes \emph{spherical symmetry}, and \citet{yoshikawa} mention
the inclusion of non-spherical effects as an important next step.

\citet{reblinsky} developed a non parametric algorithm for the
simultaneous deprojection of X-ray, weak lensing, and SZ data. They specify
a geometrical model for the cluster assuming \emph{axial symmetry}.
Using gas-dynamical simulations, they demonstrated the quality of the
deprojections.

Later, \citet{puchwein} 
proposed a method based on Richardson-Lucy deconvolution to reconstruct the 
three-dimensional gas density and temperature distributions in galaxy clusters from 
combined X-ray and SZ observations. They tested their algorithm against synthetic 
observations using both analytically and fully numerically simulated clusters. They found 
that their method reconstructs the gas density and temperature distributions accurately in
three-dimensions, even if observational noise is present. Additionally, they discussed
a method to constrain the cluster inclination along the line of sight using X-ray 
temperature maps. They came to the conclusion that the method allows to reach a level of 
accuracy of $\sim 15\%$.

\citet{foxpen} considered the problem of deprojecting aspherical clusters.
They first constructed a parametrised 3D \emph{axisymmetric} cluster model,
and determined the 3D cluster shapes using a $\chi^2$ fitting between
the model predictions and the simulated data.

\citet{phil2003} presented a Bayesian joint analysis of cluster weak lensing
and SZ data, assuming \emph{spherical symmetry}.
This methodology was applied to two sets of simulated SZ and weak lensing
data sets.

\citet{lee2004} considered a deprojection method combining SZ and X-ray data
and applied it to analytical cluster models, considering \emph{triaxial
haloes} with constant axis ratios.

\citet{ameglio07,ameglio09} presented deprojection methods
(parametric as well as non parametric)
in order to recover the three dimensional density and temperature profiles
by combining SZ and X-ray surface brightness maps, assuming the cluster
to be \emph{spherically symmetric}.
They apply their techniques to a set of hydrodynamical simulations of
galaxy clusters in order to estimate the biases and scatters on the recovered
masses.
%See also \citet{chakrabarty}.

\citet{allison11} proposed a parametrised \emph{spherical} model of the 
intracluster medium aimed
for jointly analysing SZ and X-ray data. This entropy based model which relies on
the assumption of hydrostatic equilibrium is tested against mock observations
of clusters from N-body/hydrodynamic simulations.

\citet{samsing} presented a novel method for measuring a radially dependent shape
along the line of sight of the ICM from the X-ray observations only. The method
hinges on the assumption that the shapes, temperature and density profiles can be
described by parametrised functions. This model generates fake spectra to be 
compared with the observed spectral information.
A clear advantage of this approach is that it does not require any combination
with independent measurements of \emph{e.g.} the cluster mass or density profile.
The major downside for the current observations is that it requires data of impressively
high quality ($\sim$ 10$^{6}$ photons) to get a 5$\sigma$ detection of the shape.

\subsubsection{Observational Studies}

As seen before, from their different dependencies on the density,
combining X-ray and SZ allows us to directly infer the elongation of the
gas distribution.
No assumption is needed about hydrostatic equilibrium. The method exploits 
X-ray spectroscopic and photometric data plus measurements of the Sunyaev-Zel'dovich 
effect. 
One needs to deproject the X-ray and SZ data.
Both data sets are supposed to trace the same temperature (remember that SZ gives a mass
weighted temperature, unlike X-ray), and one assumes that no clumpiness and no 
contamination from structures along the line of sight might bias the SZ data.
The gas distribution is modelled with an ellipsoidal parametric profile 
which can fit the observed X-ray surface-brightness and temperature. Comparison with the SZ 
amplitude fixes the elongation along the line of sight $e_{\Delta}$. For an isothermal 
plasma \citep{filippis05},
\begin{equation}
e_\Delta \propto D_\mathrm{d} \frac{SB}{\Delta T_{SZ}^2}\frac{T^2}{\Lambda_X}
\end{equation}
where $D_\mathrm{d}$ is the angular diameter distance to the cluster, $SB$ is the 
surface brightness, $\Delta T_{SZ}$ the SZ temperature decrement, $T$ the temperature,
and $\Lambda_X$ the emissivity. Finally, Bayesian inference allows us to deproject the 
measured elongation and the projected ellipticity measured in the plane
of the sky, in order to constrain the intrinsic shape and orientation of the cluster.

This approach was first employed in \citet{filippis05}, who considered a sample of 25 
X-ray luminous clusters. They used parametric \emph{ellipsoidal profiles} of constant
eccentricities aligned along the 
line of sight, assuming an isothermal beta model parametrisation for the ICM.
They found that the spherical hypothesis is strongly rejected for most of
the 25 clusters of the X-ray selected sample studied.
Considering the same data but assuming \emph{axial symmetry}, \citet{filippis06} 
showed that this sample is composed of a mixed population of prolate and
oblate haloes, with prolate shapes preferred in $\sim$ 60-76\% of the cases.
They observe an excess of clusters elongated along the line of sight, with
respect to what is expected from a randomly oriented cluster population.
They claim that a more general triaxial morphology might better describe
the morphology of these clusters.
Both studies acknowledged that adding
gravitational lensing data would help further to constrain the shape.
Indeed elongation strongly enters in the lensing properties of a haloe. Therefore
gravitational lensing features brings information on the elongation of the
total matter distribution.

\citet{jaco} provided a framework for the joint analysis of cluster observations
(\textsc{JACO}) which fits the mass models simultaneously to X-ray, SZ, and
weak lensing data. They applied this method to Abell~478, assuming 
\emph{spherical symmetry}.
They do find a good agreement between all data sets, which points out that the
spherical symmetry hypothesis was well motivated in the case of Abell~478.

However, we stress that it can be questionable to combine different data sets
assuming spherical symmetry: if a given galaxy cluster exhibits a significant mass
discrepancy between X-ray and lensing estimates assuming spherical symmetry,
then a simultaneous fit of both data sets using a spherical model is not appropriate.

The non parametric method based on Abel's integral discussed in the previous
Section has been applied to some clusters in order to study the temperature and
density profiles of the ICM, assuming \emph{spherical symmetry}:
\citet{yuan} and \citet{kitayama} for RXJ\,1347; 
\citet{nord09} for Abell~2163; and \citet{basu10} for Abell~2204.

Recently, \cite{mahdavi} derived a model-independent expression for the minimum
line of sight extent of the hot plasma in a cluster of galaxies.
No a priori assumptions regarding equilibrium or geometry are required, and the
inputs are X-ray and SZ data. They applied this method to the Bullet Cluster, and constrained a 
\emph{minimum} line of sight / plane of the sky axial ratio of $\sim$ 1.

In deprojecting, assumptions regarding the three-dimensional structure of the
DM or the ICM are often required, as well as the assumption regarding hydrostatic
equilibrium.
Although these assumptions could be justified in relaxed  and uni-modal clusters
(i.e. clusters for which the mass distribution can be described using a single mass clump), 
they are likely to be too strong for the vast majority of the systems. Indeed, numerical 
simulations often show that clusters deviate systematically from hydrostatic equilibrium,
especially at large radii ($R>R_{500}$). The main interpretation for this result is that 
clusters, being relatively young structures, are dynamically active and continue to 
accrete mass from their outskirts. Bulk motions of the gas lead to non-thermal pressure 
support and thus alter the state of hydrostatic equilibrium \citep{piffaretti08,lau09}. 
In fact, as shown by \citet{rasia06}, X-ray mass estimates of simulated clusters  based 
on the hydrostatic equilibrium assumptions are biased low by $10-15\%$. The bias is 
however dependent on the gas physics implemented in the simulation. For example, 
\citet{rasia2012} find that the bias is sensitive to the model assumed to include thermal conduction in hydrodynamical simulations. 
Observational results by the CCCP and the LoCuss collaborations 
\citep[\emph{e.g.}][]{mahdavi08,zhang2010}, based on the comparison between X-ray and 
lensing masses of large samples of clusters, seem to support the view emerging from the 
simulations: if we believe that the lensing masses are nearly un-biased on average 
\citep{massimo2010}, the ratio between X-ray and lensing masses should give an 
indication of the lack of hydrostatic equilibrium in these systems. From the above mentioned observational projects, it emerges that X-ray masses are generally smaller than the lensing masses by $\sim 15\%$.

%We will review later in more details recent works proposed by Andrea
%Morandi and collaborators, and by Mauro Sereno and collaborators.

\section{Priors}
\label{priors}

Besides combining different observational data sets
and performing a simultaneous fit of a triaxial mass model,
it is also possible to consider well motivated priors in order to narrow down the
parameter space that will be explored when performing the fit.
We refer the reader to \citet{corless08} and \citet{corless09} for a thorough discussion.
Use of priors applies when the fit is performed on a single given data set 
or when the fit is applied to complementary data sets.
Of course, the fewer data sets one is using and/or the poorer the quality of these
data sets, the more one relies on priors.
Priors sometimes rely on the results from N-body simulations which reasonably capture the
physics of the DM, being dominated by the gravity and hence relatively simple.
However, predictions are coming from dissipationless N-body simulations, and
the physics of baryons may modify predictions, specially in the inner part of galaxy clusters
where most of the baryons are found.

For example, \citet{corless08} presented a Bayesian MCMC (Monte Carlo Markov Chain)
method to fit fully
triaxial NFW haloes to weak lensing data. Their method allows to combine weak lensing
data with prior probability functions on the model haloe parameters to return
parameter and error estimates that reflect the true uncertainties of the problem.

A similar approach has been proposed by Mauro Sereno and collaborators
\citep{sereno_umetsu,sereno_zitrin}.
Using weak and strong lensing data sets, they deprojected the surface density maps to
infer the triaxial structure of the cluster. 
Priors were propagated through the inversion by means of the Bayes theorem.
The method proceeds as follows.
As a first step, they obtain the surface mass density of a galaxy cluster by
strong-lensing modelling of multiple images or/and weak lensing analysis of shear and 
magnification. As a second step, a projected ellipsoidal NFW haloe with 
arbitrary orientation is fitted to the convergence map. Finally, the measured 
distributions of projected parameters are recovered using Bayesian statistics. 

Given the measured convergence map $k_\mathrm{obs}$, the weak lensing 
$\chi^2_\mathrm{WL}$ function can be expressed as \citep{oguri}, 
\begin{equation}
\label{like_wl}
\chi^2_\mathrm{WL}=\sum_{i,j}\left[ k_\mathrm{obs}(\bf{r}_i)- k(\bf{r}_i) \right] \left( V^{-1}\right)_{i,j} \left[ k_\mathrm{obs}(\bf{r}_j)- k(\bf{r}_j) \right] 
\end{equation}
where $\bf{V}^{-1}$ is the inverse of the pixel-pixel covariance matrix. The corresponding likelihood is ${\cal{L}}_\mathrm{WL} \propto \exp(-\chi^2_\mathrm{WL}/2)$. 

As far as strong lensing is concerned, one can use either a parametric or a non-parametric
approach in order to retrieve the surface mass density map. 

The final likelihood to be employed in the Bayes' theorem is then 
${\cal L}(\kappa_\mathrm{s}, r_\mathrm{s_P}, \epsilon, \theta_\epsilon)$,
where $\kappa_\mathrm{s}, r_\mathrm{s_P}, \epsilon, \theta_\epsilon$ corresponds
to the lensing strength, the projected length scale, the projected ellipticity and
the ellipticity in the plane of the sky of an ellipsoidal NFW haloe respectively.
Each projected parameter is on turn a function of the intrinsic shape and 
orientation. According to the relevant data-set, the likelihood is 
${\cal L}_\mathrm{WL}$, ${\cal L}_\mathrm{SL}$ or 
${\cal L}_\mathrm{All} \propto {\cal L}_\mathrm{WL} \times {\cal L}_\mathrm{SL}$ 
for weak, strong or combined strong plus weak lensing analyses, respectively. 

We emphasise that strong lensing clusters are typically more elliptical in the core
than in the weak lensing probed region \citep[see, \emph{e.g}][]{massimo2010}.
This may have important consequences when the likelihoods are combined.
%XXXX CLEARLY relevant to study/probe on simulations XXXXXXXXXXX

\subsection{Priors on the Axis Ratio}
Using a spherical mass model is equivalent to putting $\delta$-function priors on
both axis ratios, which is a strong prior.
The opposite would be to impose a weak flat prior on both axis ratios.

A prior on the axis ratio of a triaxial model has been used in recent works
by Mauro Sereno and collaborators. It is detailed below.
We note that \citet{oguri} were the first to use this prior.

The distribution of minor to major axis ratios ($\eta_{{\rm DM},a}$) can be approximated as 
\citep{jing2002,lee05},
\begin{equation}
\label{nbod3}
p(\eta_{{\rm DM},a}) \propto \exp \left[ -\frac{(\eta_{{\rm DM},a}-\eta_{\mu}/r)^2}{2\sigma_\mathrm{s}^2}\right]
\end{equation}
where the parameters of these distributions were obtained from numerical simulations:
$\eta{_\mu}=0.54$, $\sigma_\mathrm{s}=0.113$ and
\begin{equation}
r = (M_\mathrm{vir}/M_*)^{0.07 \Omega_\mathrm{M}(z)^{0.7}},
\end{equation}
with $M_*$ the characteristic nonlinear mass at redshift $z$ and $M_\mathrm{vir}$ the 
virial mass. 

The conditional probability for intermediate to major axis ratio $\eta_{{\rm DM},b}$ goes as
\begin{equation}
p(\eta_{{\rm DM},a}/\eta_{DM,b} | \eta_{{\rm DM},a})=\frac{3}{2(1-r_\mathrm{min})}
\left[ 1-\frac{2 \eta_{{\rm DM},a} / \eta_{{\rm DM},b} -1-r_\mathrm{min}}{1-r_\mathrm{min}}\right]
\end{equation}
for $\eta_{{\rm DM},a}/\eta_{{\rm DM},b} \geq r_\mathrm{min} \equiv \max[\eta_{{\rm DM},a} ,0.5]$, whereas 
is null otherwise. The distribution of axial ratios of the lensing population mimics that of 
the total cluster population \citep{hennawi07}.

As prior for the intrinsic shape, we consider a flat distribution for the axial ratios 
in the range $\eta{_\mathrm{min}}<\eta_{{\rm DM},a} \le 1$ and $\eta_{{\rm DM},a} \le \eta_{{\rm DM},b} \le 1$.
Probabilities are defined such that the marginalised probability 
$P(\eta_{{\rm DM},a})$ and the conditional probability $P(\eta_{{\rm DM},b}|\eta_{{\rm DM},a})$ are constant.
The probabilities can then be expressed as

\begin{equation}
\label{flat1}
p(\eta_{{\rm DM},a}) =1/(1-\eta{_\mathrm{min}})
\end{equation}
for the full range $\eta{_\mathrm{min}} < \eta_{{\rm DM},a} \le 1$ and
\begin{equation}
\label{flat2}
p(\eta_{{\rm DM},b}|\eta_{{\rm DM},a}) = (1-\eta_{{\rm DM},a})^{-1}
\end{equation}
for $\eta_{{\rm DM},b} \ge \eta_{{\rm DM},a}$ and zero otherwise. The resulting probability for
$\eta_{{\rm DM},b}$ is then

\begin{equation}
p(\eta_{{\rm DM},b}) =\frac{1}{1-\eta{_\mathrm{min}}} \ln \left(\frac{1-\eta{_\mathrm{min}}}{1-\eta_{{\rm DM},b}}\right).
\end{equation}

A flat distribution allows also for very triaxial clusters 
($\eta_{{\rm DM},a} \le \eta_{{\rm DM},b} \ll1$), which are preferentially excluded 
by $N$-body simulations. Therefore, $\eta{_\mathrm{min}}$ is fixed to 0.1.

%one can use the distribution of axis ratio inferred from $\Lambda$CDM
%simulations as prior distribution, such as those of \citet{shaw06} or \citet{bett07}.

\subsection{Strong Lensing Clusters}
\label{slbias}
Different authors, based on simulations \citep{hennawi07,corless07,oguri09a,massimo2010},
investigated to which degree strong lensing (SL) clusters constitute a biased 
population of galaxy clusters.
\citet{hennawi07} found that strong lensing clusters have 3D concentrations 18\% higher than the
typical cluster with similar mass.
Besides, strong lensing clusters are found to be triaxial
and viewed preferentially along their major axis. Therefore, we expect an additional
bias to exist in the distribution of 2D concentrations (the quantity to which lensing is sensitive to).
Indeed, \citet{hennawi07} found that they have 2D concentrations which are 34\% higher
than the typical cluster.
The bias in concentration of strong lensing clusters may be even higher than the
18\% found by \citet{hennawi07}. \citet{massimo2010} showed that the strongest gravitational
lenses (i.e. characterised by large lensing cross sections) are typically affected by much
larger concentration bias (up to 100\%). The bias also depends on the cluster redshift,
being stronger at those redshifts that are least favourable for strong lensing.
Indeed, at these redshifts, only clusters very elongated along the line of sight seem able
to produce strong lensing events.

It is now well established that strong lensing clusters constitute a
biased population of triaxial haloes whose major axis is preferentially 
aligned with the line of sight,
boosting the lensing efficiency.
Therefore, when studying galaxy clusters presenting strong lensing features,
using a prior on the angle between its major axis and the line of sight is
reasonable and has been used in a number of studies \citep[\emph{e.g.}][]{corless09,sereno_umetsu,sereno_zitrin}.

If $\theta$ represents the angle between the major axis of the haloe and our line of sight,
this orientation bias can be expressed as:
\begin{equation}
\label{nbod5}
p(\cos \theta) \propto \exp \left[-\frac{(\cos \theta -1)^2}{2\sigma_\theta^2}\right].
\end{equation}
A value of $\sigma_\theta=0.115$ can be representative of the orientation bias for massive strong lensing clusters \citep{corless09}.

%sereno_zitrin analysed an xray selected sample, so the elongation bias was expected
%to be smaller; in fact, they found ( but did not assume it ) a mild elongation bias
%for the high_z MACS sample/ note, however, the relevance of using a triaxial model
%to study such messy clusters, expect for MACS 1423 for which we should compare our
%results (A1835 as well)

Apart from this strong orientation bias, it is worth noting that strong lensing clusters 
do not exhibit a significant excess of triaxiality and
display nearly the same distribution
of axis ratios as the total cluster population \citep{hennawi07,massimo2010}.

Having mentioned the existence of this orientation bias, it is worth noting that SL clusters 
do not constitute an homogeneous population: all SL clusters are not triaxial haloes with their

major axes pointing right at us.
Actually, observing deep enough at any X-ray luminous cluster, it is very likely that
SL features will be detected.
As mentionned by \citet{massimo2010}, the orientation bias is a growing function of the
lensing cross section. Large Einstein radii clusters will tend to be more biased in their
orientation compared to smaller Einstein radii clusters.
For example, in the sample of 7 strong lensing clusters studied by \citet{newman2012a},
only 1 appears to present an orientation bias. These clusters have Einstein radii of order
$\sim$ 20$''$.

Another way to 'boost' the lensing efficiency is through merger. Indeed, mergers
provide an efficient mechanism to substantially increase the strong lensing efficiency of individual
clusters \citep{zitrin12,redlich12}.
Therefore, both merging and elongated clusters along the line of sight are overrepresented in strong lensing
cluster samples.
In most cases, the optical properties should be able to disentangle both populations.

\subsection{Priors from the Mass-Concentration Relation}
$N$-body simulations \citep{oguri09a,mac+al08,gao+al08,duffy08,prada11} have provided a 
picture of the expected properties of dark matter haloes. Results may 
depend on parameters such as the overall normalisation of the power spectrum, the mass 
resolution, and the simulation volume. The dependence of haloe
concentration on mass and redshift can be adequately described by a power law,

\begin{equation}
\label{nbod1}
c =A(M/M_\mathrm{pivot})^B(1+z)^C.
\end{equation}
As reference, we follow \citet{duffy08}, who used the cosmological parameters from 
WMAP5 and found $\{A,B,C\}=\{ 5.71 \pm0.12, -0.084 \pm 0.006, -0.47\pm0.04\}$ for a 
pivotal mass $M_\mathrm{pivot}=2\times10^{12}M_\odot/h$ in the redshift range $0-2$ for 
their full sample of clusters. The scatter in the concentration about the median $c(M)$ 
relation is lognormal,
\begin{equation}
\label{nbod2}
p(\ln c | M)=\frac{1}{\sigma\sqrt{2\pi}}\exp \left[ -\frac{1}{2} \left(  \frac{\ln c - \ln c(M)}{\sigma}\right) \right],
\end{equation}
with a dispersion $\sigma (\log_{10} c_{200})=0.15$ for a full sample of clusters 
\citep{duffy08}. Recently, \citet{prada11} claimed that the dependence of concentration 
on haloe mass and its evolution can be obtained from the root-mean-square fluctuation 
amplitude of the linear density field. They noticed a flattening and upturn of the 
relation with increasing mass and estimated concentrations for galaxy clusters 
substantially larger than results reported in Eq.~(\ref{nbod1}).
However, more recently, \citet{ludlow} studied how the dynamical state of dark matter haloes affects the relation between mass and concentration. 
When considering only dynamically relaxed haloes, they find that the aforementioned
upturn disappears.
Finally, Meneghetti \& Rasia (submitted) show that the high amplitude and the upturn
of the \citet{prada11} mass-concentration relation can be explained in terms of:
i) the different method (compared to other works in the literature) to measure the
concentration (from the measured circular velocity) and ii) the different selection applied
to haloes for building the mass-concentration relation (clusters are selected by maximum
circular velocity).

In the literature, there is no consensus on the evolution of the $c(M)$ relation with 
redshift, in particular for massive haloes. While several authors predict a strong redshift 
evolution of the concentrations at all mass scales 
\citep[see, \emph{e.g.}][]{bullock01,eke01}, \citet{zhao03} find that the evolution of the 
concentration of individual haloes is not just a function of redshift but is tightly 
connected to their mass growth rate \citep[see also][]{wechsler02}.
In particular, the faster the mass grows, the slower the concentration increases. 
Since most of the massive haloes are in a fast mass accretion phase at high redshift, 
they find that the cluster $c(M)$ relation has a very slow redshift evolution. Similar
results were found recently by \citet{munozcuartas}, who confirm that the growth rate of 
the concentration depends on the haloe mass, with low-mass haloes experiencing a faster 
concentration evolution. These authors also find that the evolution of the $c(M)$ relation 
is faster at lower redshifts than at higher redshifts.

\subsection{On the Choice of Priors}
The choice of priors used on the parameters of
a triaxial model is very important because of the inherently under-constrained nature
of the problem.
This choice should be carefully made with respect to the particular problem at hand.
For example, if one wants to test the mass-concentration relation predicted by the
$\Lambda$CDM scenario, adopting a prior based on that relation would not
be relevant.
On the other hand, if one aims to model a sample of dark matter haloes to calculate a mass
function, a loose mass-concentration prior may be appropriate in order to take into account
existing knowledge of the cluster and group population into the model.
The work by \citet{corless09} further illustrates this point: their weak lensing analysis of Abell~2204
is able to constrain the ellipticity in the plane of the sky, which is found to be slightly
larger than what is found statistically in $\Lambda$CDM simulations.
Therefore, imposing a prior on the axis ratio derived from simulations
for this cluster suppresses real information and is not appropriate.
This highlights the fact that a prior may be well adapted to determine the statistics of a large population
of dark matter haloes, but it may not be relevant for an individual cluster.

When considering priors within a Bayesian framework, one can compute the
Bayesian evidence with or without the prior and see which hypothesis performs
better.
It is worth noting that, the better the data, the lesser the impact of priors in Bayesian
methods.

\section{Three-Dimensional Structure of Galaxy Clusters: Triaxiality}
\label{morandi_method}

We present in this Section a general parametric framework intended to
simultaneously fit X-ray, SZ, and gravitational lensing (both weak and strong)
data sets. It is based on the works published by Andrea Morandi and 
collaborators. More detail can be found in the relevant
publications \citep{morandi2010a,morandi2011a,morandi2011b,morandi2012a,morandi2012b}.

The lensing effects and the X-ray/SZ emission both depend on the properties of the DM gravitational potential well, the former being a direct probe of the two-dimensional mass map via the lensing equation and the latter an indirect proxy of the three-dimensional mass profile through the 
hydrostatic equilibrium (HE)
equation applied to the gas temperature and density. In order to infer the model parameters of both the IC gas and of the underlying DM density profile, we perform a joint analysis of lensing and X-ray/SZ data. We briefly outline the methodology in order to infer physical properties in triaxial galaxy clusters: 
\begin{itemize}
\item We start with a generalised Navarro, Frenk and White (gNFW) triaxial model of the DM as described in \cite{jing2002}, which is representative of the total underlying mass distribution and depends on a few parameters to be determined, namely the concentration parameter $c_{200}$, the scale radius $R_{\rm s}$, the inner slope of the DM $\gamma$ , the two axis ratios ($\eta_{{\rm DM},a}$ and $\eta_{{\rm DM},b}$) and the Euler angles $\psi$, $\theta$ and $\phi$
\item following \cite{lee2003,lee2004}, we recover the gravitational potential and two-dimensional surface mass ${\bf \Sigma}$ (Equation \ref{convergence}) of a dark haloe with such triaxial density profile
\item we solve the generalised HE equation, i.e. including the non-thermal
pressure $P_{\rm nt}$ (Equation \ref{aa4}), for the density of the IC gas
sitting in the gravitational potential well previously calculated, in order to
infer the theoretical three-dimensional temperature profile $T$ 
\item we calculate the SZ temperature decrement map $\Delta T(\nu)$ (Equation \ref{eq:deltai}) and the surface brightness map $S_{\rm X}$ (Equation \ref{1.em.x.eq22}) related to the triaxial ICM haloe
\item the joint comparisons of $T$ with the observed temperature, of $S_{\rm X}$ with the observed brightness image, of $\Delta T(\nu)$ with the observed SZ temperature decrement, and of ${\bf \Sigma}$ with the observed two-dimensional mass map give us the parameters of the triaxial ICM and DM density model.
\end{itemize}

\subsection{ICM \& DM Triaxial Haloes}

\begin{figure}
\begin{center}
\includegraphics[width=7cm,height=7cm]{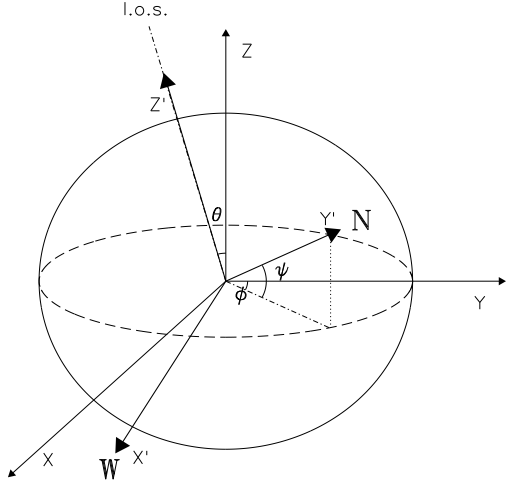}
\caption{The orientations of the coordinate systems. The Cartesian axes ($x,y,z$) represent the DM and ICM haloe principal coordinate system while the axes ($x',y',z'$) represent the observer's coordinate system with the $z'$-axis aligned with the line of sight (l.o.s.) direction. We define $\psi$, $\theta$ and $\phi$ as the rotation angles about the $x$, $y$ and $z$ axes, respectively. The labels {\bf N} and {\bf W} indicate the position of the North and West, respectively, on the plane of the sky.}
\label{fig1}
\end{center}
\end{figure}

We describe the DM and ICM as ellipsoids oriented in an arbitrary
direction on the sky.
We introduce two Cartesian coordinate systems, ${\bf x} = (x,y,z)$ and ${\bf x'} = (x',y',z')$, which represent respectively the principal coordinate system of the triaxial 
dark haloe and the observer's coordinate system, with the origins set at the centre of the haloe. We assume that the $z'$-axis points along the line of sight direction of the observer and that the $x',y'$ axes identify the directions of West and North, respectively, on the plane of the sky. We also assume that the $x,y,z$-axes point along the minor, intermediate and major axes, respectively, of the DM haloe. We define $\psi$, $\theta$ and $\phi$ as the rotation angles about the $x$, $y$ and $z$ axis, respectively (see Figure \ref{fig1}). Then the relation between the two coordinate systems can be expressed in terms of the rotation matrix $M$ as
\begin{equation}
{\bf x'}=M{\bf x},
\end{equation}
where \# denotes matrix multiplication and $M$ represents the orthogonal matrix corresponding to counter-clockwise/right-handed rotations $M_x(\psi),M_y(\theta),M_z(\phi)$ with Euler angles $\psi,\theta,\phi$, and it is given by:
\begin{equation}
M=M_x(\psi)\#M_y(\theta)\#M_z(\phi)\,,
\end{equation}
where
\begin{equation}
\begin{array}{c}  \\ M_x(\psi)= \\ \end{array}
\left[\begin{array}{ccc}
1  & 0  & 0 \\
0 & \cos\psi & -\sin\psi \\
0 & \sin\psi & \cos\psi\\
\end{array}\right]\nonumber ;
\end{equation}
\begin{equation}
\begin{array}{c}  \\ M_y(\theta)= \\ \end{array}
\left[\begin{array}{ccc}
\cos\theta  & 0  & \sin\theta \\
0 & 1 & 0\\
-\sin\theta & 0 & \cos\theta\\
\end{array}\right] ;
\end{equation}

\begin{equation}
\begin{array}{c}  \\ M_z(\phi)= \\ \end{array}
\left[\begin{array}{ccc}
\cos\phi  & -\sin\phi  & 0 \\
\sin\phi & \cos\phi & 0\\
0 & 0 & 1\\
\end{array}\right]\nonumber.
\end{equation}

Figure \ref{fig1} represents the relative orientation between the observer's coordinate system and the haloe principal coordinate system.

In order to parametrise the cluster mass distribution, we consider a triaxial generalised Navarro, Frenk \& White model \citep[gNFW, \emph{e.g.}][]{jing2002}:
\begin{equation}\label{aa33344}
\rho(R) = \frac{\delta_{c}\rho_{\rm c, z}}{\left(R/R_{\rm s}\right)^{\gamma}
\left(1 + R/R_{\rm s}\right)^{3-\gamma}} ,
\end{equation}
where $R_{\rm s}$ is the scale radius, $\delta_{c}$ is the dimensionless characteristic density contrast with respect to the critical density of the Universe $\rho_{\rm c, z}$ at the redshift $z$ of the cluster, and $\gamma$ represents the inner slope of the density profile; $\rho_{\rm c, z}\equiv 3H(z)^2/ 8 \pi G$ is the critical density of the universe at redshift $z$, $H_z\equiv E_z\,H_0$, $E_z \!=\left[\Omega_M (1+z)^3 + \Omega_{\Lambda}\right]^{1/2}$, and
\begin{equation}\label{aqrt}
\delta_{\rm c} = \frac{200}{3} \frac{c_{200}^3}{ F(c_{200},\gamma)} \ ,
\end{equation}
where $c_{200} \equiv R_{200}/R_{\rm s}$ is the concentration parameter, 
with \citep{wyithe2001}:
\begin{equation}
F(y,\gamma)\equiv\int_0^y s^{2-\gamma}(1+s)^{\gamma-3}ds.
\end{equation}

The radius $R$ can be regarded as the major axis length of the iso-density surfaces:
\begin{eqnarray}
\label{eq:isodensity}
R^2= c^{2}\left(\frac{x^2}{a^2} + 
\frac{y^2}{b^2} + \frac{z^2}{c^2}\right), \qquad (a \le b \le c).
\label{rsuto1}
\end{eqnarray}
We have previously defined $\eta_{{\rm DM},a}=a/c$ and $\eta_{{\rm DM},b}=b/c$ as the minor-major and 
intermediate-major axis ratios of the DM haloe, respectively.

The gravitational potential of a dark haloe with the triaxial density profile (Equation \ref{aa33344}) can be written in terms of complex implicit integrals \citep{binney1987}. While numerical integration is required in general to obtain the triaxial gravitational potential, \cite{lee2003} retrieved the following approximation (which holds for small eccentricities for the gravitational potential $\Phi$ under the assumption of triaxial gNFW model for the DM (Equation \ref{aa33344}):
\begin{equation}
%\begin{split}
\Phi({\bf u}) \simeq C_0\;{{ F_{1}(u) +C_0\;\frac{e_{b}^{2}+e_{c}^2}{2}F_{2}(u)}} 
{{+ C_0\;\frac{e_{b}^{2}\sin^{2}\theta\sin^{2}\phi +  e_{c}^{2} \cos^{2}\theta}{2} F_{3}(u)} ,}
%\end{split}
\label{aa44425}
\end{equation}
with ${\bf u} \equiv {\bf r}/R_{\rm s}$, $C_0 = 4\pi G\delta_{c}\rho_{\rm c}(z)R_{\rm s}^{2}$, and the three functions, $F_{1}(u), F_{2}(u)$, and $F_{3}(u)$ have been defined in \cite{morandi2010a}, $e_{b}$ ($\epsilon_{b}$) and $e_{c}$ ($\epsilon_{c}$) are the eccentricity of DM (IC gas) with respect to the major axis (e.g. $e_{b}=\sqrt{1-(b/c)^2}$).

The work of \cite{lee2003} showed that the iso-potential surfaces of the triaxial dark haloe
are well approximated by a sequence of concentric triaxial distributions of radius $R_{\rm icm}$ with different eccentricity ratio. 
For $R_{\rm icm}$ a similar definition as for $R$ holds (Equation \ref{rsuto1}), but with eccentricities $\epsilon_{b}$ and $\epsilon_{c}$. Note that $\epsilon_{b}=\epsilon_{b}(e_{b},u,\gamma)$ and
$\epsilon_{c}=\epsilon_{c}(e_{c},u,\gamma)$, unlike the constant $e_{b},e_{c}$ for the adopted DM
haloe profile. In the whole range of $u$, $\epsilon_{b}/e_{b}$
($\epsilon_{c}/e_{c}$) is less than unity ($\sim 0.7$ at the
centre), i.e., the intracluster gas is altogether more spherical than
the underlying DM haloe (see \cite{morandi2010a} for further details).

The iso-potential surfaces of the triaxial dark haloe
coincide also with the iso-density (pressure, temperature) surfaces of
the intracluster gas. This is simply a direct consequence of the {\it
X-ray shape theorem} \citep{buote1994}; the HE
equation (\ref{aa4}) yields
\begin{equation}\label{eqn:ecc}
\nabla P \times \nabla\Phi = \nabla \rho_{\rm gas} \times \nabla\Phi = 0 .
\end{equation}

\subsection{X-ray, SZ and lensing equations}\label{xray}

For the X-ray analysis we rely on a generalisation of the HE equation \citep{morandi2011b}, which accounts for the non-thermal pressure $P_{\rm nt}$ and reads:
\begin{equation}\label{aa4}
\nabla P_{\rm tot} = -\rho_{\rm gas} \nabla \Phi
\end{equation}
where $\rho_{\rm gas}$ is the gas mass density, $\Phi$ is the gravitational potential, $P_{\rm tot}= P_{\rm th}+ P_{\rm nt}$. We implemented a model where $P_{\rm nt}$ is a fraction of the total pressure $P_{\rm tot}$, and we set this fraction to be a power law with the radius \citep{shaw2010}:
\begin{equation}
\frac{P_{\rm nt}}{P_{\rm tot}} =\xi \,(R/R_{200})^n \ .
\label{pnt12}
\end{equation}
Note that X-ray data probe only the thermal component of the gas $P_{\rm th}=n_e\, {\bf k}  T$, ${\bf k}$ being the Boltzmann constant. From Equations (\ref{aa4}) and (\ref{pnt12}) we point out that neglecting $P_{\rm nt}$ (i.e. $P_{\rm tot} = P_{\rm th}$) systematically biases low the determination of cluster mass profiles.  
This effect increases at larger radii, where the contribution of the gas
motion is larger.

Given that Equation (\ref{aa4}) is a first order differential equation, we need a boundary condition on the pressure, $\tilde P$, which represents the pressure at $R_{200}$, and it is an unknown parameter to be determined.

To model the electron density profile in the triaxial ICM haloe, we use the following fitting function,
which corresponds to a simplified version of the function given by \citet{vikhlinin06}:
%\begin{eqnarray}
%n_e(R_{\rm icm}) = {n_0\; (R_{\rm icm}/r_{c_1})^{-\delta}}
%{(1+R_{\rm icm}^2/r_{c_1}^2)^{-3/2 \, \varepsilon+\delta/2}}\cdot\nonumber\\
%\cdot(1+R_{\rm icm}^4/r_{c_2}^4)^{-\upsilon/4}
%\label{eq:density:model}
%\end{eqnarray}

\begin{equation}
n_e(R_{\rm icm}) = {n_0\; (R_{\rm icm}/r_{c_1})^{-\delta}}
{(1+R_{\rm icm}^2/r_{c_1}^2)^{-3/2 \, \varepsilon+\delta/2}}
(1+R_{\rm icm}^4/r_{c_2}^4)^{-\upsilon/4}
\label{eq:density:model}
\end{equation}

with parameters ($n_0,r_{c_1},\varepsilon,\delta,r_{c_2},\upsilon$). 
We computed the theoretical three-dimensional temperature $T$ by numerically integrating the equation of the HE (Equation \ref{aa4}), assuming triaxial geometry and a functional form of the gas density given by Equation (\ref{eq:density:model}).

The observed X-Ray surface brightness $S_{\rm X}$ is given by:
\begin{equation}
S_{\rm X} = \frac{1}{4 \pi (1+z)^4} \Lambda(T^*_{\rm proj},Z) \int n_{\rm e} n_{\rm p}\, dz'\;\;,
\label{1.em.x.eq22}
\end{equation}
where $\Lambda(T^*_{\rm proj},Z)$ is the cooling function. Since the projection on the sky of the plasma emissivity gives the X-ray surface brightness, the latter can be geometrically fitted with the model $n_e(R_{\rm icm})$ of the assumed distribution of the electron density (Equation \ref{eq:density:model}) by applying Equation (\ref{1.em.x.eq22}). This has been accomplished via simulated \emph{Chandra} spectra, where the current model is folded through response curves (ARF and RMF) and then added to a background file, and with absorption, temperature and metallicity measured in that neighbouring ring in the spectral analysis. 
In order to calculate $\Lambda(T^*_{\rm proj},Z)$, we adopted a 
MEKAL model for the emissivity.

The thermal SZ effect is expressed as a small variation in the temperature $\Delta T(\nu)$ of the CMB as a function of the observation frequency:
\begin{eqnarray}
\frac{\Delta T(\nu)}{T_{\rm cmb}} = \frac{\sigma_{T}}{m_e c^2} \int P_e({\bf r})\, f(\nu;T({\bf r}))  \, dz' 
\label{eq:deltai}
\end{eqnarray}
where $\sigma_T$ is the Thomson cross-section, $P_e({\bf r}) \equiv n_e({\bf r})\, {\bf k} \, T_e({\bf r})$ is the pressure of the electrons of the ICM at the volume element of coordinate {\bf r}, ${\bf k}$  is the Boltzmann constant, and 
$T_{\rm cmb} = 2.725$ K.

$f(\nu;T({\bf r}))$ takes into account the spectral shape of the SZ effect and it reads:
\begin{equation}
f(\nu;T({\bf r})) = ( x \frac{e^x + 1}{e^x - 1} - 4) ( 1 + o_f(x; T) )\ ,
\label{eq:deltai2}
\end{equation}
where $x = h \nu / k T_{\rm cmb}$ accounts for the frequency dependence of the SZ effect, and for the relativistic corrections related to the term $o_f(x, T)$ \citep{itoh1998}. Note that in Equation (\ref{eq:deltai}) we account for the implicit dependence of $f(\nu;T({\bf r}))$ on the radius.

Next, the two-dimensional SZ model $\Delta T(\nu)$ is convolved with the instrumental point-spread function and the measured transfer function. In practise, the transfer function convolution is performed via multiplication in the Fourier domain. This filtering significantly reduces the peak decrement of the cluster and creates a ring of positive flux at $r\sim 2$ arcmin. This filtered model is then compared to the observed SZ temperature decrement map. We also calculated the 
noise covariance matrix ${\bf C}$ 
among all the pixels of the observed SZ temperature decrement map through 1000 jackknife realisations of our cluster noise. In this perspective we assumed that the noise covariance matrix for SZ data is diagonal, as this was shown to be a good assumption in \cite{sayers2011a}.

For the lensing analysis the two-dimensional surface mass
density ${\bf \Sigma}$ can be expressed as:
\begin{equation}
{\bf \Sigma}=\int_{-\infty}^{\infty}\rho(R)dz'
\label{convergence}
\end{equation}
We also calculated the covariance matrix ${\bf C}$ among all the pixels of the reconstructed surface mass (see \cite{morandi2011b} for further details).

\subsection{Joint X-ray+SZ+Lensing Analysis}\label{sryen2}
The probability distribution function of model parameters has been evaluated via Markov Chain Monte Carlo (MCMC) algorithm, by using as proposal density a likelihood ${\mathcal{L}}$ and a standard method for rejecting proposed moves. This allows to compare observations and predictions, and to infer the desired physical parameters. The likelihood has been constructed by performing a joint analysis for lensing and X-ray/SZ data. More specifically, the system of equations we simultaneously rely on in our joint X-ray\,+\,SZ\,+\,lensing analysis is:

\begin{eqnarray}
T{(c_{200},R_{\rm s},\gamma,\eta_{{\rm DM},a},\eta_{{\rm DM},b},\psi, \theta,\phi,n_0,r_{c_1},\varepsilon,\delta,r_{c_2},\upsilon,\xi,n,\tilde P)}\nonumber \\
S_{\rm X}(c_{200},R_{\rm s},\gamma,\eta_{{\rm DM},a},\eta_{{\rm DM},b},\psi, \theta,\phi,n_0,r_{c_1},\varepsilon,\delta,r_{c_2},\upsilon)\nonumber\\
\Delta T(c_{200},R_{\rm s},\gamma,\eta_{{\rm DM},a},\eta_{{\rm DM},b},\psi, \theta,\phi,n_0,r_{c_1},\varepsilon,\delta,r_{c_2},\upsilon,\xi,n,\tilde P)\nonumber\\
{\bf \Sigma}(c_{200},R_{\rm s},\gamma,\eta_{{\rm DM},a},\eta_{{\rm DM},b},\psi, \theta,\phi)
\end{eqnarray}
where the parameters $c_{200}$ (concentration parameter), $R_{\rm s}$ (scale radius), $\gamma$ (inner DM slope), $\eta_{{\rm DM},a}$ (minor-major axis ratio), $\eta_{{\rm DM},b}$ (intermediate-major axis ratio), and $\psi,\theta,\phi$ (Euler angles) refer to the triaxial DM haloe (Equation \ref{aa33344}); the parameters $n_0,r_{c_1},\varepsilon,\delta,r_{c_2},\upsilon$ refer to the IC gas density (Equation \ref{eq:density:model}); $\xi,n$ (normalisation and slope, respectively) refer to the non-thermal pressure (Equation \ref{pnt12}); and $\tilde P$ to the pressure at $R_{200}$, which is a boundary condition of the generalised HE equation (Equation \ref{aa4}).

In this triaxial joint analysis the three-dimensional model temperature $T$ is recovered by solving equation (\ref{aa4}) and constrained by the observed temperature profile; the surface brightness is recovered via projection of the gas density model (Equation \ref{1.em.x.eq22}) and constrained by the observed brightness; the SZ signal is deduced via projection of the three-dimensional pressure (Equation \ref{eq:deltai}) and constrained by the observed SZ temperature decrement; and the model two-dimensional mass density ${\bf \Sigma}$ is recovered via Equation (\ref{convergence}) and constrained by the observed surface mass density.

Hence the likelihood ${\mathcal{L}}\propto \exp(-\chi^2/2)$, and $\chi^2$ reads:
\begin{equation}\label{chi2wwf}
\chi^2=\chi^2_{\rm x,T}+\chi^2_{\rm x,S}+\chi^2_{\rm SZ}+\chi^2_{\rm lens}
\end{equation}
with $\chi^2_{\rm x,T}$, $\chi^2_{\rm x,S}$, $\chi^2_{\rm SZ}$ and $\chi^2_{\rm lens}$ being the $\chi^2$ coming from the X-ray temperature, X-ray brightness, SZ temperature decrement and lensing data, respectively.
We note that, when both weak (lensing WL) and strong lensing (SL) data are available 
\citep{morandi2011b}, $\chi^2_{\rm lens} = \chi^2_{\rm WL} + \chi^2_{\rm SL} $.

For the spectral analysis, $\chi^2_{\rm x,T}$ is equal to:
\begin{equation}\label{chi2wwe}
\chi^2_{\rm x,T}= \sum_{i=1}^{n^*} {\frac{{ (T_{\rm proj,i}-T^*_{\rm proj,i})}^2 }{\sigma^2_{T^*_{\rm proj,i}}  }}\
\end{equation}
$T^*_{\rm proj,i}$ being the observed projected temperature profile in the $i$th circular ring and $T_{\rm proj,i}$ the azimuthally-averaged projection \citep[following][]{mazzotta2004} of the theoretical three-dimensional temperature $T$; the latter is the result of solving the HE equation, with the gas density $n_e(R_{\rm icm})$.

For the X-ray brightness, $\chi^2_{\rm x,S}$ reads:
\begin{equation}\label{chi2wwe2}
\chi^2_{\rm x,S}=  \sum_j \sum_{i=1}^{N_j} {\frac{{ (S_{X,i}-S^*_{X,i})}^2 }{\sigma^2_{S,i}}  }\
\end{equation}
with $S_{X,i}$ and $S^*_{X,i}$ theoretical and observed counts in the $i$th pixel of the $j$th image.
Given that the number of counts in each bin might be small ($ <$ 5), then we cannot assume that the Poisson distribution from which the counts are sampled has a nearly Gaussian shape. The standard deviation (i.e., the square-root of the variance) for this low-count case has been derived by \cite{gehrels1986}:
\begin{equation}\label{chi2wwe3}
\sigma_{S,i}= 1+\sqrt{S^*_{X,i}+0.75}
\end{equation}
which has been proved to be accurate to approximately one percent. Note that we added background to $S_{X,i}$ as measured locally in the brightness images, and that the vignetting has been removed in the observed brightness images.

For the SZ (lensing) constraint ${\bf D}$, the $\chi^2_{{\bf D}}$ contribution is:
\begin{equation}\label{aa2w2q}
\chi^2_{{\bf D}}={{[ {\bf D}-{\bf D}^*]}^{\rm t}  {\bf C}^{-1} [{\bf D}-{\bf D}^*]}\ ,
\end{equation}
where ${\bf C}$ is the covariance matrix of the two-dimensional SZ temperature decrement (projected mass density), ${\bf {\bf D}^*}$ are the observed measurements of the two-dimensional SZ temperature decrement (projected mass density) in the $i$th pixel, and ${\bf {\bf D}}$ is the theoretical 2D model. 

Errors on the individual parameters have been evaluated by considering average value and standard deviation on the marginal probability distributions of the same parameters.

\begin{figure*}
\begin{center}
\includegraphics[width=14cm,height=15cm]{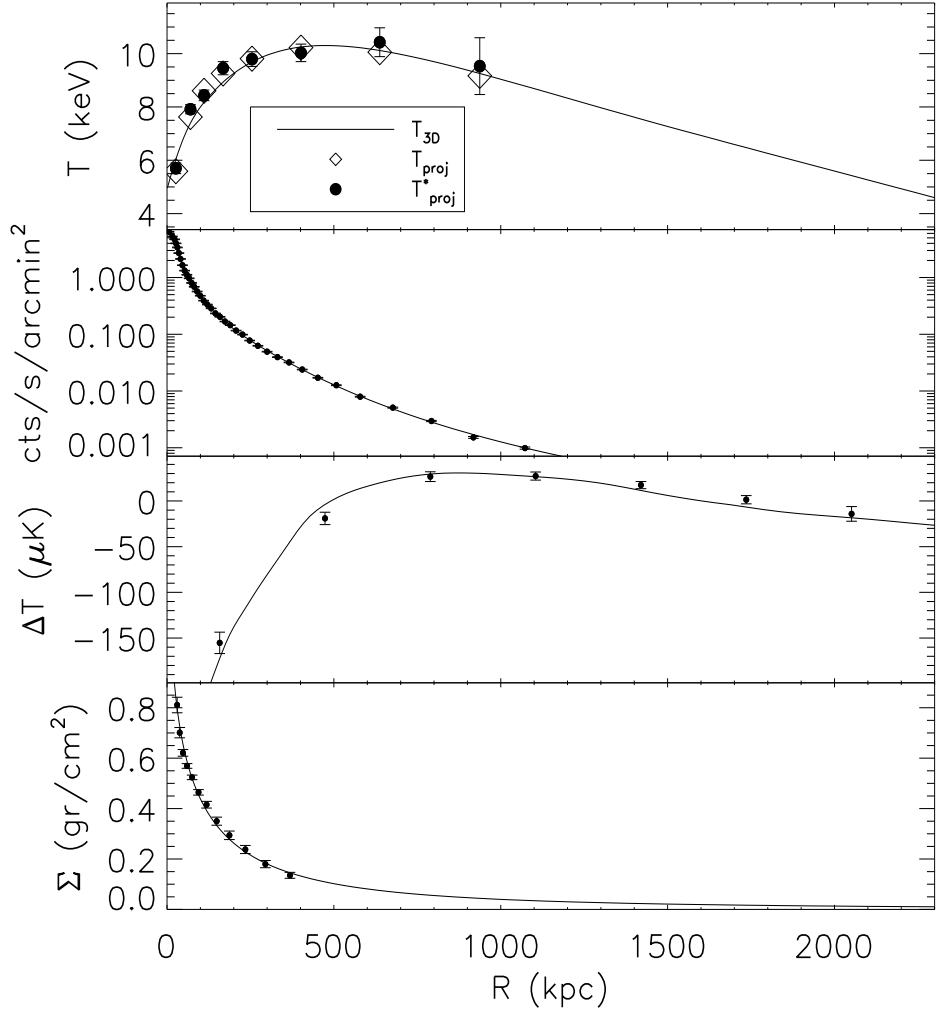}
\caption[]{Example of the joint analysis for $T$, $S_{\rm X}$, $\Delta T(\nu)$ and 
${\bf \Sigma}$, from \citet{morandi2012b}.
In the upper panel we display the two quantities which enter in the 
X-ray analysis (Equation \ref{chi2wwe}): the observed spectral projected temperature $T^*_{\rm proj,m}$ (big points with errorbars) and the theoretical projected temperature $T_{\rm proj,m}$ (diamonds). We also show the theoretical 3D temperature (solid line), which generates $T_{\rm proj,m}$ through convenient projection techniques. In the second panel from the top we display the two quantities which enter in the X-ray brightness analysis (Equation \ref{chi2wwe2}): the observed surface brightness profile $S_{\rm X}^*$ (points with errorbars) and the theoretical one $S_{\rm X}$ (solid line). In the third panel from the top  we display the two quantities which enter in the SZ temperature decrement analysis (Equation \ref{eq:deltai}): the observed SZ temperature decrement profile (points with errorbars) and the theoretical one $\Delta T(\nu)$ (solid line). Both observed and theoretical SZ temperature decrement are convolved with the transfer function: note that this filtering significantly reduces the peak decrement of the cluster and creates a ring of positive flux at $r\sim 2$ arcmin. In the lowest panel we display the two quantities which enter in the lensing analysis (Equation (\ref{aa2w2q})): the observed surface mass profile $\Sigma^*$ (points with error bars) and the theoretical one ${\bf \Sigma}$ (solid line). Note that for surface brightness (surface mass) and the SZ data the 1D profile 
is presented for visualisation purpose only, the fit being applied on the 2D data. Moreover, for the surface brightness we plotted data referring to the observation ID 6880. The virial radius corresponds to a scale length on the plane of the sky of $\sim \eta_{{\rm DM},a}\cdot R_{200}\approx 2240$ kpc.}
\label{joint}
\end{center}
\end{figure*}

So we can determine the physical parameters of the cluster, for example the 3D temperature $T$, the shape of DM and ICM, just by relying on the generalised HE equation and on the robust results of the hydrodynamical simulations of the DM profiles. In Fig.~\ref{joint} we present an example of a joint analysis for $T$, $S_{\rm X}$, $\Delta T(\nu)$ and ${\bf \Sigma}$: for $S_{\rm X}$, $\Delta T(\nu)$ and ${\bf \Sigma}$ the 1D profile has been presented only for visualisation purpose, the fit being applied on the 2D X-ray brightness/SZ/surface mass data. Note that in the joint analysis both X-ray, SZ and lensing data are well fitted by our model, with a $\chi^2_{\rm red}=1.04$.

\subsection{Limitations}

\paragraph{The BCG:}
The centre of galaxy clusters is usually populated by a bright central galaxy (BCG).
Actually, the BCG has an influence on the formation of multiples images \citep{massimo03,donnarumma2011},
and the physical processes taking place in the BCG substantially influence the X-ray gas
\citep{gitti2012}.
In the triaxial framework described in this Section, we have removed the central 25\,kpc of the 
data in the joint analysis, to avoid the contamination from the BCG.
However, modelling properly the BCG contribution is essential in order to probe the dark matter
distribution in the very centre.

In galaxy cluster Abell~1703, the strong lensing analysis by \citet{mypaperV} takes into account the
stellar contribution of the BCG in a parametric mass modelling aimed at constraining the underlying
smooth dark matter component distribution, in particular the inner slope of the dark matter distribution.

A more advanced approach which requires high quality spectroscopic data is to combine lensing 
observations with the stellar kinematics of the BCG
\citep{sand02,sand04,sand07,newman,newman2012a,newman2012b}.
These studies found density profiles shallower than canonical NFW models at radii $<$ 30\,kpc, 
comparable to the effective radii of the BCG.

Numerical simulations by \citet{dubinski98} suggested that the measured velocity dispersion profile
strongly depends on the line of sight, where the central value peaks between 300 and 450\,km\,s$^{-1}$
depending on the considered line of sight.
However, recent observations do not confirm this claim.
In the sample of 7 massive strong lensing clusters studied by \citet{newman2012a}, the
observed velocity profiles display a very homogeneous shape which are mutually consistent.
Similar conclusions is reached within the SAURON project \citep{sauron}, which found that
in giant slowly/non rotating ellipticals (including BCGs), the average shape of the velocity
ellipsoid does not differ by more than $\sim$ 10 per cent from a spherical shape.

\paragraph{The gas mass component:}
When studying the dark matter distribution, 
we did not subtract the mass contribution from the X-ray gas to the total mass.
Nevertheless, the contribution of the gas to the total matter 
is small: the measured gas fraction is 0.06-0.07 in the spatial range 30-400\,kpc, and the
slope of the density profile is very similar to that of the DM beyond a characteristic scale
$\sim$ 20-30\,kpc, a self-similar property of the gas common to cool core clusters
\citep{morandi_ettori07}, suggesting that the assumption to model the total mass as a gNFW is
reliable. Similar conclusions have been reached by \citet{bradac08a} and 
\citet{jesper}.

\subsection{Degeneracies \& Priors}

\begin{figure}
\begin{center}
\includegraphics[width=7cm,height=7cm]{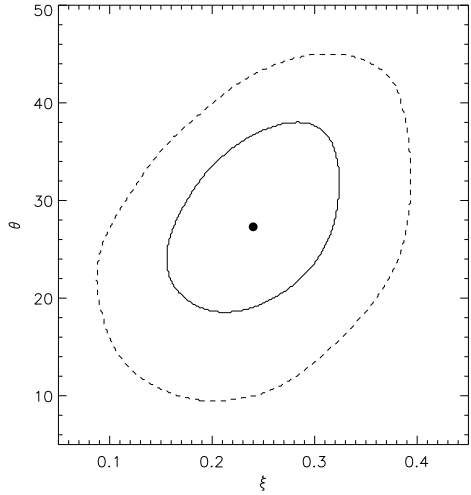}
\caption[]{Degeneracies between $\theta$ and $\xi$.
$\theta$ represents the angle between the major axis of the haloe and our line of sight,
and $\xi$ is related to the amount of non-thermal component considered in the model.
The solid (dashed) line represent the 1 (2)-$\sigma$ error region, while the point represents the best fit value.}
\label{degen2}
\end{center}
\end{figure}

\begin{figure*}
\begin{center}
\includegraphics[width=6cm,height=6cm]{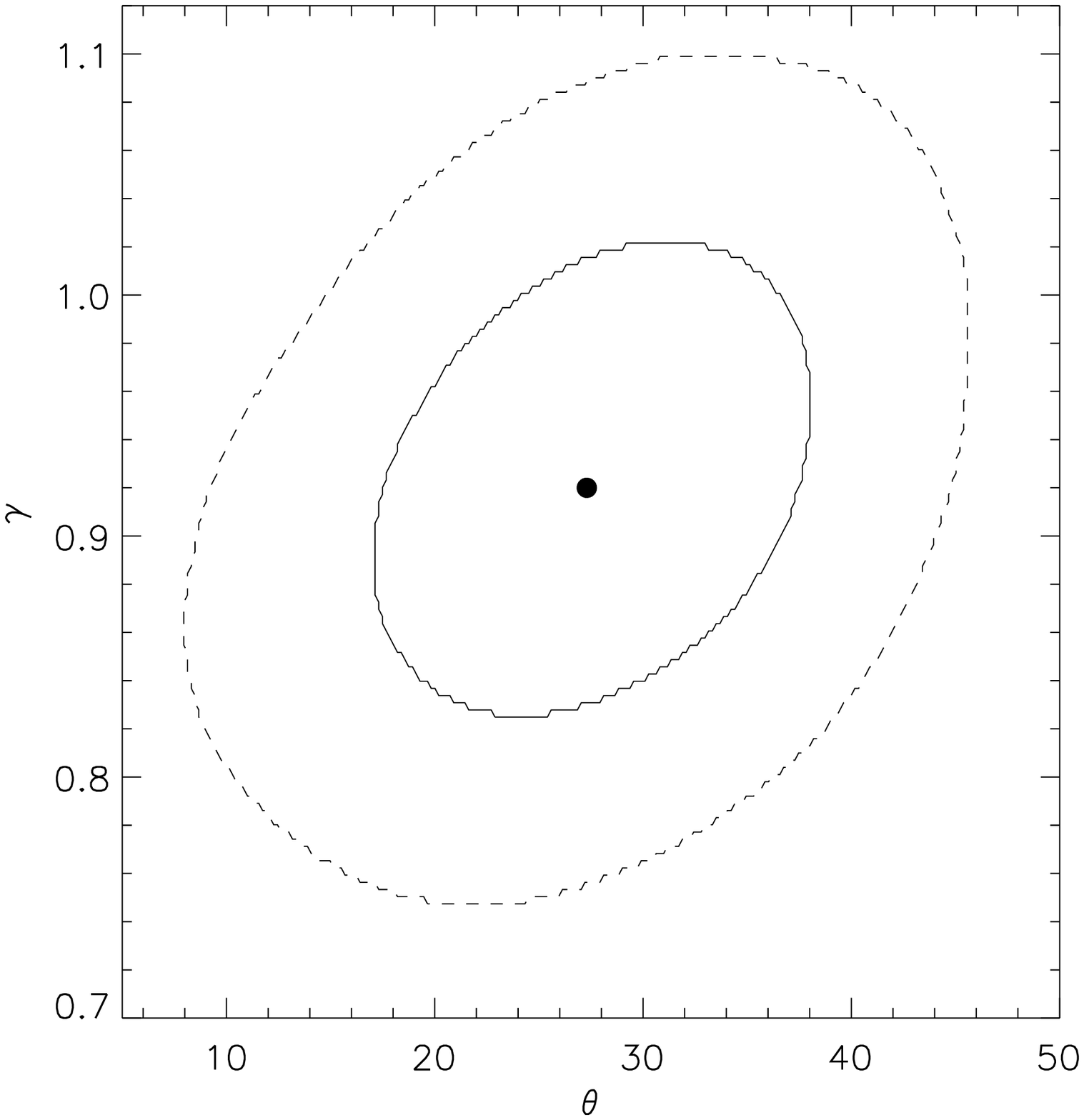}
\includegraphics[width=6cm,height=6cm]{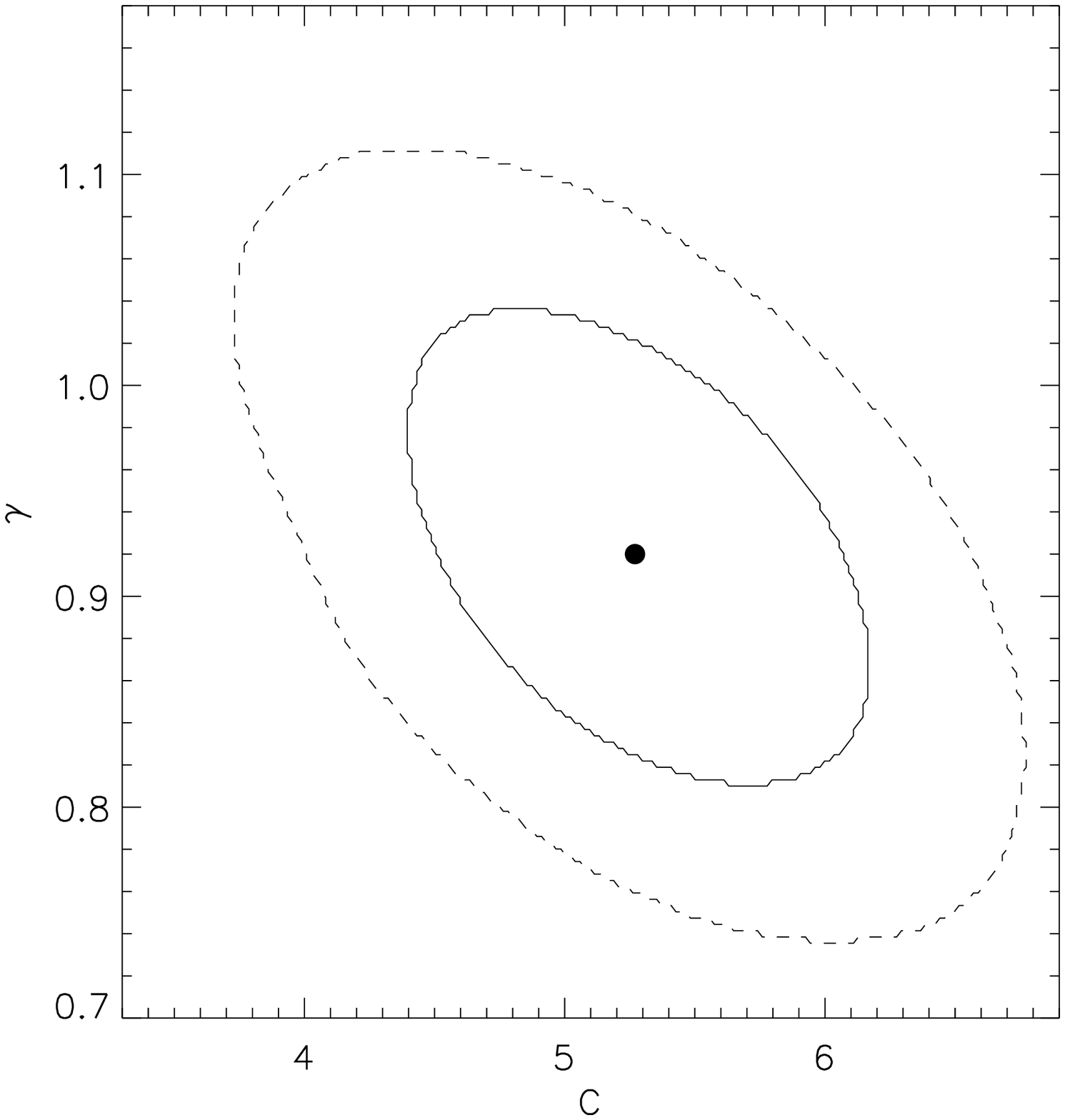}\\
\includegraphics[width=6cm,height=6cm]{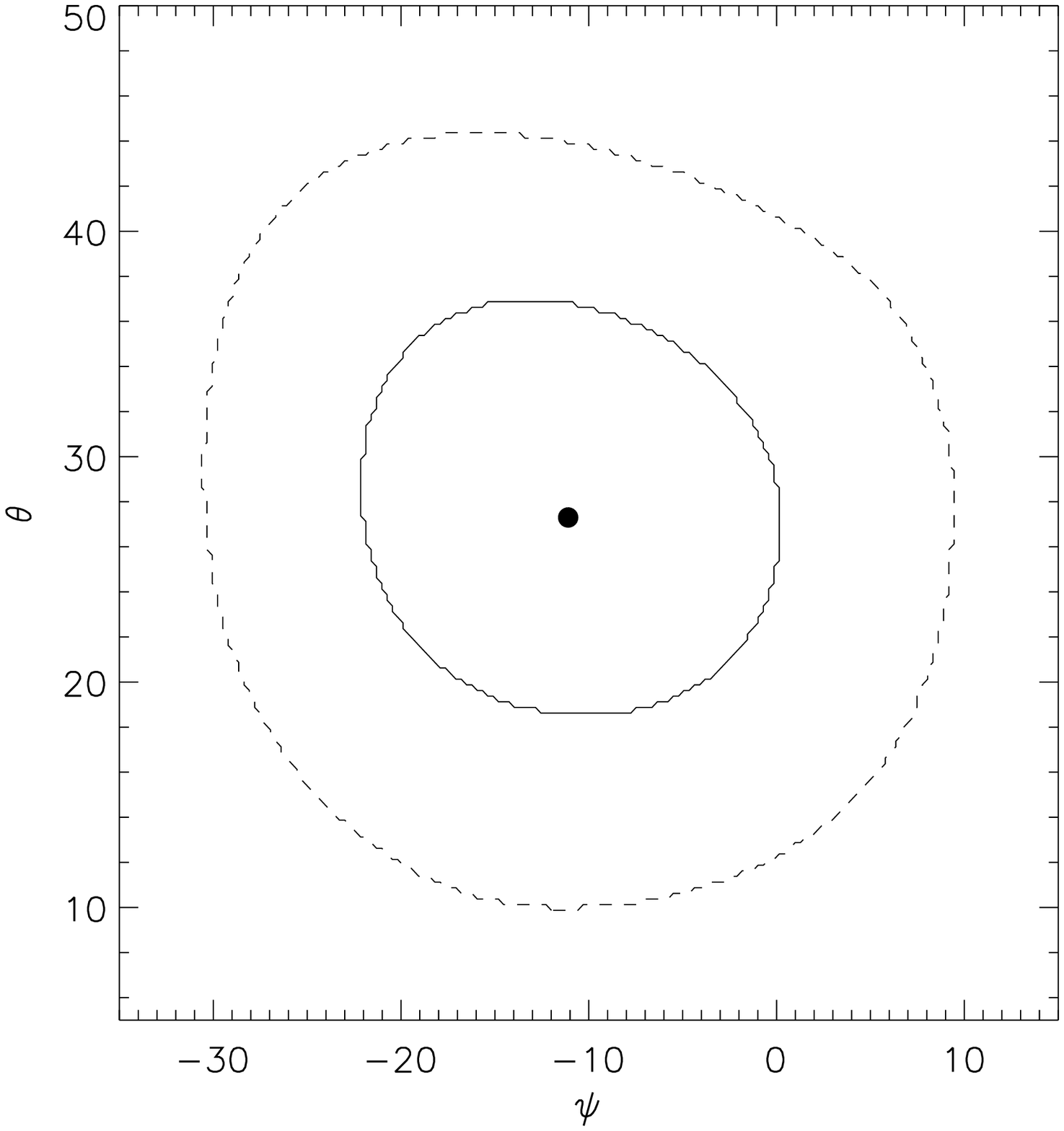}
\includegraphics[width=6cm,height=6cm]{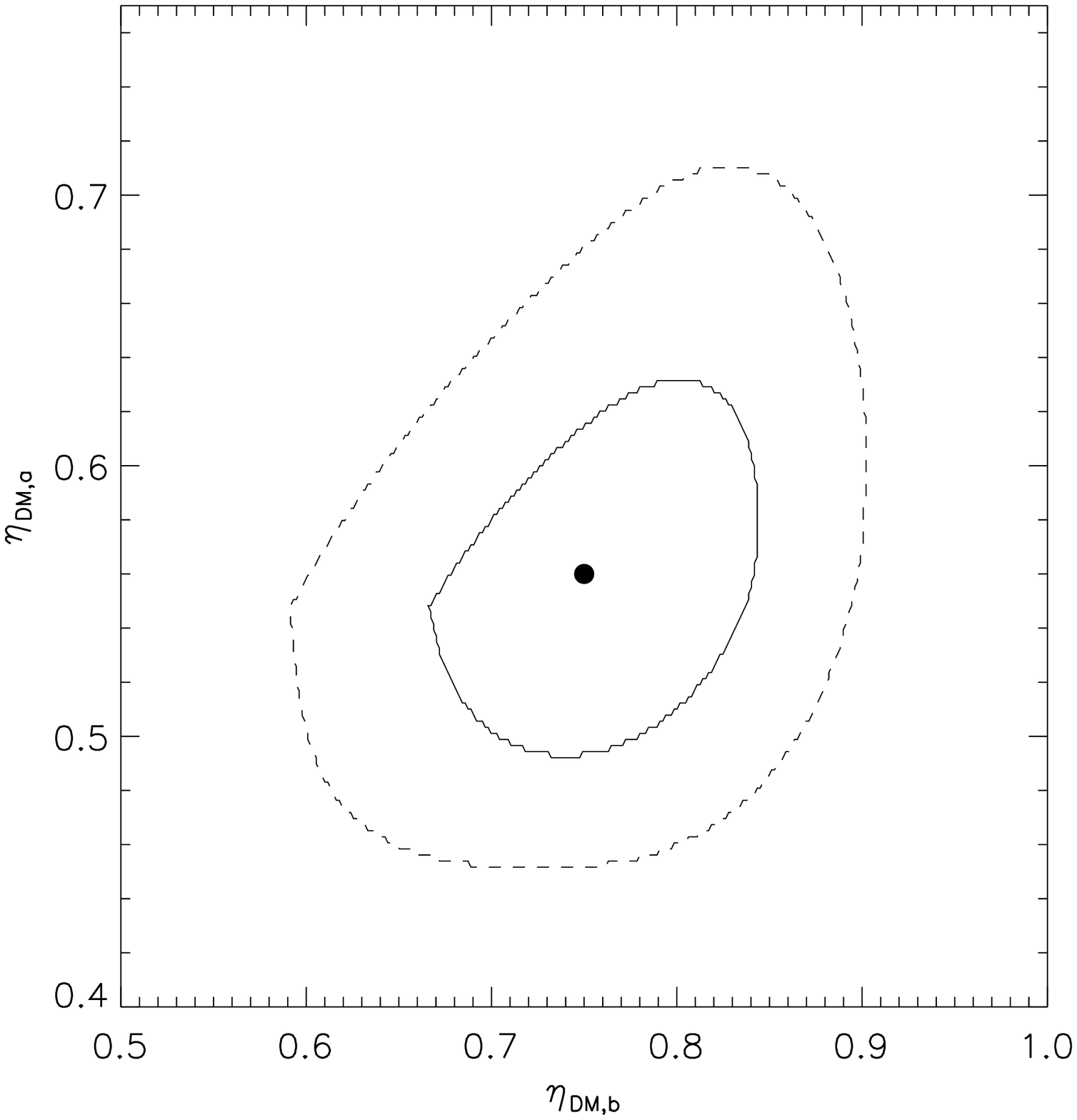}\\
\includegraphics[width=6cm,height=6cm]{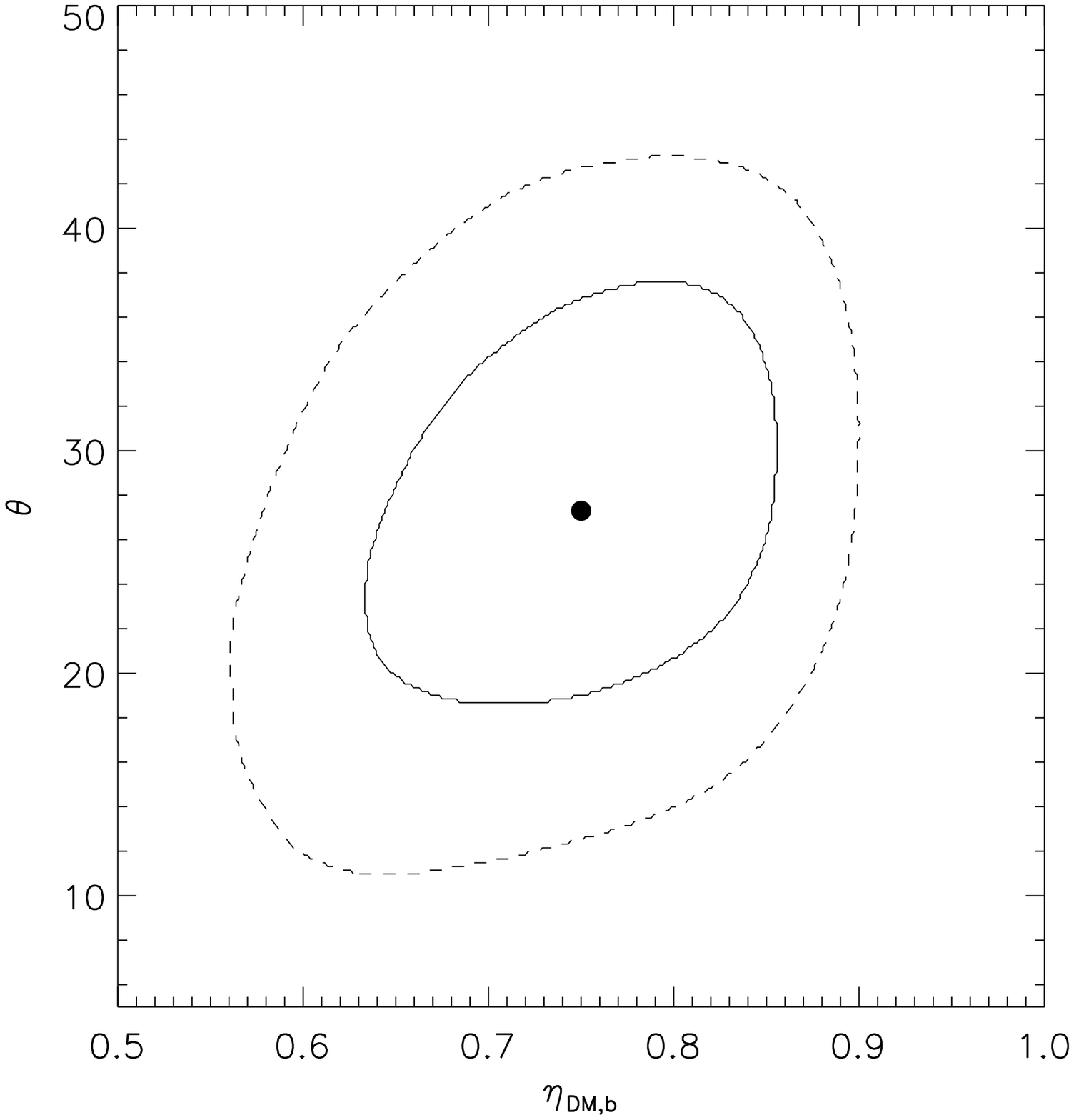}
\includegraphics[width=6cm,height=6cm]{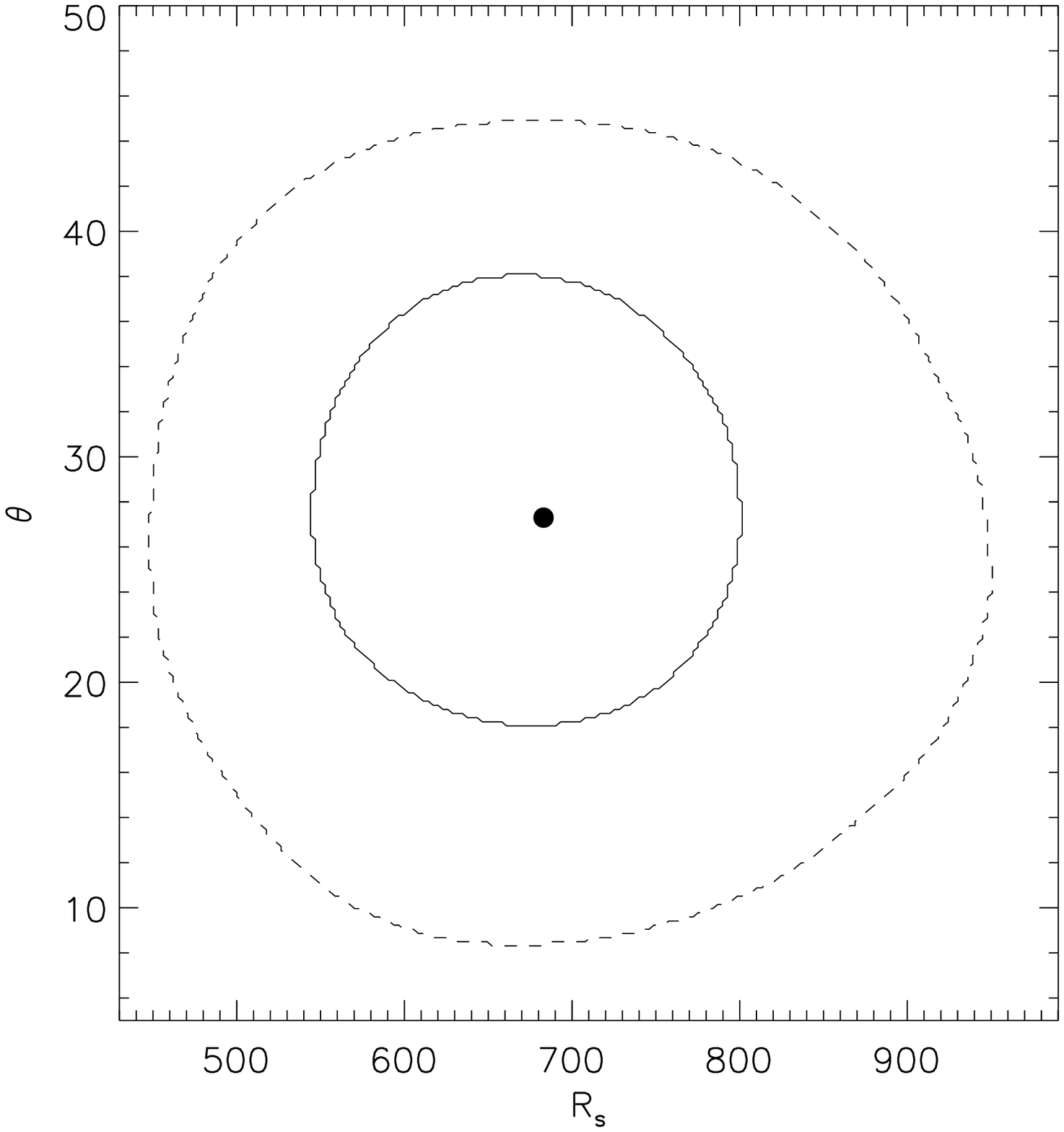}
\includegraphics[width=6cm,height=6cm]{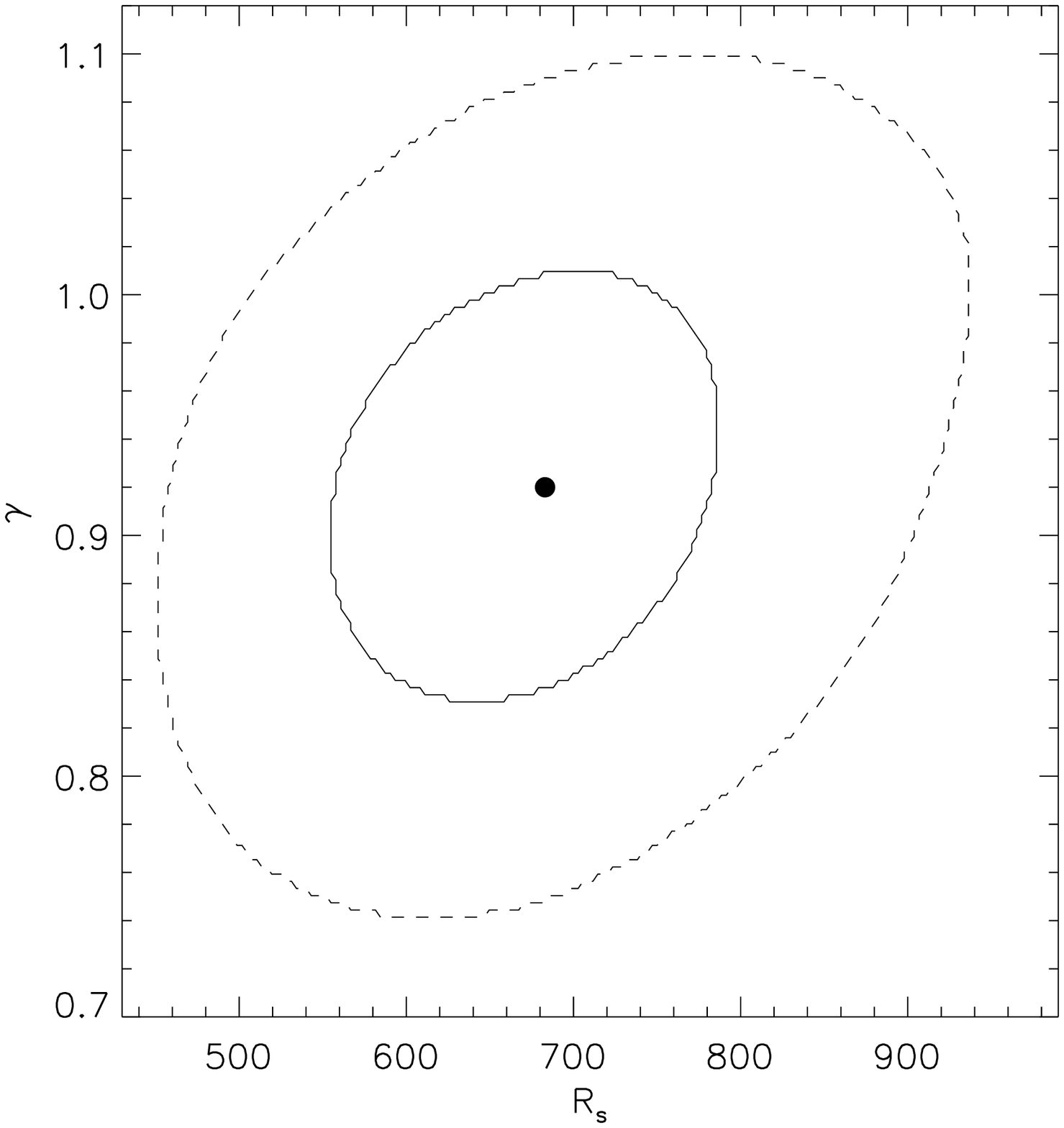}
\includegraphics[width=6cm,height=6cm]{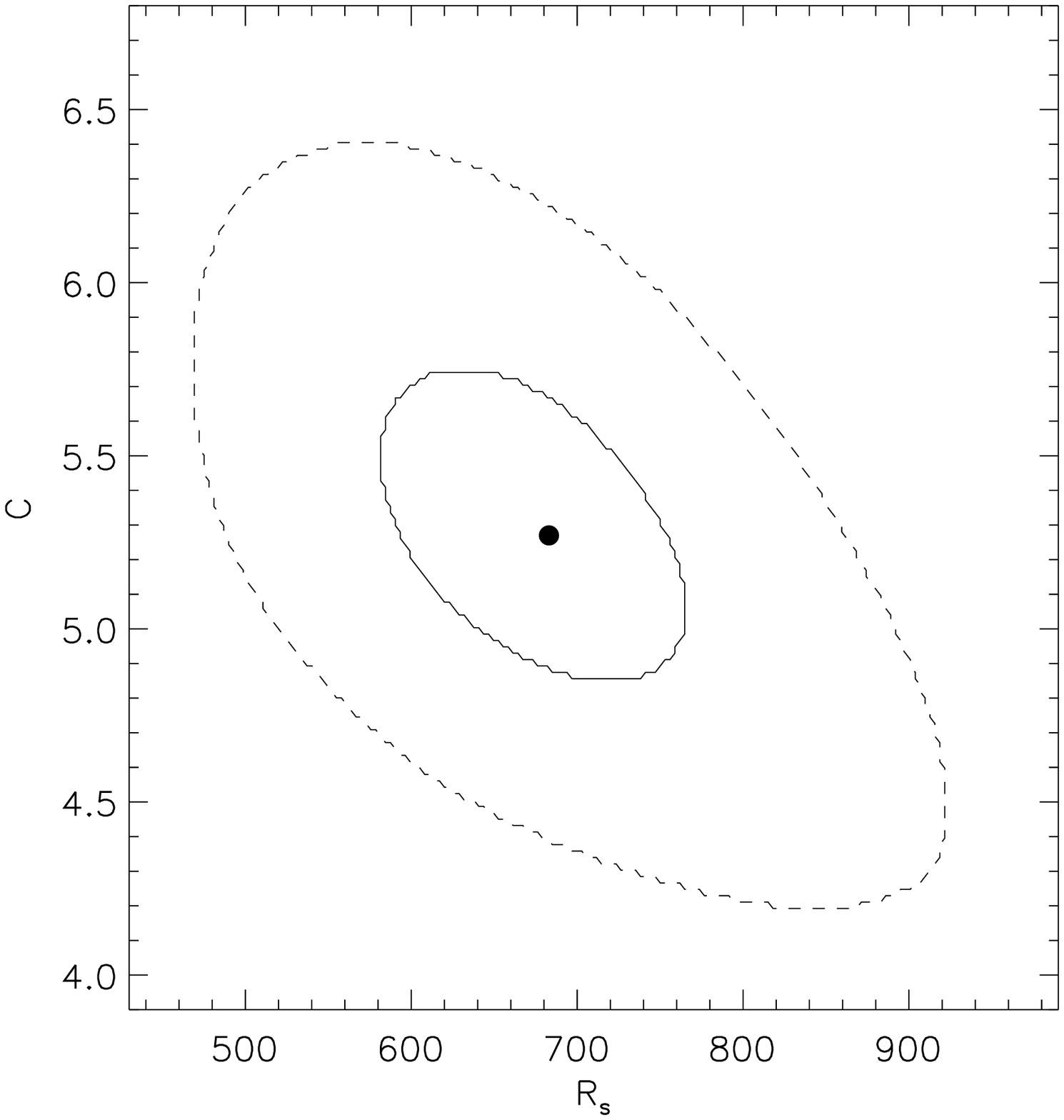}
\caption[]{Marginal probability distribution among different parameters in the triaxial model for Abell~1689 \citep[results from][but updated within the full
triaxial framework presented in this Section]{morandi2011b}.
The solid (dashed) line represent the 1 (2)-$\sigma$ error region, while the point represents the best fit value.}
\label{degen}
\end{center}
\end{figure*}

Degeneracies arise between the different parameters involved in the
modelling. 
Fig.~\ref{degen2} presents the degeneracies expected between $\theta$ and $\xi$.
$\theta$ represents the angle between the major axis of the haloe and our line of sight,
and $\xi$ is related to the amount of non-thermal component considered in the model. 
In Fig.~\ref{degen} we present the joint probability distribution
among different parameters in our triaxial model for Abell~1689.

Regarding the use of priors, we point out that, when combining complementary
data sets as described in this Section, we do not need to rely on any priors
like those discussed in Section~\ref{priors}. 
This is welcome since priors may be potentially biased due to our
incomplete understanding of cluster physics.

\clearpage

\section{Three-Dimensional Mass Distribution in A1689 (and Other Clusters)}
\label{1689comparison}

We aim to illustrate how spherical and triaxial modelling can lead
to different haloe parameters. For this purpose, we focus on galaxy
cluster Abell~1689 which has been extensively studied at different
wavelengths.
Abell~1689 is a massive galaxy cluster at redshift 0.18 with a very large 
Einstein radius, around 45$''$ for a source redshift $\sim 2$. 

Before going into detail, we would like to emphasise the complexity of this
structure and try to justify why we treat such a complicated structure using
a single mass clump.

\cite{lokas} used spectroscopic redshifts to study the kinematics of
about 200 galaxies in the cluster. They showed that the cluster is
probably surrounded by a few structures aligned along the line of sight.
\cite{olliturin} reported redshifts for 525 galaxies, spanning from the 
centre outward to 3$h^{-1}$\,Mpc. They found only one apparently distinct
group of galaxies that lies about 350 kpc to the northeast of the cluster
centre. It corresponds to a group of bright galaxies well identified
in optical images of this cluster. All strong lensing studies
have taken into account the gravitational perturbation it generates,
but its contribution to the total mass budget is found to be small.
The redshift distribution of these galaxies is skewed toward slightly
higher redshifts. On larger scales (R\,$> 1 h^{-1}$\,Mpc), no evidence for
any substructures is found: the outskirts of Abell~1689 look rather
homogeneous. However, deep X-ray observations obtained with Suzaku
\citep{suzaku1689} reveal anisotropic gas temperature and entropy
distributions in the cluster outskirts.

If Abell~1689 appears as a complex structure, the main mass clump
seems to be dominant in the mass budget: all strong lensing studies
find the mass centre to coincide with the brightest cluster galaxy,
which also coincides with the peak of the circular X-ray emission \citep{lemze}.
Besides, deep Chandra data have revealed the presence of a cool core 
\citep{signe}.

\subsection{Spherical Analysis: the Abell~1689 Puzzle}

While it has been proposed as a standard example of a relaxed object in 
hydrostatic equilibrium, Abell~1689 has been the subject of some controversies.
To summarise, as long as spherical symmetry has been assumed, 
this cluster has been problematic in two ways:
\begin{itemize}
\item High concentration parameters (up to $\sim 30$) have been derived from lensing analyses. These large values represent a major inconsistency with the theoretical 
$\Lambda$CDM expectations \citep[$c\sim 3-4$,][]{neto,duffy08}.
\item The 2D mass derived from X-ray data is only half of the mass found using SL estimates.
\end{itemize}

In the following, we discuss these two points and then we turn to triaxial 
models and how they
have allowed to resolve these issues.

The concentration parameter of Abell~1689 has been constrained extensively by different
authors, finding very different results.
In weak lensing analyses, first studies reported 
$c_{200}$= 4.8 \citep{king02a};
3.5$^{+0.5}_{-0.3}$ \citep{bardeau05},
but these analyses likely suffered from dilution of the
inner shear profile by foreground cluster members.
On the other hand, very high concentrations were also inferred from weak lensing
analyses based on Subaru data:
$c_{200}$=30.4 \citep{halkola};
22.1$^{+2.9}_{-4.7}$ \citep{elinor}.
Using the same Subaru data but a different algorithm, other authors
reported a smaller concentration parameter, between 10 and 15
\citep{umetsu09,umetsu11}.
Weak and strong lensing analyses converged to:
$c_{200}$=7.9 \citep{clowe03};
7.6$^{+0.3}_{-0.5}$ \citep{halkola}; 7.6$\pm$1.6 \citep{mypaperIII}, values which 
are still high
compared to the theoretical expectations. Actually, only a few haloes formed in 
the Millennium
simulation could reach this value \citep{neto}.
These values where found in agreement with X-ray analyses: $c_{200}=7.7^{+1.7}_{-2.6}$
\citep{andersson}; $6.6\pm0.4$ \citep{peng}.

Besides the large variance in the concentration parameter, these studies agree on the 
fact that the mass derived from X-ray measurement is half of that found
from strong gravitational lensing at most radii \citep[see also][]{lemze}.

\cite{signe} showed that this discrepancy is reduced if we exclude a cool clump plus
some substructure in the North-Eastern part of the cluster; nevertheless a 
discrepancy still remains in the strong lensing region.

A way to reconcile the mass derived from X-ray and lensing measurement within a
spherical mass distribution is to add the contribution from non-thermal
pressure. In the case of Abell~1689, \citet{molnar} found a contribution
of about 40\%. This is larger than the theoretical expectations by
\citep{shaw2010} but consistent with the set of simulations by \citet{molnar}.
% \cite{meneghetti, nagai}

\subsection{Triaxial Models: Solving the Puzzle}

\cite{oguri}, using a triaxial mass model, found that weak lensing 
measurements in Abell~1689 based on Subaru data are indeed compatible with $\Lambda$CDM
if Abell~1689 represents a rare population ($\sim$\,6\% by number) of cluster-scale haloes. 
\cite{corless09} constrained the triaxial shape of the total mass distribution of Abell~1689 via weak lensing data and under a range of 
Bayesian priors derived from theory, though large errors accompany their 
triaxial parameter estimates.

\cite{morandi2011a} presented the determination of the intrinsic 
shapes and the physical parameters of both DM and ICM in Abell~1689 by 
combining X-ray and strong lensing data. They showed that Abell~1689 can be 
described as 
elongated along the line of sight, with a minor-major principal axis
ratio equal to $0.42\pm 0.02$.
They assumed that the triaxial ellipsoid is oriented
along the line of sight, an assumption justified in light of the 
"orientation bias" of strong lensing clusters (see Section~\ref{slbias}).
A subsequent re-analysis of \cite{morandi2011b}, by jointly 
analysing also weak lensing data and accounting for the non-thermal 
pressure of the IC gas, strengthened the view of a triaxial cluster 
elongated along the line of sight, though a bit larger value of the minor-major
principal axis ratio has been inferred
($0.50\pm 0.01$).
Besides, it was shown in this work that the large Einstein radius observed
was reproduced by a triaxial model with $\Lambda$CDM friendly parameters.
In the present review we extend these
previous works, by allowing the DM and ICM ellipsoids to be oriented in
 an arbitrary direction on the sky
(see Section~\ref{morandi_method} for further details). Our work
indicates that Abell~1689 is a triaxial galaxy cluster with DM haloe axial
ratios $\eta_{{\rm DM},a}=0.56\pm0.07$ and $\eta_{{\rm DM},b}=0.75\pm0.08$,
$c_{200}=5.27\pm 0.46$, and with the major axis slightly inclined with respect
to the line of sight of $\theta=27.3\pm7.1$ deg, in agreement with the
predictions of \cite{oguri09a}.

\cite{sereno_umetsu} developed a method for a three-dimensional analysis of 
the DM haloe via SL and WL data. 
They re-analysed the weak lensing convergence map of A1689 obtained by 
\citet{um+br08} and \citet{umetsu09} on wide-field 
($\sim 30' \times 24'$) Subaru data of A1689. For strong lensing, 
they employed a parametric lensing analysis method based on the {\it gravlens}
kernel evaluation of $\chi^2_\mathrm{SL}$ values \citep{kee01b,kee01a}. 
The SL likelihood values were computed by comparing observed to predicted 
image positions. 
Several priors were considered. For the axial ratios $\eta_{{\rm DM},a}$ and 
$\eta_{{\rm DM},b}$, they considered either the $N$-body predictions or a flat 
distribution. For the alignment angle $\theta$, they considered either the 
biased distribution for $p(\theta)$ (see Equ.~\ref{nbod5}) or a random distribution. 
For the azimuthal 
angle they always used a random flat distribution. For the mass, they always 
used a flat prior $p(M_{200}) = const.$, whereas the a priori PDF for the 
concentration was flat in the range $0 < c_{200} \le 30$ and null otherwise. 
The theoretical $c(M)$ relation from \citet{duffy08} was either enforced 
or neglected. 
%Results are quite independent of priors.

\begin{figure*}
\begin{center}
\includegraphics[width=6cm,height=6cm]{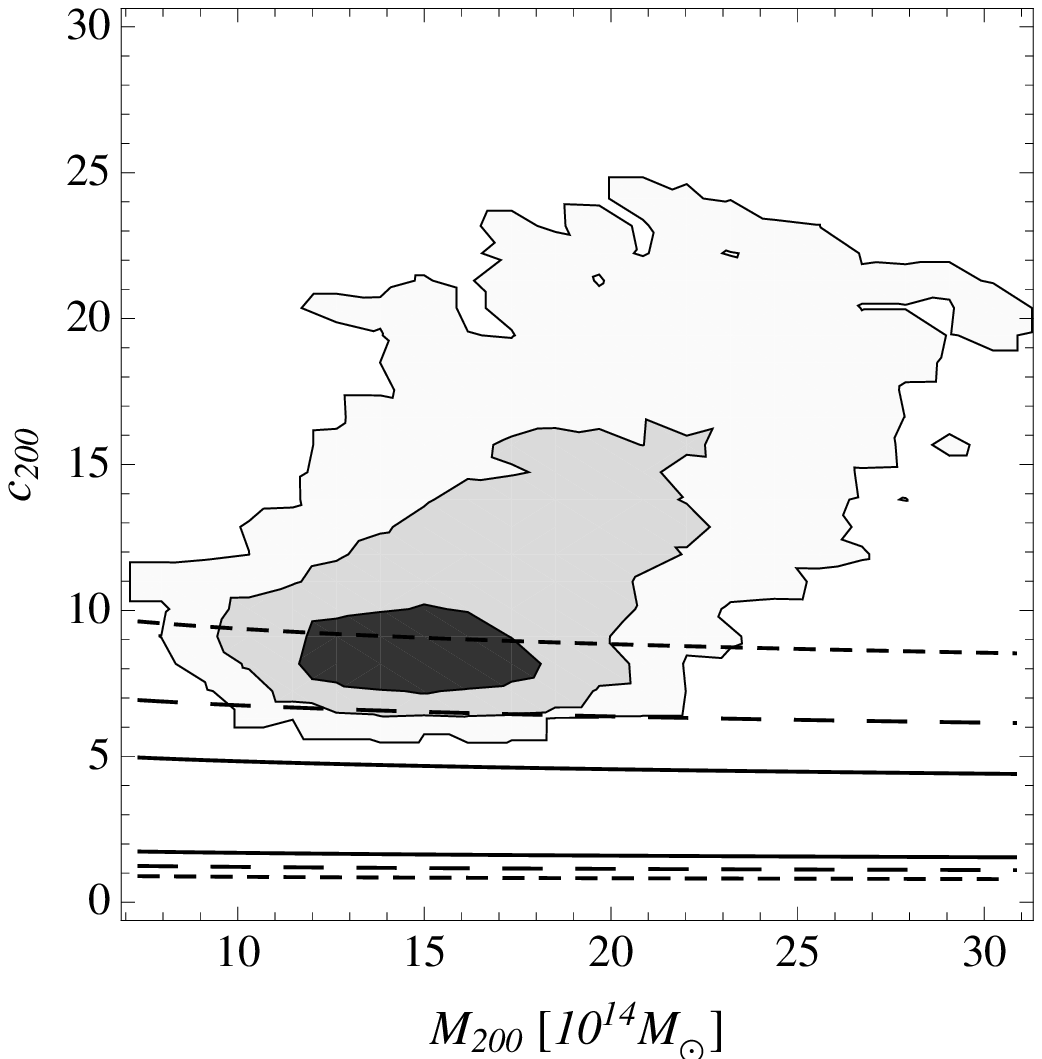}
\includegraphics[width=6cm,height=6cm]{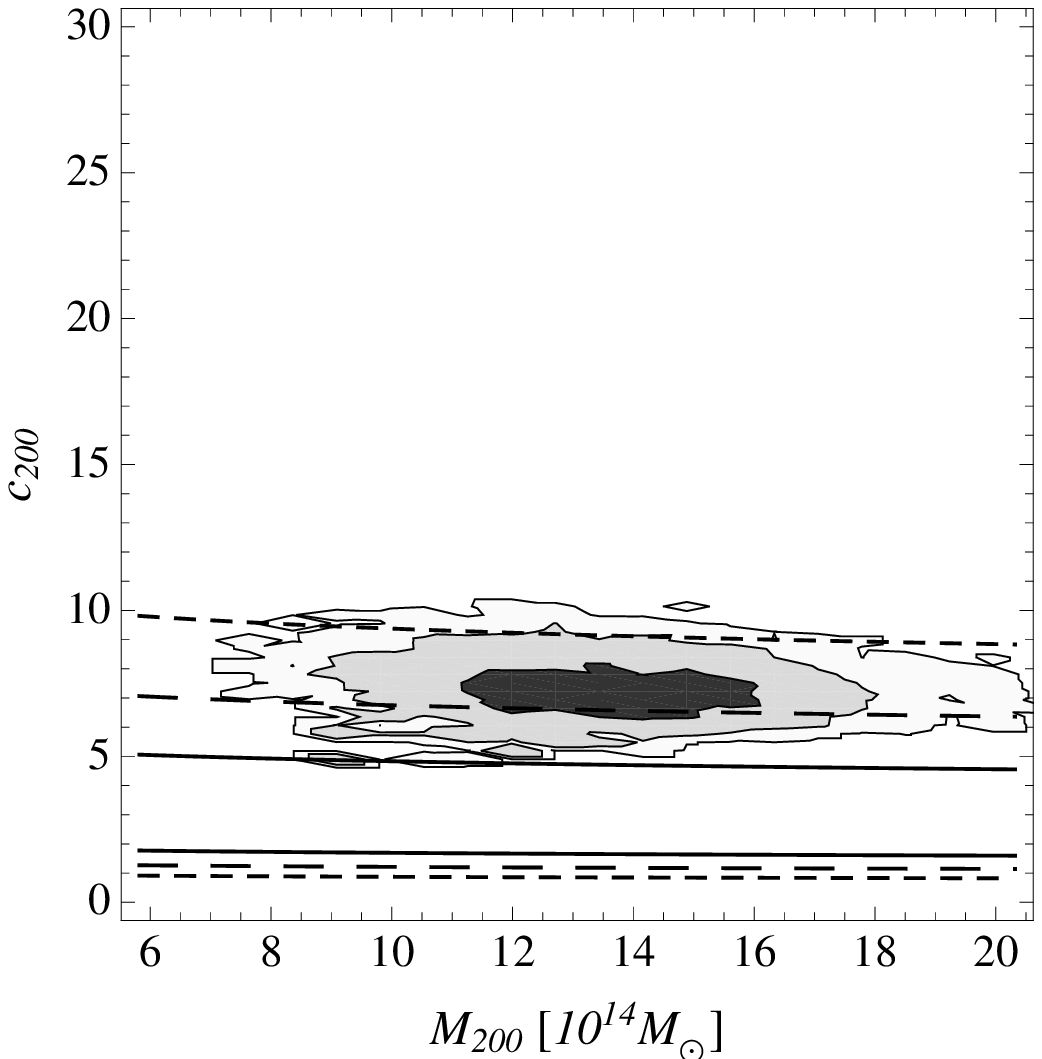}
\caption[]{Results of the combined strong and weak lensing analysis. Contour plot of the 
marginalised PDF for $M_{200}$ and $c_{200}$ for the dark matter haloe as 
derived under the prior assumptions of either flat axis ratio distribution and
random orientation angles (left panel) or $N$-body like axis ratio 
distribution and biased orientation angles (right panel). Contours are 
plotted at fraction values $\exp (-2.3/2)$, $\exp(-6.17/2)$, and 
$\exp(-11.8/2)$ of the maximum, which denote confidence limit regions of 1, 2
and $3\sigma$ in a maximum likelihood investigation, respectively. 
The full, long-dashed and dashed lines enclose the 1, 2 and $3\sigma$ regions
for the predicted conditional probability $c(M)$ by 
\citet{duffy08}, respectively.
Figure adapted from \citet{sereno_umetsu}.}
\label{fig_M200_c200_flat_random_noprior}
\end{center}
\end{figure*}

Whatever the assumptions on either orientation or shape, the concentration is a 
bit larger but still compatible with theoretical predictions 
(Fig.~\ref{fig_M200_c200_flat_random_noprior}). A1689 appears to be a quite 
typical massive cluster with a concentration in agreement with the tail at 
large values of the expected population of clusters of that given mass. 
Independently of the priors, the inferred $c_{200}$ are only $\ge$ 1$\sigma$ 
away from the predicted median value. Priors from $N$-body simulations also 
help to put an upper bound on the concentrations. 

Mildly triaxial haloes do a better job in fitting data than nearly spherical 
lenses. Values of $0.4 \le \eta_{{\rm DM},a} \le 0.8$ are more likely than either 
extremely triaxial or nearly spherical shapes.
Triaxial shapes predicted by $N$-body simulations are in good agreement with 
these results. Axial ratios derived assuming a flat distribution are 
compatible at $1\sigma$ confidence level with predictions from $N$-body 
simulations. Furthermore, they can exclude nearly spherical shapes 
($\eta_{{\rm DM},a} \sim \eta_{{\rm DM},b} \sim 1$) at the $3\sigma$ confidence level. 
Finally, they find indications for an orientation bias.

The minor-major axis ratio is found
in agreement with the findings by \cite{morandi2011b},
confirming a triaxial shape; yet, the value of the concentration 
parameter ($c_{200}=7.3\pm0.8$) is a bit larger than the values presented in 
this review (Table~\ref{sample}).

\begin{figure}
\begin{center}
\includegraphics[width=9cm,height=9cm]{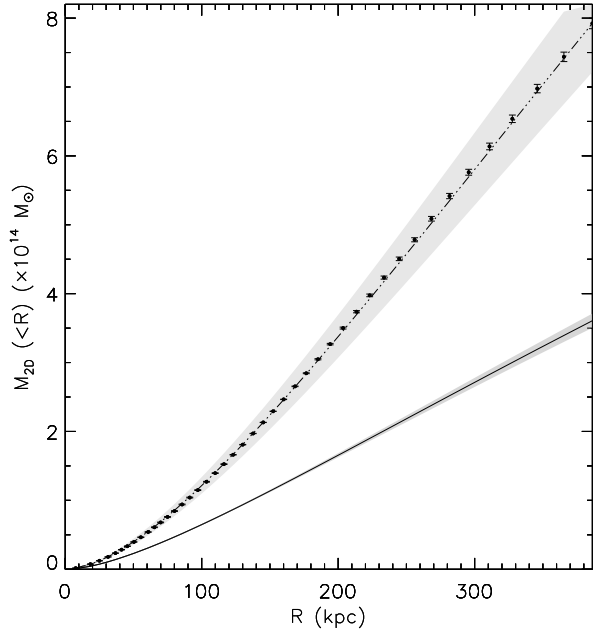}
\caption{2D mass enclosed within a circular aperture of radius R
from lensing data (points with error bars), from an X-ray only analysis under the 
assumption
of spherical geometry (solid line with the 1$\sigma$ error grey
shaded region), and from a joint X-ray+lensing analysis taking
into account the 3D geometry (dot-dashed line with the 1$\sigma$ error cray shaded
region). In this latter case, we see that both estimates
agree with each other.
\citep[Result from][but updated within the full triaxial framework presented in Section~\ref{morandi_method}]{morandi2011a}.}
\label{mass1689}
\end{center}
\end{figure}

Regarding the mass discrepancy, the triaxial model proposed by 
\citet{morandi2011b}, as well as its
extension presented in this review, is able to
solve the X-ray/lensing mass discrepancy.
We plot in Fig.~\ref{mass1689} the comparison between the 2D masses inferred
from X-ray and lensing under spherical and triaxial models.

We have illustrated that a triaxial mass model for Abell~1689 is able to
reconcile mass estimates from different probes, as well as to 
reproduce the large Einstein radius using $\Lambda$CDM friendly
parameters.
It is worth noting that the value of the inner slope of the dark matter
density profile, $\gamma$, also depends on the adopted geometry.
Considering Abell~1689, a standard spherical modelling leads to $\gamma=1.16\pm0.04$,
whereas we find $\gamma=0.92\pm0.07$ using a triaxial mass model (Table~\ref{sample}).

\subsection{The ICM Properties in a Triaxial Framework}
The ICM of Abell~1689 within a triaxial model has also been the subject of interest.

\cite{filippis05} and \cite{filippis06} investigated the ICM shape by
 combining X-ray and SZ observations.
Given the data sets used (in particular, the absence of lensing data), 
error bars were pretty large and a wide range of
geometries was possible:  
they concluded that these data sets are compatible both with a 
prolate and an oblate shape.

More recently, \cite{sereno_ettori} 
implemented some significant improvements by using Bayesian inference to 
determine the intrinsic form: $i)$ The method was still parametric but did not
rely anymore on the simple isothermal $\beta$ model for the X-ray data. 
The employed profiles can
mimic complex features in either the electronic density or the temperature 
profile for the X-ray data. $ii)$ Instead of the central Compton parameter $y_0$, they considered
the more reliable integrated Compton parameter. $iii)$ Even if astronomical 
deprojection is an under-constrained problem \citep{ser07}, 
\citet{sereno_ettori} could infer the 3D structure of
the cluster using a Bayesian method without assuming any specific configuration. 
On the other hand in 
\citet{filippis05} and \citet{filippis06}, the 3D distribution was assumed to
be either triaxial and aligned with the line of sight or prolate or oblate. 

This method was applied to Abell~1689, where SZ and X-ray observations cover
in detail a region R\,$\le 1~$Mpc. 
The 3D electron density and temperature were modelled with parametric 
profiles. Distributions were assumed to be coaligned and ellipsoidal, with 
constant eccentricity and orientation. Intrinsic profiles were taken from 
\citet{vik+al06}, and \citet{ett+al09}. The metallicity was fixed to the mean observed 
value. The elongation $e_\Delta$  enters when 3D profiles are projected into 
the plane of the sky, so that fitting at once X-ray surface brightness, 
temperature, and the integrated Compton parameter, one
can infer $e_\Delta$ as well as the parameters describing the distribution. 

The combined X-ray plus SZ analysis allows to infer the width of the cluster 
in the plane of the sky (parametrised in terms of the ellipticity $\epsilon$)
and its size along the line of sight (expressed as the elongation $e_\Delta$).
These two observational constraints have to be used to infer the intrinsic 
shape of the cluster and its orientation.

\citet{sereno_ettori} found a minor to major axial ratio for the ICM of 0.7$\pm$0.15,
preferentially elongated along the line of sight.
The hydrostatic equilibrium is not involved in their analysis since they do not derive 
the mass of the cluster.
This value of the axis ratio is in good agreement with the one 
inferred by \cite{morandi2011b} (assuming generalised hydrostatic equilibrium), 
who quoted a ratio for the ICM between 0.66 and 0.77 (since the method includes a
radial variation of the axial ratios).
This suggests that both approaches are consistent, and that the ICM in
Abell~1689 is not far from being in hydrostatic equilibrium.
Note that \citet{sereno_ettori} use XMM data for the temperature, whereas \citet{morandi2011b}
use a Chandra temperature, which is lower than the XMM temperature by 10-20\% due
to calibration reasons.

It is worth noting that several processes like radiative cooling and
turbulences are affecting the shape of the gas distribution more than
the dark matter shape.

%According to \cite{sereno_ettori} and \cite{sereno_ettori}, the ICM
%in hydrostatic equilibrium under the gravity of a triaxial dark matter
%halo with an axis ratio of 0.5$\pm$0.2
%\citep[as consistently inferred by][]{sereno_umetsu,morandi2011a} 
%should have an axis ratio of 0.8$\pm$0.1.

\subsection{Overview of the Present Results on Clusters Triaxiality}
\begin{figure}
\begin{center}
\includegraphics[width=6cm,height=6cm]{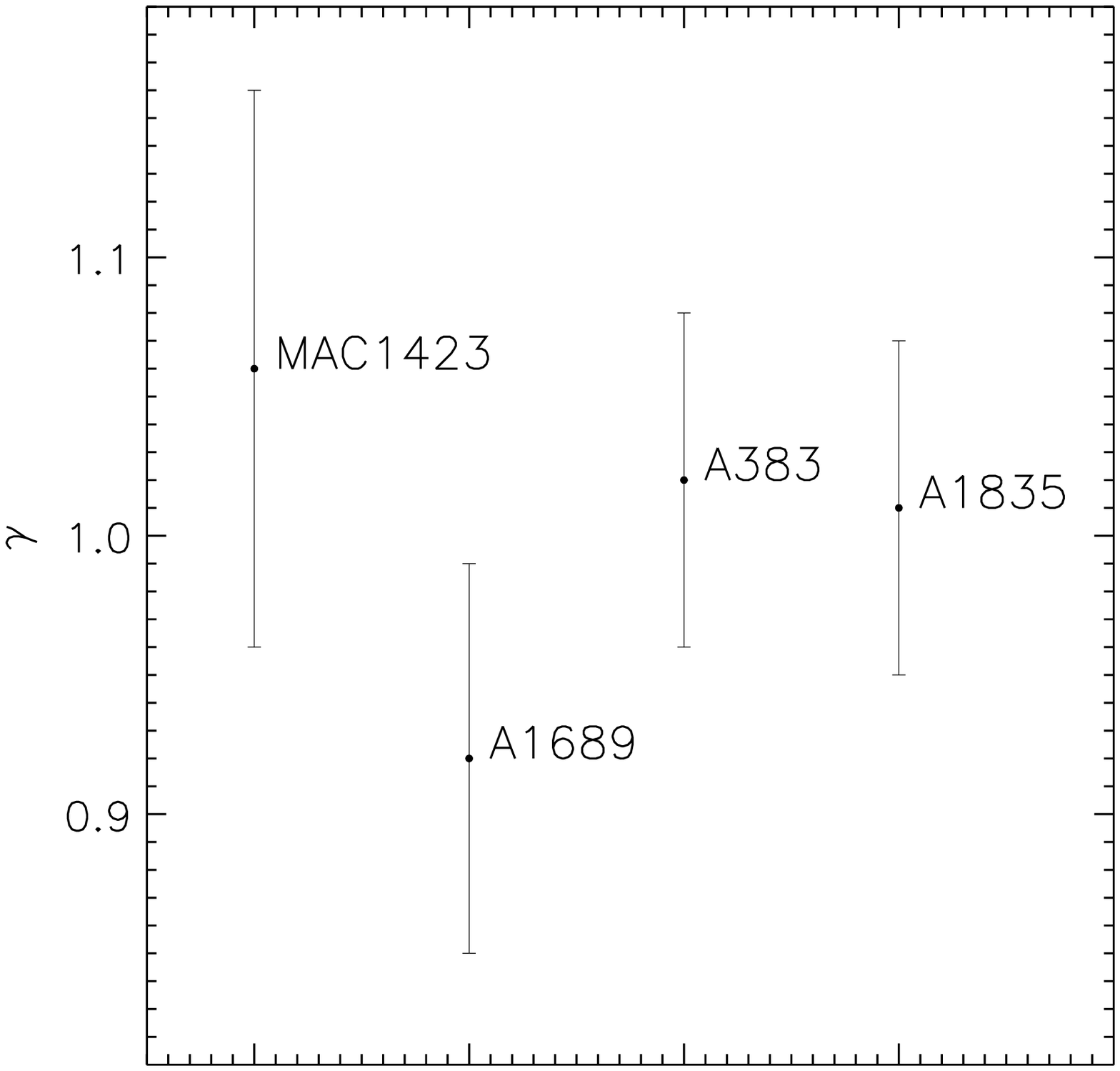}
\includegraphics[width=6cm,height=6cm]{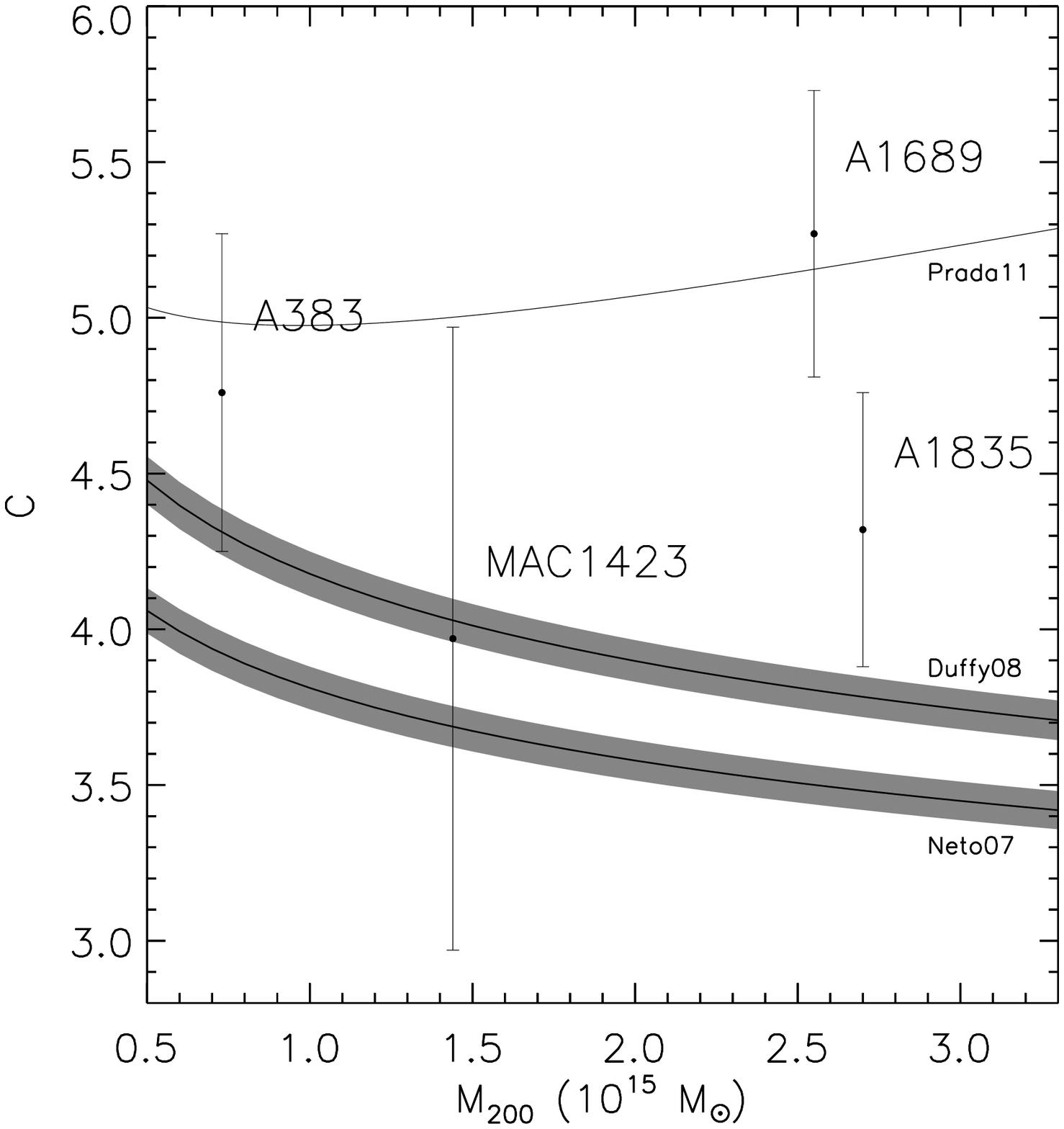}
\caption{Properties of the four clusters studied within the full triaxial 
framework described in Section~\ref{morandi_method}. \emph{Left:} inner 
slope of the dark matter density profile. 
\emph{Right:} Masses and concentrations. Results from $\Lambda$CDM based
simulations \citep{neto,duffy08,prada11} are shown.
Error bars correspond to 1$\sigma$ confidence levels.
We caution that this comparison can be misleading since results from numerical
simulations are based on a spherical assumption.}
\label{samplefigs}
\end{center}
\end{figure}

So far, four clusters have been studied within the full
triaxial framework described in Section~\ref{morandi_method}: 
MACS\,1423, Abell~1689, Abell~383, and Abell~1835. 
The results published for MACS\,1423 \citep{morandi2010a} 
and Abell~1689 \citep{morandi2011a,morandi2011b} were derived using
a triaxial model where the haloe's major axis was aligned with our line of
sight. Since then, the algorithm was improved and we report
the results for MACS\,1423 and Abell~1689 derived using the full triaxial
framework.
Results of the parameters for each cluster are given in
Table~\ref{sample}.
As discussed in the relevant papers, the parameters obtained through a
triaxial model can be very different from the one obtained through a
spherical model.
In particular, regarding the concentration parameter, similar to what has been
found for Abell~1689,
lower values are inferred within a triaxial framework.
Shown on Fig.~\ref{samplefigs} are some key properties of these four clusters:
the inner slope of the dark matter density profile, and the masses and
concentrations.
The comparison with results from numerical simulations will be discussed
in Section~\ref{limitations}.

\begin{table*}
\begin{center}
\caption{Best-fit model parameters for the four clusters for which a full triaxial
modelling exists.
Error bars correspond to 1$\sigma$ confidence level.
The lines $1-8$ refer to the best fit parameters of the DM haloe:
$c_{200}$ (concentration parameter), $R_{\rm s}$ (scale radius), $\gamma$ (inner DM slope), 
$\eta_{{\rm DM},a}$ (minor-major axis ratio), $\eta_{{\rm DM},b}$ (intermediate-major axis ratio), 
and $\psi,\theta,\phi$ (Euler angles).
The lines $9-14$ refer to the best fit parameters $n_0,r_{c_1},\varepsilon,\delta,r_{c_2},\upsilon$ of the IC gas density, 
while the lines $15-16$ to the best fit parameters $\xi,n$ 
(normalisation and slope, respectively) of the non-thermal pressure. Finally, 
the last line refers to the best fit parameter $\tilde P$ of the pressure at $R_{200}$, 
which is a boundary condition of the generalised HE equation (see relevant Equations in
Section~\ref{morandi_method}).
Note that only in the case of Abell~1835 all these parameters are constrained. This is due to
the inclusion of SZ data.
}
\begin{tabular}[h!]{ccccc}
\hline
\noalign{\smallskip}
Cluster & Abell~1835 & Abell~383 & Abell~1689 & MACS\,1423 \\
\noalign{\smallskip}
\hline
\noalign{\smallskip}
$c_{200}$ & $4.32\pm 0.44$ & $4.76\pm 0.51$  & $5.27\pm 0.46$  &  $3.97\pm 1.0$\\
\noalign{\smallskip}
$R_{\rm s}$ (kpc) & $891.0\pm114.3$   &  $511.2\pm 73.6$  & $683.1\pm84.7$  &  $644.7\pm162.1$\\
\noalign{\smallskip}
$\gamma$ &  $1.01\pm0.06$  & $1.02\pm 0.06$  & $0.92\pm0.07$  & $1.06\pm0.1$\\
\noalign{\smallskip}
$\eta_{{\rm DM},a}$ &  $0.59\pm0.05$  & $0.55\pm0.06$  & $0.56\pm0.07$  & $0.62\pm0.04$\\
\noalign{\smallskip}
$\eta_{{\rm DM},b}$ &  $0.71\pm0.08$  & $0.71\pm 0.10$  & $0.75\pm0.08$  & $0.72\pm0.06$\\
\noalign{\smallskip}
$\psi$ (deg) &  $3.8\pm4.6$  & $-13.6\pm5.5$  & $-35.5\pm13.7$  & $-34.4\pm5.4$\\
\noalign{\smallskip}
$\theta$ (deg) & $18.3\pm5.2$   &  $21.1\pm10.1$ & $27.3\pm7.1$  & $34.7\pm8.7$\\
\noalign{\smallskip}
$\phi$ (deg)  & $-55.0\pm6.9$   &  $-16.9\pm15.9$ & $-11.1\pm6.7$  & $-72.3\pm8.3$ \\
\noalign{\smallskip}
$n_0$ (cm$^{-3}$) & $0.018\pm0.002$   &  $0.063\pm0.003$ & $0.017\pm0.001$  & $0.15\pm0.02$\\
\noalign{\smallskip}
$r_{c_1}$ (kpc)  & $117.7\pm10.1$   & $26.4\pm1.7$  & $119.3\pm5.3$  & $20.6\pm3.1$\\
\noalign{\smallskip}
$\varepsilon$ & $0.68\pm0.02$ &  $0.55\pm0.01$  & $0.72\pm0.02$  & $0.55\pm0.02$\\
\noalign{\smallskip}
$\delta$ &  $0.82\pm0.03$  & $0.02\pm0.01$  & $0.33\pm0.01$  & $0.02\pm0.01$\\
\noalign{\smallskip}
$r_{c_2}$ (kpc) & $1674.3\pm266.7$   & -  & -  & -\\
\noalign{\smallskip}
$\upsilon$ & $0.44\pm0.04$ &  - & -  & -\\
\noalign{\smallskip}
$\xi$ & $0.177\pm0.065$   & $0.11\pm0.05$  & $0.24\pm0.05$  & $0.08\pm0.03$\\
\noalign{\smallskip}
$n$  & $0.77\pm0.21$   & 0  & 0 & 0\\
\noalign{\smallskip}
$\tilde P$ (erg/cm$^{3}$) & $(2.7\pm0.7)\times 10^{-13}$   & -  & -  & -\\
\noalign{\smallskip}
\hline \\
\end{tabular}
\label{sample}
\end{center}
\end{table*}

\section{Discussion}

\subsection{(Limited) Comparison with Simulations}
\label{limitations}

The parameters of a triaxial model obtained for galaxy clusters are to be
compared to results from numerical simulations in order to test cosmological
models.
Such comparisons are routinely performed in the \emph{spherical} case.
In order to build a density profile, simulators usually do a spherical fit to their
elongated profiles: they determine the enclosed mass or density assuming spherical
symmetry even though the haloes are found to be triaxial and asymmetric
(Section~\ref{not_spherical}).
Since triaxial analyses will have a growing importance on the observational side
in the future, we advocate the need for simulations to be analysed within a
triaxial framework, allowing reliable and meaningful comparisons.

{\emph{
Actually, the bias induced by fitting a spherical analytic formula to an elongated
mass profile should be compared to the intrinsic scatter in the concentration
of NFW haloes.
}}

Having stressed the possible limitations of comparing the triaxial observational results with
spherical numerical results,
we can go back to Fig.~\ref{samplefigs} where some key properties of four clusters
studied within a full triaxial framework are presented:
the inner slope of the dark matter density profile, and the masses and
concentrations.
(Spherical) theoretical expectations \citep{neto,duffy08,prada11} are shown on top
of the triaxial results.

Regarding inner slopes, one should keep in mind that, from the theoretical side,
numerical simulations did not converge on a proper treatment of the baryonic 
component which is a difficult task.
In this respect, when the baryonic component is
taken into account, different simulations infer either a steepening or a 
flattening of the dark matter inner density profile 
(see the revue on cluster cores in this volume).

We have mentioned that simulations do not provide the relevant material to compare triaxial results to theoretical expectations.
Moreover, as far as massive haloes are concerned, statistic is poor.
\cite{oguri09b} argued that the probability distribution of the concentration parameter
for very massive haloes (larger that 10$^{15}$\,M$_{\odot}$) and its redshift evolution had not
been studied with great statistics by N-body simulations.
It is worth noting that this statement also applies to extreme axis ratios in
different mass and redshift bins.
Therefore, we rely on extrapolations from smaller haloes. 
More reliable theoretical predictions at the high mass end are needed.
Recently, \cite{prada11} argued that previous simulation works underestimated the mean
concentration at the high mass end and find an upturn in the (c,M) relation.
\cite{bhattacharya}, in their analysis of an even larger simulation, find no evidence for 
such an upturn but they do find a (c,M) relation that differs in normalisation and shape
from previous studies that have limited statistics in the upper mass range.
Moreover, as already mentioned earlier, this upturn may be due to the inclusion
of dynamically disturbed merging clusters \citep{ludlow}.

The situation is changing and the most recent simulations contain
a few hundreds of such haloes. For example,
\cite{angulo}, based on the Millenium-XXL dark matter only simulation, reported 464 haloes
more massive than 2$\times$10$^{15}$\,M$_{\odot}$, in agreement with analytical
calculations \citep[see, \emph{e.g.}][]{mortonson}.

Another difficulty is that the expected abundance of massive haloes strongly depends
on the algorithm used to find them.
\cite{angulo} show that this abundance changes by a
factor of $\sim$ 2 when Friends-of-Friends \citep{davis85} selected subhaloes and self-bound
subhaloes are compared.
They argue this is in part a consequence of large haloes not forming an homogeneous population.
Indeed, clusters displaying similar virial masses present a considerable diversity in shape,
concentration, and amount of substructure.

%Coe2261:
%they do a spherical fit to their elongated profile; a method questionable
%but consistent with that generally used to measure the mass profile of simulated
%clusters:
%enclosed mass or density is determined assuming spherical symmetry even
%though the haloes are triaxial and asymmetric.
%Ideally, simu should be analysed in the same way allowing for direct comparisons.

To draw reliable conclusions regarding the properties of clusters and their evolution
with cosmic time, we would need to study within a triaxial framework a large number
of (massive) galaxy clusters.
These results should be compared to predictions from simulations inferred within a
triaxial framework.

\subsection{On Which Scales do we Measure Triaxiality?}
The scale on which a triaxial model is constrained depends on the data sets used.
In the case of Abell~1835, the triaxial model has been constrained up to the virial
radius thanks to the inclusion of SZ data.
This question is linked with the following one:
out to which radius do we have observational data with a sufficiently high signal to
noise ratio?
Strong lensing is limited to the very central part of a galaxy cluster.
Regarding X-ray data obtained with current facilities, one can expect to reach
R$_{500}$. 
Regarding weak lensing, beyond R$_{500}$, the signal to noise ratio becomes small, and
the lensing signal estimated for a single cluster is likely to be dominated
by mass not associated with this cluster, either correlated or uncorrelated
\citep{henk2011}.
This question is also relevant when we compare with triaxial parameters
measured on simulated clusters. Ideally, we aim to compare measurements performed
on the same scales.

\subsection{Implementing Triaxiality in the Mass Distribution or in the
Gravitational Potential?}
\label{massorpotential}

There are two possible approaches when introducing a triaxial model for a galaxy cluster:
one can either implement triaxiality directly in the mass distribution, or in the
gravitational potential.

An ellipsoidal mass distribution generates isopotential surfaces which are approximated 
by concentric ellipsoids of decreasing axial ratio towards the outer volumes, and not of constant shape.
Vice versa, if the potential is described by a triaxial ellipsoid, then the mass
distribution giving birth to this potential will not be described by a triaxial
ellipsoid.

Both approaches have their advantages and inconveniences and have been followed by different authors.

\subsubsection{A Triaxial Mass Distribution}

We have discussed in the introduction that
%either from first principles, either
%from results of numerical simulations,
the DM distribution behaves as an ellipsoid, being collisionless and dominating in term of mass.
The dynamics of the gas (which is the second largest contribution to the mass after the DM) 
should be driven by the gravitational potential 
well of the DM.
Therefore, it seems well motivated to implement triaxiality in the mass distribution as proposed in
Section~\ref{morandi_method}, being aware that it has its own limitations.

Recent investigations of galaxy scale dark matter haloes forming in the
Aquarius simulation \citep{simu_aquarius} show that the shape of a haloe depends
strongly on its environment, the time at which we consider it, and the radius
at which we measure its shape \citep{aquarius}. At redshift 0, haloes exhibit a
variation from prolate in the inner regions to triaxial/oblate in the outskirts,
which clearly complicates the modelling. If these results are derived for galaxy
scale dark matter haloes, we can expect that some of these complications also
arise for cluster scale dark matter haloes.
Indeed, \cite{munozcuartas} found that the asphericity is more pronounced in the
haloe's central region (i.e. for radii smaller than 30\% of the virial radius).
However, internal regions are more likely to be affected by baryonic physics and
therefore the inclusion of baryons is needed in order to draw more quantitative
conclusions.
 
It is worth noting that models which involve a gravitational potential 
generated by an ellipsoidal mass distribution are computationally expensive. 

\subsubsection{A Triaxial Potential}

\citet{buote11a,buote11b} investigated a different type of model where the potential,
rather than the underlying mass distribution, is an ellipsoid of constant shape and orientation. 
They inferred analytical formulae for galaxy clusters with triaxial potential. 
These triaxial models 
themselves lead to a straightforward generalisation of analytic spherical models, and hence they are 
computationally fast.

In a similar fashion, \citet{sereno_ettori} modelled the gas density distribution 
as ellipsoid of constant shape and orientation. 
Note that they did not make any assumption about the equilibrium of the gas.

There is both observational and theoretical work suggesting that this 
modelling has its limitations
and which shows that the gas distribution follows isopotential surfaces
approximated by concentric ellipsoids of decreasing axial ratio towards the outer volumes
\citep{buote1994,buote96,lee2003,hayashi07,morandi2010a,kawahara,deb12}.

\subsubsection{So ?}

It appears that, when the accuracy of the simulations or the observations increases, none
of the mass distribution nor the potential can be rigorously described by a triaxial model.
Indeed, \cite{deb12} showed in Abell~1689 that the axial ratio of both the DM and the ICM is not
strictly constant with radius.

It is worth noting that, in Abell~1689, Morandi et~al. and Sereno et~al., using different modelling 
approaches find results for the shape of the ICM
in pretty good agreement, which proves that both approaches are consistent and deserve additional
investigation (see comparison in Section~\ref{1689comparison}).

\subsection{Large Einstein Radii \& the Overconcentration Problem}

The first claim regarding the fact that large Einstein radii may be a problem for
the $\Lambda$CDM scenario 
was made by \cite{largeRe}.
This claim was very constructive since it led to a number of interesting studies 
investigating if strong lensing
clusters were over-concentrated with respect to theoretical expectations, and if
the large Einstein radii observed for some extreme clusters were challenging 
the $\Lambda$CDM scenario.
The problems of large Einstein radii and of overconcentrated clusters are somehow linked to
each other: a very elongated haloe with its major axis aligned along the line of sight
can lead to a highly concentrated projected surface mass density profile resulting in a
large tangential critical curve, hence a large observed Einstein radius.

\cite{oguri09b} studied the mass profiles of four clusters by combining strong and
weak lensing data. The mass profiles were found to be well described by an NFW
profile. They found values for the concentration parameter that are slightly
higher than the $\Lambda$CDM predictions, even after taking into account the lensing
bias which includes the projection effect.
Taking into account the error bars they derived, the value of the concentration
for each cluster is marginally consistent with theory, and the excess is not so strong
compared to earlier claims.
Then they add to their sample six clusters from the literature which have strong plus weak
lensing analyses available. Considering the 10 clusters, they claim a 7$\sigma$
excess of the concentration parameter compared with the $\Lambda$CDM predictions.

\cite{okabe10} conducted a weak lensing analysis of a sample of 30 X-ray
luminous galaxy clusters and found a mean concentration 
$c_{\rm vir} = 3.48^{+1.65}_{-1.15}$ for clusters with 
M$_{\rm vir}\sim 10^{15} h^{-1}$\,M$_{\odot}$, displaying no over-concentration.
However, it is worth noting that
the inclusion of strong lensing data is important for an
unbiased value of the concentration parameter.
Indeed, background galaxy catalogues may have an uncontrolled degree of contamination
from unlensed cluster members. This will dilute the shear profile in the inner region and
bias low the inferred concentration parameter.

\cite{sereno_overconcentrated} investigated the over-concentration problem 
on a sample of 10 strong lensing clusters.
They derive which elongation along the line of
sight is needed by the data in order to be compatible with the mass-concentration
relation.
Half of the clusters of their sample support the expectation from $\Lambda$CDM
simulations, being triaxial haloes with a strong orientation bias.
The other half would fit the mass-concentration relation only if they were 
characterised by a filamentary structure extremely elongated along the line of sight,
which the authors consider very unlikely considering standard scenarios of structure
formations.

\cite{oguri11} studied the mass distribution of 28 galaxy
clusters using strong and weak lensing observations. The sample is made of clusters
selected via their strong lensing properties: the Sloan Giant Arc Survey.
They found the inferred mass-concentration relation for these clusters to be in 
reasonable agreement with the simulations for very massive haloes 
(M$_{\rm vir} \sim 10^{15} h^{-1}$\,M$_{\odot}$).
However, they found that the observed concentrations are much higher than theoretical
expectations for less massive haloes
(M$_{\rm vir} \sim 10^{14} h^{-1}$\,M$_{\odot}$),
even after taking into account the mass dependence of the lensing bias.

%It appears that triaxiality is an important ingredient to take into
%account regarding the over-concentration problem and helps to reconcile 
%theory with observations (see the case of Abell~1689 discussed in Section~\ref{1689comparison})

We have seen that, even after accounting for projection biases, 
an over-concentration problem 
remains in some lensing clusters \citep[see also][]{kerenSZ}.

Another problem concerns the \emph{number} of over-concentrated
clusters one can find in a given survey.
This is an open question worth investigating in detail.
We have seen that Abell~1689 by itself does not pose a severe challenge to the $\Lambda$CDM model,
but such high values of the concentration parameter appear to be common
in the combined strong plus weak lensing analyses of massive clusters.

We now come back to the Einstein radius, which provides a relatively model-independent measure
of the mass density of a cluster core, and is observationally easier to infer than 
the concentration parameter.
To date, the largest Einstein radius has been observed in MACS\,J0717.5+3745, hereafter
MACS\,J0717 \citep{adi0717,my0717}. Its effective Einstein radius was estimated to be
55$\pm$3" for a source redshift of $z\sim2.5$.
\citet{adi0717} claimed that the probability to find such a system in a $\Lambda$CDM Universe is
very unlikely, of the order of 10$^{-7}$.
Recently, \citet{waizmann12} modeled the distribution function of the single largest
measured value of the
Einstein radius in a given cosmological volume based on a Monte Carlo
approach. Results are fitted with the general extreme
value distribution. They showed that the large Einstein radius in MACS\,J0717 does not 
exhibit tension with $\Lambda$CDM, even if they neglect the impact of dynamical merging
which is clearly established for MACS\,J0717, and which would allow large Einstein radii
more likely to be found \citep{redlich12}.
This finding is at odds with the claims by \citet{adi0717}, and we refer the reader to
\citet{waizmann12} for a discussion of the difference of calculations between 
each studies.
They concluded that, for an observed Einstein radius to challenge $\Lambda$CDM, one should
observe Einstein radii larger than 100".

Besides, \citet{waizmann12} investigated the influence of triaxiality on the resulting
extreme value distributions of the largest Einstein radii, and find that it is very sensitive
to very elongated objects. In agreement with \citet{oguri09a}, they confirm that the 
single largest Einstein radius has not necessarily its origin in the most massive clusters.
Instead, triaxiality, together with the haloe orientation, has a stronger impact than the mass
of the cluster itself.

\subsection{Triaxiality \& Self Interacting DM}

The asphericity of galaxy cluster scale DM haloes demonstrates that the
self-interaction of DM particles cannot be too large.
\cite{yoshida2000} investigated how the internal structure of dark haloes
is affected if cold dark matter particles are assumed to have a large cross
section for elastic collisions.
It results in a cluster that is more nearly spherical at all radii compared
to the collisionless case.
More recently, \citet{peter12} and \citet{rocha12} presented cosmological
simulations with self-interacting dark matter (SIDM). They showed that the currently observed
asphericity allow dark matter self-interaction cross section at least as large as
$\sigma/m = 0.1$\,cm$^{2}$/g.

\subsection{Dynamical Information}

A spectroscopic campaign targeting cluster members is highly relevant when
studying a galaxy cluster and its 3D geometry.
Indeed, it can give us some clues about the dynamical state of the system.
This is particularly relevant since the more relaxed and unimodal the cluster, the 
more legitimate
to describe it with a single mass clump (whatever the geometry considered).

For example, in the galaxy cluster Cl\,0024, an apparently relaxed cluster in X-ray and
lensing data sets, \citet{olli00241} discovered that this cluster is composed of two
mass clumps aligned along the line of sight and argued that a high-speed collision
had taken place \citep{olli00242}.

\section{Conclusions: Galaxy Clusters are \emph{not} Triaxial}

We have discussed that a triaxial model is clearly an important step forward 
to describe galaxy clusters
more realistically than using a spherical model.

However, as any model, it has its own limitations we aim to mention.
We have seen in Section~\ref{massorpotential} that, when looking at details,
haloes exhibit departures from a simple triaxial geometry.
Both observations and simulations show the presence of substructures
which are not accounted for by a triaxial model for the galaxy cluster.
If the substructures are small compared to the main cluster's haloe, then
the triaxial approximation may be accurate enough.
On the other hand, unrelaxed haloes often have shapes that are not adequately
described by ellipsoids, making shape parameters ill-defined.
Indeed, if there is no clear dominant haloe but a superposition
of sub-haloes with comparable masses, the triaxial approximation may be questionable.
For example, the high redshift MACS clusters \citep{highzmacs} which are found
to be highly disturbed \citep{harald0717,macslargescale,1149,zitrin1149,adi0717,adisample,mannebeling,my0717} are unlikely to be well described by a triaxial haloe, as proposed
by \citet{sereno_zitrin}. We note, however, that the latter study constitutes a step
forward with respect to a spherical analysis.
Therefore, we emphasise the fact that, in order to apply a triaxial mass model,
one should concentrate on unimodal galaxy clusters with little substructures and 
as close as possible to virialization.

We have presented in this review results obtained in a triaxial framework for four strong 
lensing clusters.
To draw serious conclusions regarding the clusters' properties and how they might
evolve with time, or with any characteristic of the cluster,
we need to study within a triaxial framework a sample of clusters as large as
possible.
Current cluster samples, which already have complementary data sets, 
can allow such an ambitious project to be started:
the Local Cluster Substructure Survey \citep[LoCuSS][]{locussbcg}\footnote{http://www.sr.bham.ac.uk/locuss/};
the MAssive Cluster Survey \citep[MACS][]{EbelingMacsCat2001}; 
the Cluster Lensing And Supernova survey with Hubble \citep[CLASH][]{clash}\footnote{http://www.stsci.edu/~postman/CLASH/Home.html};
and the Dark energy American French Team sample \citep[DAFT][]{guennou10,guennou12}\footnote{http://cencos.oamp.fr/DAFT/}.

As mentioned earlier, a triaxial model is still a simplification of what
a galaxy cluster might be. 
It is not clear how departures from a strict triaxial geometry can bias results based
on the assumption of a triaxial model.
Actually, this ought to be tested using numerically simulated clusters:
mock lensing, X-ray and SZ observations can be generated 
\citep[\emph{e.g.}][]{rasia2012} for galaxy clusters presenting different dynamical states,
and analysed in the same way as observational data.
Comparing the inferred triaxial parameters to those measured on the simulated haloes
is essential to test the methods and to quantify how the presence of substructures
and deviations from equilibrium can bias triaxial reconstructions.
This pioneering work will also pave the road to more refined mass models beyond
the triaxial framework.

%\begin{table}
%\begin{tabular}{lll}
%\hline\noalign{\smallskip}
%\end{tabular}
%\end{table}

\begin{acknowledgements}
 ML acknowledges the Centre National de la Recherche Scientifique (CNRS) for its support.
The Dark Cosmology Centre is funded by the Danish National Research Foundation.
T.V. acknowledges support from CONACYT grant 165365 through the program “Estancias posdoctorales y sab\'aticas al extranjero para la consolidaci\'on de grupos de investigaci\'on.
We thank the International Space Science Institute at Berne, Switzerland 
(ISSI) for providing to our group very good working conditions.
\end{acknowledgements}

\bibliographystyle{aps-nameyear}      % basic style, author-year citations
%\bibliography{references}   % name your BibTeX data base
\bibliography{main}   % name your BibTeX data base

\begin{thebibliography}{197}
% BibTex style file: aps.bst (nameyear), 2011-02-21
\ifx \bisbn   \undefined \def \bisbn  #1{ISBN #1}\fi
\ifx \binits  \undefined \def \binits#1{#1} \fi
\ifx \bauthor  \undefined \def \bauthor#1{#1} \fi
\ifx \bjtitle  \undefined \def \bjtitle#1{\textrm{#1}}\fi
\ifx \batitle  \undefined \def \batitle#1{#1} \fi
\ifx \bctitle  \undefined \def \bctitle#1{#1} \fi
\ifx \bvolume  \undefined \def \bvolume#1{\textbf{#1}}\fi
\ifx \byear  \undefined \def \byear#1{#1} \fi
\ifx \bissue  \undefined \def \bissue#1{#1} \fi
\ifx \bfpage  \undefined \def \bfpage#1{#1} \fi
\ifx \blpage  \undefined \def \blpage #1{#1} \fi
\ifx \burl  \undefined \def \burl#1{#1} \fi
\ifx \doiurl  \undefined \def \doiurl#1{#1} \fi
\ifx \betal  \undefined \def \betal{et al.} \fi
\ifx \binstitute  \undefined \def \binstitute#1{#1} \fi
\ifx \beditor  \undefined \def \beditor#1{#1} \fi
\ifx \bpublisher  \undefined \def \bpublisher#1{#1} \fi
\ifx \bbtitle  \undefined \def \bbtitle#1{\textit{#1}} \fi
\ifx \bedition  \undefined \def \bedition#1{#1} \fi
\ifx \bseriesno  \undefined \def \bseriesno#1{#1} \fi
\ifx \blocation  \undefined \def \blocation#1{#1} \fi
\ifx \bsertitle  \undefined \def \bsertitle#1{#1} \fi
\ifx \bsnm \undefined \def \bsnm#1{#1} \fi
\ifx \bsuffix \undefined \def \bsuffix#1{#1} \fi
\ifx \bparticle \undefined \def \bparticle#1{#1} \fi
\ifx \barticle \undefined \def \barticle#1{#1} \fi
\ifx \botherref \undefined \def \botherref #1{#1} \fi
\ifx \url \undefined \def \url#1{#1} \fi
\ifx \bchapter \undefined \def \bchapter#1{#1} \fi
\ifx \bbook \undefined \def \bbook#1{#1} \fi
\ifx \bcomment \undefined \def \bcomment#1{#1} \fi
\ifx \oauthor \undefined \def \oauthor#1{#1} \fi
\ifx \citeauthoryear \undefined \def \citeauthoryear#1{#1} \fi
\ifx \texttildelow  \undefined \def \texttildelow{\symbol{126}} \fi
\def \endbibitem {}
\ifx \bconflocation  \undefined \def \bconflocation#1{#1} \fi

\bibitem[\protect\citeauthoryear{{Alam} and {Ryden}}{2002}]{al+ry02}
\begin{barticle}
\bauthor{\binits{S.M.K.} \bsnm{{Alam}}},
\bauthor{\binits{B.S.} \bsnm{{Ryden}}},
\batitle{{The Shapes of Galaxies in the Sloan Digital Sky Survey}}.
\bjtitle{ApJ}
\bvolume{570},
\bfpage{610}--\blpage{617}
(\byear{2002}).
doi:\doiurl{10.1086/339790}
\end{barticle}
\endbibitem

\bibitem[\protect\citeauthoryear{{Allgood} et~al.}{2006}]{allgood}
\begin{barticle}
\bauthor{\binits{B.} \bsnm{{Allgood}}},
\bauthor{\binits{R.A.} \bsnm{{Flores}}},
\bauthor{\binits{J.R.} \bsnm{{Primack}}},
\bauthor{\binits{A.V.} \bsnm{{Kravtsov}}},
\bauthor{\binits{R.H.} \bsnm{{Wechsler}}},
\bauthor{\binits{A.} \bsnm{{Faltenbacher}}},
\bauthor{\binits{J.S.} \bsnm{{Bullock}}},
\batitle{{The shape of dark matter haloes: dependence on mass, redshift, radius
  and formation}}.
%\bjtitle{\mnras}
\bvolume{367},
\bfpage{1781}--\blpage{1796}
(\byear{2006}).
doi:\doiurl{10.1111/j.1365-2966.2006.10094.x}
\end{barticle}
\endbibitem

\bibitem[\protect\citeauthoryear{{Allison} et~al.}{2011}]{allison11}
\begin{barticle}
\bauthor{\binits{J.R.} \bsnm{{Allison}}},
\bauthor{\binits{A.C.} \bsnm{{Taylor}}},
\bauthor{\binits{M.E.} \bsnm{{Jones}}},
\bauthor{\binits{S.} \bsnm{{Rawlings}}},
\bauthor{\binits{S.T.} \bsnm{{Kay}}},
\batitle{{A parametric physical model for the intracluster medium and its use
  in joint SZ/X-ray analyses of galaxy clusters}}.
%\bjtitle{\mnras}
\bvolume{410},
\bfpage{341}--\blpage{358}
(\byear{2011}).
doi:\doiurl{10.1111/j.1365-2966.2010.17447.x}
\end{barticle}
\endbibitem

\bibitem[\protect\citeauthoryear{{Altay} et~al.}{2006}]{altay06}
\begin{barticle}
\bauthor{\binits{G.} \bsnm{{Altay}}},
\bauthor{\binits{J.M.} \bsnm{{Colberg}}},
\bauthor{\binits{R.A.C.} \bsnm{{Croft}}},
\batitle{{The influence of large-scale structures on halo shapes and
  alignments}}.
%\bjtitle{\mnras}
\bvolume{370},
\bfpage{1422}--\blpage{1428}
(\byear{2006}).
doi:\doiurl{10.1111/j.1365-2966.2006.10555.x}
\end{barticle}
\endbibitem

\bibitem[\protect\citeauthoryear{{Ameglio} et~al.}{2007}]{ameglio07}
\begin{barticle}
\bauthor{\binits{S.} \bsnm{{Ameglio}}},
\bauthor{\binits{S.} \bsnm{{Borgani}}},
\bauthor{\binits{E.} \bsnm{{Pierpaoli}}},
\bauthor{\binits{K.} \bsnm{{Dolag}}},
\batitle{{Joint deprojection of Sunyaev-Zeldovich and X-ray images of galaxy
  clusters}}.
%\bjtitle{\mnras}
\bvolume{382},
\bfpage{397}--\blpage{411}
(\byear{2007}).
doi:\doiurl{10.1111/j.1365-2966.2007.12384.x}
\end{barticle}
\endbibitem

\bibitem[\protect\citeauthoryear{{Ameglio} et~al.}{2009}]{ameglio09}
\begin{barticle}
\bauthor{\binits{S.} \bsnm{{Ameglio}}},
\bauthor{\binits{S.} \bsnm{{Borgani}}},
\bauthor{\binits{E.} \bsnm{{Pierpaoli}}},
\bauthor{\binits{K.} \bsnm{{Dolag}}},
\bauthor{\binits{S.} \bsnm{{Ettori}}},
\bauthor{\binits{A.} \bsnm{{Morandi}}},
\batitle{{Reconstructing mass profiles of simulated galaxy clusters by
  combining Sunyaev-Zeldovich and X-ray images}}.
%\bjtitle{\mnras}
\bvolume{394},
\bfpage{479}--\blpage{490}
(\byear{2009}).
doi:\doiurl{10.1111/j.1365-2966.2008.14324.x}
\end{barticle}
\endbibitem

\bibitem[\protect\citeauthoryear{{Andersson} and {Madejski}}{2004}]{andersson}
\begin{barticle}
\bauthor{\binits{K.E.} \bsnm{{Andersson}}},
\bauthor{\binits{G.M.} \bsnm{{Madejski}}},
\batitle{{Complex Structure of Galaxy Cluster A1689: Evidence for a Merger fr
  om X-Ray Data?}}
\bjtitle{\apj}
\bvolume{607},
\bfpage{190}--\blpage{201}
(\byear{2004}).
doi:\doiurl{10.1086/383258}
\end{barticle}
\endbibitem

\bibitem[\protect\citeauthoryear{{Angulo} et~al.}{2012}]{angulo}
\begin{botherref}
\oauthor{\binits{R.E.} \bsnm{{Angulo}}},
\oauthor{\binits{V.} \bsnm{{Springel}}},
\oauthor{\binits{S.D.M.} \bsnm{{White}}},
\oauthor{\binits{A.} \bsnm{{Jenkins}}},
\oauthor{\binits{C.M.} \bsnm{{Baugh}}},
\oauthor{\binits{C.S.} \bsnm{{Frenk}}},
{Scaling relations for galaxy clusters in the Millennium-XXL simulation}.
ArXiv e-prints
(2012)
\end{botherref}
\endbibitem

\bibitem[\protect\citeauthoryear{{Arag{\'o}n-Calvo} et~al.}{2007}]{aragon07}
\begin{barticle}
\bauthor{\binits{M.A.} \bsnm{{Arag{\'o}n-Calvo}}},
\bauthor{\binits{R.} \bsnm{{van de Weygaert}}},
\bauthor{\binits{B.J.T.} \bsnm{{Jones}}},
\bauthor{\binits{J.M.} \bsnm{{van der Hulst}}},
\batitle{{Spin Alignment of Dark Matter Halos in Filaments and Walls}}.
%\bjtitle{\apjl}
\bvolume{655},
\bfpage{5}--\blpage{8}
(\byear{2007}).
doi:\doiurl{10.1086/511633}
\end{barticle}
\endbibitem

\bibitem[\protect\citeauthoryear{{Bailin} and {Steinmetz}}{2005}]{bailin05}
\begin{barticle}
\bauthor{\binits{J.} \bsnm{{Bailin}}},
\bauthor{\binits{M.} \bsnm{{Steinmetz}}},
\batitle{{Internal and External Alignment of the Shapes and Angular Momenta of
  {$\Lambda$}CDM Halos}}.
\bjtitle{\apj}
\bvolume{627},
\bfpage{647}--\blpage{665}
(\byear{2005}).
doi:\doiurl{10.1086/430397}
\end{barticle}
\endbibitem

\bibitem[\protect\citeauthoryear{{Bardeau} et~al.}{2005}]{bardeau05}
\begin{barticle}
\bauthor{\binits{S.} \bsnm{{Bardeau}}},
\bauthor{\binits{J.-P.} \bsnm{{Kneib}}},
\bauthor{\binits{O.} \bsnm{{Czoske}}},
\bauthor{\binits{G.} \bsnm{{Soucail}}},
\bauthor{\binits{I.} \bsnm{{Smail}}},
\bauthor{\binits{H.} \bsnm{{Ebeling}}},
\bauthor{\binits{G.P.} \bsnm{{Smith}}},
\batitle{{A CFH12k lensing survey of X-ray luminous galaxy clusters. I. Weak
  lensing methodology}}.
\bjtitle{\aap}
\bvolume{434},
\bfpage{433}--\blpage{448}
(\byear{2005}).
doi:\doiurl{10.1051/0004-6361:20041643}
\end{barticle}
\endbibitem

\bibitem[\protect\citeauthoryear{{Basilakos} et~al.}{2000}]{bas+al00}
\begin{barticle}
\bauthor{\binits{S.} \bsnm{{Basilakos}}},
\bauthor{\binits{M.} \bsnm{{Plionis}}},
\bauthor{\binits{S.J.} \bsnm{{Maddox}}},
\batitle{{The apparent and intrinsic shape of the APM galaxy clusters}}.
\bjtitle{MNRAS}
\bvolume{316},
\bfpage{779}--\blpage{785}
(\byear{2000})
\end{barticle}
\endbibitem

\bibitem[\protect\citeauthoryear{{Basu} et~al.}{2010}]{basu10}
\begin{barticle}
\bauthor{\binits{K.} \bsnm{{Basu}}},
\bauthor{\binits{Y.-Y.} \bsnm{{Zhang}}},
\bauthor{\binits{M.W.} \bsnm{{Sommer}}},
\bauthor{\binits{A.N.} \bsnm{{Bender}}},
\bauthor{\binits{F.} \bsnm{{Bertoldi}}},
\bauthor{\binits{M.} \bsnm{{Dobbs}}},
\bauthor{\binits{H.} \bsnm{{Eckmiller}}},
\bauthor{\binits{N.W.} \bsnm{{Halverson}}},
\bauthor{\binits{W.L.} \bsnm{{Holzapfel}}},
\bauthor{\binits{C.} \bsnm{{Horellou}}},
\bauthor{\binits{V.} \bsnm{{Jaritz}}},
\bauthor{\binits{D.} \bsnm{{Johansson}}},
\bauthor{\binits{B.} \bsnm{{Johnson}}},
\bauthor{\binits{J.} \bsnm{{Kennedy}}},
\bauthor{\binits{R.} \bsnm{{Kneissl}}},
\bauthor{\binits{T.} \bsnm{{Lanting}}},
\bauthor{\binits{A.T.} \bsnm{{Lee}}},
\bauthor{\binits{J.} \bsnm{{Mehl}}},
\bauthor{\binits{K.M.} \bsnm{{Menten}}},
\bauthor{\binits{F.P.} \bsnm{{Navarrete}}},
\bauthor{\binits{F.} \bsnm{{Pacaud}}},
\bauthor{\binits{C.L.} \bsnm{{Reichardt}}},
\bauthor{\binits{T.H.} \bsnm{{Reiprich}}},
\bauthor{\binits{P.L.} \bsnm{{Richards}}},
\bauthor{\binits{D.} \bsnm{{Schwan}}},
\bauthor{\binits{B.} \bsnm{{Westbrook}}},
\batitle{{Non-parametric modeling of the intra-cluster gas using APEX-SZ
  bolometer imaging data}}.
\bjtitle{\aap}
\bvolume{519},
\bfpage{29}
(\byear{2010}).
doi:\doiurl{10.1051/0004-6361/200913334}
\end{barticle}
\endbibitem

\bibitem[\protect\citeauthoryear{{Bett} et~al.}{2007}]{bett07}
\begin{barticle}
\bauthor{\binits{P.} \bsnm{{Bett}}},
\bauthor{\binits{V.} \bsnm{{Eke}}},
\bauthor{\binits{C.S.} \bsnm{{Frenk}}},
\bauthor{\binits{A.} \bsnm{{Jenkins}}},
\bauthor{\binits{J.} \bsnm{{Helly}}},
\bauthor{\binits{J.} \bsnm{{Navarro}}},
\batitle{{The spin and shape of dark matter haloes in the Millennium simulation
  of a {$\Lambda$} cold dark matter universe}}.
%\bjtitle{\mnras}
\bvolume{376},
\bfpage{215}--\blpage{232}
(\byear{2007}).
doi:\doiurl{10.1111/j.1365-2966.2007.11432.x}
\end{barticle}
\endbibitem

\bibitem[\protect\citeauthoryear{{Bhattacharya} et~al.}{2011}]{bhattacharya}
\begin{botherref}
\oauthor{\binits{S.} \bsnm{{Bhattacharya}}},
\oauthor{\binits{S.} \bsnm{{Habib}}},
\oauthor{\binits{K.} \bsnm{{Heitmann}}},
{Dark Matter Halo Profiles of Massive Clusters: Theory vs. Observations}.
ArXiv e-prints
(2011)
\end{botherref}
\endbibitem

\bibitem[\protect\citeauthoryear{{Binggeli}}{1980}]{bin80}
\begin{barticle}
\bauthor{\binits{B.} \bsnm{{Binggeli}}},
\batitle{{On the intrinsic shape of elliptical galaxies}}.
\bjtitle{A\&A}
\bvolume{82},
\bfpage{289}--\blpage{294}
(\byear{1980})
\end{barticle}
\endbibitem

\bibitem[\protect\citeauthoryear{{Binggeli}}{1982}]{bingelli82}
\begin{barticle}
\bauthor{\binits{B.} \bsnm{{Binggeli}}},
\batitle{{The shape and orientation of clusters of galaxies}}.
\bjtitle{\aap}
\bvolume{107},
\bfpage{338}--\blpage{349}
(\byear{1982})
\end{barticle}
\endbibitem

\bibitem[\protect\citeauthoryear{{Binney} and {de Vaucouleurs}}{1981}]{bi+de81}
\begin{barticle}
\bauthor{\binits{J.} \bsnm{{Binney}}},
\bauthor{\binits{G.} \bsnm{{de Vaucouleurs}}},
\batitle{{The apparent and true ellipticities of galaxies of different Hubble
  types in the Second Reference Catalogue}}.
\bjtitle{MNRAS}
\bvolume{194},
\bfpage{679}--\blpage{691}
(\byear{1981})
\end{barticle}
\endbibitem

\bibitem[\protect\citeauthoryear{{Binney} and {Tremaine}}{1987}]{binney1987}
\begin{bbook}
\bauthor{\binits{J.} \bsnm{{Binney}}},
\bauthor{\binits{S.} \bsnm{{Tremaine}}},
\bbtitle{{Galactic Dynamics}}
(\bpublisher{Princeton, NJ, Princeton University Press, 1987, 747 p.},
  \blocation{???}, \byear{1987})
\end{bbook}
\endbibitem

\bibitem[\protect\citeauthoryear{{Brada{\v c}} et~al.}{2008}]{bradac08a}
\begin{barticle}
\bauthor{\binits{M.} \bsnm{{Brada{\v c}}}},
\bauthor{\binits{T.} \bsnm{{Schrabback}}},
\bauthor{\binits{T.} \bsnm{{Erben}}},
\bauthor{\binits{M.} \bsnm{{McCourt}}},
\bauthor{\binits{E.} \bsnm{{Million}}},
\bauthor{\binits{A.} \bsnm{{Mantz}}},
\bauthor{\binits{S.} \bsnm{{Allen}}},
\bauthor{\binits{R.} \bsnm{{Blandford}}},
\bauthor{\binits{A.} \bsnm{{Halkola}}},
\bauthor{\binits{H.} \bsnm{{Hildebrandt}}},
\bauthor{\binits{M.} \bsnm{{Lombardi}}},
\bauthor{\binits{P.} \bsnm{{Marshall}}},
\bauthor{\binits{P.} \bsnm{{Schneider}}},
\bauthor{\binits{T.} \bsnm{{Treu}}},
\bauthor{\binits{J.-P.} \bsnm{{Kneib}}},
\batitle{{Dark Matter and Baryons in the X-Ray Luminous Merging Galaxy Cluster
  RX J1347.5-1145}}.
\bjtitle{\apj}
\bvolume{681},
\bfpage{187}--\blpage{196}
(\byear{2008}).
doi:\doiurl{10.1086/588377}
\end{barticle}
\endbibitem

\bibitem[\protect\citeauthoryear{{Broadhurst} and {Barkana}}{2008a}]{largeRe}
\begin{barticle}
\bauthor{\binits{T.J.} \bsnm{{Broadhurst}}},
\bauthor{\binits{R.} \bsnm{{Barkana}}},
\batitle{{Large Einstein radii: a problem for {$\Lambda$}CDM}}.
%\bjtitle{\mnras}
\bvolume{390},
\bfpage{1647}--\blpage{1654}
(\byear{2008}a).
doi:\doiurl{10.1111/j.1365-2966.2008.13852.x}
\end{barticle}
\endbibitem

\bibitem[\protect\citeauthoryear{{Broadhurst} and {Barkana}}{2008b}]{tom08}
\begin{barticle}
\bauthor{\binits{T.J.} \bsnm{{Broadhurst}}},
\bauthor{\binits{R.} \bsnm{{Barkana}}},
\batitle{{Large Einstein radii: a problem for {$\Lambda$}CDM}}.
%\bjtitle{\mnras}
\bvolume{390},
\bfpage{1647}--\blpage{1654}
(\byear{2008}b).
doi:\doiurl{10.1111/j.1365-2966.2008.13852.x}
\end{barticle}
\endbibitem

\bibitem[\protect\citeauthoryear{{Brunino} et~al.}{2007}]{brunino07}
\begin{barticle}
\bauthor{\binits{R.} \bsnm{{Brunino}}},
\bauthor{\binits{I.} \bsnm{{Trujillo}}},
\bauthor{\binits{F.R.} \bsnm{{Pearce}}},
\bauthor{\binits{P.A.} \bsnm{{Thomas}}},
\batitle{{The orientation of galaxy dark matter haloes around cosmic voids}}.
%\bjtitle{\mnras}
\bvolume{375},
\bfpage{184}--\blpage{190}
(\byear{2007}).
doi:\doiurl{10.1111/j.1365-2966.2006.11282.x}
\end{barticle}
\endbibitem

\bibitem[\protect\citeauthoryear{{Bullock} et~al.}{2001}]{bullock01}
\begin{barticle}
\bauthor{\binits{J.S.} \bsnm{{Bullock}}},
\bauthor{\binits{T.S.} \bsnm{{Kolatt}}},
\bauthor{\binits{Y.} \bsnm{{Sigad}}},
\bauthor{\binits{R.S.} \bsnm{{Somerville}}},
\bauthor{\binits{A.V.} \bsnm{{Kravtsov}}},
\bauthor{\binits{A.A.} \bsnm{{Klypin}}},
\bauthor{\binits{J.R.} \bsnm{{Primack}}},
\bauthor{\binits{A.} \bsnm{{Dekel}}},
\batitle{{Profiles of dark haloes: evolution, scatter and environment}}.
%\bjtitle{\mnras}
\bvolume{321},
\bfpage{559}--\blpage{575}
(\byear{2001}).
doi:\doiurl{10.1046/j.1365-8711.2001.04068.x}
\end{barticle}
\endbibitem

\bibitem[\protect\citeauthoryear{{Buote} and {Canizares}}{1992}]{buote92}
\begin{barticle}
\bauthor{\binits{D.A.} \bsnm{{Buote}}},
\bauthor{\binits{C.R.} \bsnm{{Canizares}}},
\batitle{{X-ray constraints on the shape of the dark matter in five Abell
  clusters}}.
\bjtitle{\apj}
\bvolume{400},
\bfpage{385}--\blpage{397}
(\byear{1992}).
doi:\doiurl{10.1086/172004}
\end{barticle}
\endbibitem

\bibitem[\protect\citeauthoryear{{Buote} and {Canizares}}{1994}]{buote1994}
\begin{barticle}
\bauthor{\binits{D.A.} \bsnm{{Buote}}},
\bauthor{\binits{C.R.} \bsnm{{Canizares}}},
\batitle{{Geometrical evidence for dark matter: X-ray constraints on the mass
  of the elliptical galaxy NGC 720}}.
\bjtitle{\apj}
\bvolume{427},
\bfpage{86}--\blpage{111}
(\byear{1994}).
doi:\doiurl{10.1086/174123}
\end{barticle}
\endbibitem

\bibitem[\protect\citeauthoryear{{Buote} and {Canizares}}{1996}]{buote96}
\begin{barticle}
\bauthor{\binits{D.A.} \bsnm{{Buote}}},
\bauthor{\binits{C.R.} \bsnm{{Canizares}}},
\batitle{{X-Ray Constraints on the Intrinsic Shapes and Baryon Fractions of
  Five Abell Clusters}}.
\bjtitle{\apj}
\bvolume{457},
\bfpage{565}
(\byear{1996}).
doi:\doiurl{10.1086/176753}
\end{barticle}
\endbibitem

\bibitem[\protect\citeauthoryear{{Buote} and {Humphrey}}{2011a}]{buote11a}
\begin{botherref}
\oauthor{\binits{D.A.} \bsnm{{Buote}}},
\oauthor{\binits{P.J.} \bsnm{{Humphrey}}},
{Spherically averaging ellipsoidal galaxy clusters in X-ray and
  Sunyaev-Zel'dovich studies - I. Analytical relations}.
%\mnras,

2136(2011a).
doi:\doiurl{10.1111/j.1365-2966.2011.20163.x}
\end{botherref}
\endbibitem

\bibitem[\protect\citeauthoryear{{Buote} and {Humphrey}}{2011b}]{buote11b}
\begin{botherref}
\oauthor{\binits{D.A.} \bsnm{{Buote}}},
\oauthor{\binits{P.J.} \bsnm{{Humphrey}}},
{Spherically Averaging Ellipsoidal Galaxy Clusters in X-Ray and
  Sunyaev-Zel'dovich Studies: II. Biases}.

ArXiv e-prints(2011b)
\end{botherref}
\endbibitem

\bibitem[\protect\citeauthoryear{{Cappellari} et~al.}{2007}]{sauron}
\begin{barticle}
\bauthor{\binits{M.} \bsnm{{Cappellari}}},
\bauthor{\binits{E.} \bsnm{{Emsellem}}},
\bauthor{\binits{R.} \bsnm{{Bacon}}},
\bauthor{\binits{M.} \bsnm{{Bureau}}},
\bauthor{\binits{R.L.} \bsnm{{Davies}}},
\bauthor{\binits{P.T.} \bsnm{{de Zeeuw}}},
\bauthor{\binits{J.} \bsnm{{Falc{\'o}n-Barroso}}},
\bauthor{\binits{D.} \bsnm{{Krajnovi{\'c}}}},
\bauthor{\binits{H.} \bsnm{{Kuntschner}}},
\bauthor{\binits{R.M.} \bsnm{{McDermid}}},
\bauthor{\binits{R.F.} \bsnm{{Peletier}}},
\bauthor{\binits{M.} \bsnm{{Sarzi}}},
\bauthor{\binits{R.C.E.} \bsnm{{van den Bosch}}},
\bauthor{\binits{G.} \bsnm{{van de Ven}}},
\batitle{{The SAURON project - X. The orbital anisotropy of elliptical and
  lenticular galaxies: revisiting the (V/{$\sigma$}, {$\epsilon$}) diagram with
  integral-field stellar kinematics}}.
%\bjtitle{\mnras}
\bvolume{379},
\bfpage{418}--\blpage{444}
(\byear{2007}).
doi:\doiurl{10.1111/j.1365-2966.2007.11963.x}
\end{barticle}
\endbibitem

\bibitem[\protect\citeauthoryear{{Carter} and {Metcalfe}}{1980}]{carter80}
\begin{barticle}
\bauthor{\binits{D.} \bsnm{{Carter}}},
\bauthor{\binits{N.} \bsnm{{Metcalfe}}},
\batitle{{The morphology of clusters of galaxies}}.
%\bjtitle{\mnras}
\bvolume{191},
\bfpage{325}--\blpage{337}
(\byear{1980})
\end{barticle}
\endbibitem

\bibitem[\protect\citeauthoryear{{Chakrabarty} et~al.}{2008}]{chakrabarty}
\begin{barticle}
\bauthor{\binits{D.} \bsnm{{Chakrabarty}}},
\bauthor{\binits{E.} \bsnm{{de Filippis}}},
\bauthor{\binits{H.} \bsnm{{Russell}}},
\batitle{{Cluster geometry and inclinations from deprojection uncertainties.
  Cluster geometry and inclination}}.
\bjtitle{\aap}
\bvolume{487},
\bfpage{75}--\blpage{87}
(\byear{2008}).
doi:\doiurl{10.1051/0004-6361:200809510}
\end{barticle}
\endbibitem

\bibitem[\protect\citeauthoryear{{Clowe}}{2003}]{clowe03}
\begin{bchapter}
\bauthor{\binits{D.} \bsnm{{Clowe}}},
\bctitle{{Wide-Field Weak Lensing Cluster Mass Reconstructions}},
in \bbtitle{Astronomical Society of the Pacific Conference Series},
ed. by \beditor{\binits{S.} \bsnm{{Bowyer}}},
\beditor{\binits{C.-Y.} \bsnm{{Hwang}}},
\byear{2003},
p. \bfpage{271}
\end{bchapter}
\endbibitem

\bibitem[\protect\citeauthoryear{{Clowe} et~al.}{2004}]{clowe04}
\begin{barticle}
\bauthor{\binits{D.} \bsnm{{Clowe}}},
\bauthor{\binits{G.} \bsnm{{De Lucia}}},
\bauthor{\binits{L.} \bsnm{{King}}},
\batitle{{Effects of asphericity and substructure on the determination of
  cluster mass with weak gravitational lensing}}.
%\bjtitle{\mnras}
\bvolume{350},
\bfpage{1038}--\blpage{1048}
(\byear{2004}).
doi:\doiurl{10.1111/j.1365-2966.2004.07723.x}
\end{barticle}
\endbibitem

\bibitem[\protect\citeauthoryear{{Cole} and {Lacey}}{1996}]{cole96}
\begin{barticle}
\bauthor{\binits{S.} \bsnm{{Cole}}},
\bauthor{\binits{C.} \bsnm{{Lacey}}},
\batitle{{The structure of dark matter haloes in hierarchical clustering
  models}}.
%\bjtitle{\mnras}
\bvolume{281},
\bfpage{716}
(\byear{1996})
\end{barticle}
\endbibitem

\bibitem[\protect\citeauthoryear{{Cooray}}{2000}]{coo00}
\begin{barticle}
\bauthor{\binits{A.R.} \bsnm{{Cooray}}},
\batitle{{Galaxy clusters: oblate or prolate?}}
\bjtitle{MNRAS}
\bvolume{313},
\bfpage{783}--\blpage{788}
(\byear{2000})
\end{barticle}
\endbibitem

\bibitem[\protect\citeauthoryear{{Corless} and {King}}{2007}]{corless07}
\begin{barticle}
\bauthor{\binits{V.L.} \bsnm{{Corless}}},
\bauthor{\binits{L.J.} \bsnm{{King}}},
\batitle{{A statistical study of weak lensing by triaxial dark matter haloes:
  consequences for parameter estimation}}.
%\bjtitle{\mnras}
\bvolume{380},
\bfpage{149}--\blpage{161}
(\byear{2007}).
doi:\doiurl{10.1111/j.1365-2966.2007.12018.x}
\end{barticle}
\endbibitem

\bibitem[\protect\citeauthoryear{{Corless} and {King}}{2008}]{corless08}
\begin{barticle}
\bauthor{\binits{V.L.} \bsnm{{Corless}}},
\bauthor{\binits{L.J.} \bsnm{{King}}},
\batitle{{An MCMC fitting method for triaxial dark matter haloes}}.
%\bjtitle{\mnras}
\bvolume{390},
\bfpage{997}--\blpage{1013}
(\byear{2008}).
doi:\doiurl{10.1111/j.1365-2966.2008.13744.x}
\end{barticle}
\endbibitem

\bibitem[\protect\citeauthoryear{{Corless} et~al.}{2009}]{corless09}
\begin{barticle}
\bauthor{\binits{V.L.} \bsnm{{Corless}}},
\bauthor{\binits{L.J.} \bsnm{{King}}},
\bauthor{\binits{D.} \bsnm{{Clowe}}},
\batitle{{A new look at massive clusters: weak lensing constraints on the
  triaxial dark matter haloes of A1689, A1835 and A2204}}.
%\bjtitle{\mnras}
\bvolume{393},
\bfpage{1235}--\blpage{1254}
(\byear{2009}).
doi:\doiurl{10.1111/j.1365-2966.2008.14294.x}
\end{barticle}
\endbibitem

\bibitem[\protect\citeauthoryear{{Czoske}}{2004}]{olliturin}
\begin{bchapter}
\bauthor{\binits{O.} \bsnm{{Czoske}}},
\bctitle{{Wide-field Spectroscopy of A1689 and A1835 with Vimos: First
  Results}},
in \bbtitle{IAU Colloq. 195: Outskirts of Galaxy Clusters: Intense Life in the
  Suburbs},
ed. by \beditor{\binits{A.} \bsnm{{Diaferio}}},
\byear{2004},
pp. \bfpage{183}--\blpage{187}.
doi:\doiurl{10.1017/S1743921304000390}
\end{bchapter}
\endbibitem

\bibitem[\protect\citeauthoryear{{Czoske} et~al.}{2001}]{olli00241}
\begin{barticle}
\bauthor{\binits{O.} \bsnm{{Czoske}}},
\bauthor{\binits{J.-P.} \bsnm{{Kneib}}},
\bauthor{\binits{G.} \bsnm{{Soucail}}},
\bauthor{\binits{T.J.} \bsnm{{Bridges}}},
\bauthor{\binits{Y.} \bsnm{{Mellier}}},
\bauthor{\binits{J.-C.} \bsnm{{Cuillandre}}},
\batitle{{A wide-field spectroscopic survey of the cluster of galaxies
  <ASTROBJ>Cl0024+1654</ASTROBJ>. I. The catalogue}}.
\bjtitle{\aap}
\bvolume{372},
\bfpage{391}--\blpage{405}
(\byear{2001}).
doi:\doiurl{10.1051/0004-6361:20010398}
\end{barticle}
\endbibitem

\bibitem[\protect\citeauthoryear{{Czoske} et~al.}{2002}]{olli00242}
\begin{barticle}
\bauthor{\binits{O.} \bsnm{{Czoske}}},
\bauthor{\binits{B.} \bsnm{{Moore}}},
\bauthor{\binits{J.-P.} \bsnm{{Kneib}}},
\bauthor{\binits{G.} \bsnm{{Soucail}}},
\batitle{{A wide-field spectroscopic survey of the cluster of galaxies
  <ASTROBJ>Cl0024+1654</ASTROBJ>. II. A high-speed collision?}}
\bjtitle{\aap}
\bvolume{386},
\bfpage{31}--\blpage{41}
(\byear{2002}).
doi:\doiurl{10.1051/0004-6361:20020230}
\end{barticle}
\endbibitem

\bibitem[\protect\citeauthoryear{{Davis} et~al.}{1985}]{davis85}
\begin{barticle}
\bauthor{\binits{M.} \bsnm{{Davis}}},
\bauthor{\binits{G.} \bsnm{{Efstathiou}}},
\bauthor{\binits{C.S.} \bsnm{{Frenk}}},
\bauthor{\binits{S.D.M.} \bsnm{{White}}},
\batitle{{The evolution of large-scale structure in a universe dominated by
  cold dark matter}}.
\bjtitle{\apj}
\bvolume{292},
\bfpage{371}--\blpage{394}
(\byear{1985}).
doi:\doiurl{10.1086/163168}
\end{barticle}
\endbibitem

\bibitem[\protect\citeauthoryear{{De Filippis} et~al.}{2005}]{filippis05}
\begin{barticle}
\bauthor{\binits{E.} \bsnm{{De Filippis}}},
\bauthor{\binits{M.} \bsnm{{Sereno}}},
\bauthor{\binits{M.W.} \bsnm{{Bautz}}},
\bauthor{\binits{G.} \bsnm{{Longo}}},
\batitle{{Measuring the Three-dimensional Structure of Galaxy Clusters. I.
  Application to a Sample of 25 Clusters}}.
\bjtitle{\apj}
\bvolume{625},
\bfpage{108}--\blpage{120}
(\byear{2005}).
doi:\doiurl{10.1086/429401}
\end{barticle}
\endbibitem

\bibitem[\protect\citeauthoryear{{de Theije} et~al.}{1995}]{det+al95}
\begin{barticle}
\bauthor{\binits{P.A.M.} \bsnm{{de Theije}}},
\bauthor{\binits{P.} \bsnm{{Katgert}}},
\bauthor{\binits{E.} \bsnm{{van Kampen}}},
\batitle{{The shapes of galaxy clusters}}.
\bjtitle{MNRAS}
\bvolume{273},
\bfpage{30}--\blpage{46}
(\byear{1995})
\end{barticle}
\endbibitem

\bibitem[\protect\citeauthoryear{{Deb} et~al.}{2012}]{deb12}
\begin{botherref}
\oauthor{\binits{S.} \bsnm{{Deb}}},
\oauthor{\binits{A.} \bsnm{{Morandi}}},
\oauthor{\binits{K.} \bsnm{{Pedersen}}},
\oauthor{\binits{S.} \bsnm{{Riemer-Sorensen}}},
\oauthor{\binits{D.M.} \bsnm{{Goldberg}}},
\oauthor{\binits{H.} \bsnm{{Dahle}}},
{Mass Reconstruction using Particle Based Lensing II: Quantifying substructure
  with Strong+Weak lensing and X-rays}.
ArXiv e-prints
(2012)
\end{botherref}
\endbibitem

\bibitem[\protect\citeauthoryear{{Donnarumma} et~al.}{2011}]{donnarumma2011}
\begin{barticle}
\bauthor{\binits{A.} \bsnm{{Donnarumma}}},
\bauthor{\binits{S.} \bsnm{{Ettori}}},
\bauthor{\binits{M.} \bsnm{{Meneghetti}}},
\bauthor{\binits{R.} \bsnm{{Gavazzi}}},
\bauthor{\binits{B.} \bsnm{{Fort}}},
\bauthor{\binits{L.} \bsnm{{Moscardini}}},
\bauthor{\binits{A.} \bsnm{{Romano}}},
\bauthor{\binits{L.} \bsnm{{Fu}}},
\bauthor{\binits{F.} \bsnm{{Giordano}}},
\bauthor{\binits{M.} \bsnm{{Radovich}}},
\bauthor{\binits{R.} \bsnm{{Maoli}}},
\bauthor{\binits{R.} \bsnm{{Scaramella}}},
\bauthor{\binits{J.} \bsnm{{Richard}}},
\batitle{{Abell 611. II. X-ray and strong lensing analyses}}.
\bjtitle{\aap}
\bvolume{528},
\bfpage{73}
(\byear{2011}).
doi:\doiurl{10.1051/0004-6361/201014120}
\end{barticle}
\endbibitem

\bibitem[\protect\citeauthoryear{{Doroshkevich}}{1970}]{doroshkevich}
\begin{barticle}
\bauthor{\binits{A.G.} \bsnm{{Doroshkevich}}},
\batitle{{Spatial structure of perturbations and origin of galactic rotation in
  fluctuation theory}}.
\bjtitle{Astrophysics}
\bvolume{6},
\bfpage{320}--\blpage{330}
(\byear{1970}).
doi:\doiurl{10.1007/BF01001625}
\end{barticle}
\endbibitem

\bibitem[\protect\citeauthoryear{{Dubinski}}{1998}]{dubinski98}
\begin{barticle}
\bauthor{\binits{J.} \bsnm{{Dubinski}}},
\batitle{{The Origin of the Brightest Cluster Galaxies}}.
\bjtitle{\apj}
\bvolume{502},
\bfpage{141}
(\byear{1998}).
doi:\doiurl{10.1086/305901}
\end{barticle}
\endbibitem

\bibitem[\protect\citeauthoryear{{Dubinski} and {Carlberg}}{1991}]{dubinski91}
\begin{barticle}
\bauthor{\binits{J.} \bsnm{{Dubinski}}},
\bauthor{\binits{R.G.} \bsnm{{Carlberg}}},
\batitle{{The structure of cold dark matter halos}}.
\bjtitle{\apj}
\bvolume{378},
\bfpage{496}--\blpage{503}
(\byear{1991}).
doi:\doiurl{10.1086/170451}
\end{barticle}
\endbibitem

\bibitem[\protect\citeauthoryear{{Duffy} et~al.}{2008}]{duffy08}
\begin{barticle}
\bauthor{\binits{A.R.} \bsnm{{Duffy}}},
\bauthor{\binits{J.} \bsnm{{Schaye}}},
\bauthor{\binits{S.T.} \bsnm{{Kay}}},
\bauthor{\binits{C.} \bsnm{{Dalla Vecchia}}},
\batitle{{Dark matter halo concentrations in the Wilkinson Microwave Anisotropy
  Probe year 5 cosmology}}.
%\bjtitle{\mnras}
\bvolume{390},
\bfpage{64}--\blpage{68}
(\byear{2008}).
doi:\doiurl{10.1111/j.1745-3933.2008.00537.x}
\end{barticle}
\endbibitem

\bibitem[\protect\citeauthoryear{{Ebeling} et~al.}{2004}]{harald0717}
\begin{barticle}
\bauthor{\binits{H.} \bsnm{{Ebeling}}},
\bauthor{\binits{E.} \bsnm{{Barrett}}},
\bauthor{\binits{D.} \bsnm{{Donovan}}},
\batitle{{Discovery of a Large-Scale Filament Connected to the Massive Galaxy
  Cluster MACS J0717.5+3745 at z=0.551,}}.
%\bjtitle{\apjl}
\bvolume{609},
\bfpage{49}--\blpage{52}
(\byear{2004}).
doi:\doiurl{10.1086/422750}
\end{barticle}
\endbibitem

\bibitem[\protect\citeauthoryear{{Ebeling} et~al.}{2001}]{EbelingMacsCat2001}
\begin{barticle}
\bauthor{\binits{H.} \bsnm{{Ebeling}}},
\bauthor{\binits{A.C.} \bsnm{{Edge}}},
\bauthor{\binits{J.P.} \bsnm{{Henry}}},
\batitle{{MACS: A Quest for the Most Massive Galaxy Clusters in the Universe}}.
\bjtitle{\apj}
\bvolume{553},
\bfpage{668}--\blpage{676}
(\byear{2001}).
doi:\doiurl{10.1086/320958}
\end{barticle}
\endbibitem

\bibitem[\protect\citeauthoryear{{Ebeling} et~al.}{2007}]{highzmacs}
\begin{barticle}
\bauthor{\binits{H.} \bsnm{{Ebeling}}},
\bauthor{\binits{E.} \bsnm{{Barrett}}},
\bauthor{\binits{D.} \bsnm{{Donovan}}},
\bauthor{\binits{C.-J.} \bsnm{{Ma}}},
\bauthor{\binits{A.C.} \bsnm{{Edge}}},
\bauthor{\binits{L.} \bsnm{{van Speybroeck}}},
\batitle{{A Complete Sample of 12 Very X-Ray Luminous Galaxy Clusters at z
  0.5}}.
%\bjtitle{\apjl}
\bvolume{661},
\bfpage{33}--\blpage{36}
(\byear{2007}).
doi:\doiurl{10.1086/518603}
\end{barticle}
\endbibitem

\bibitem[\protect\citeauthoryear{{Ebeling} et~al.}{2010}]{Ebeling2010MACSALL}
\begin{barticle}
\bauthor{\binits{H.} \bsnm{{Ebeling}}},
\bauthor{\binits{A.C.} \bsnm{{Edge}}},
\bauthor{\binits{A.} \bsnm{{Mantz}}},
\bauthor{\binits{E.} \bsnm{{Barrett}}},
\bauthor{\binits{J.P.} \bsnm{{Henry}}},
\bauthor{\binits{C.J.} \bsnm{{Ma}}},
\bauthor{\binits{L.} \bsnm{{van Speybroeck}}},
\batitle{{The X-ray brightest clusters of galaxies from the Massive Cluster
  Survey}}.
%\bjtitle{\mnras}
\bvolume{407},
\bfpage{83}--\blpage{93}
(\byear{2010}).
doi:\doiurl{10.1111/j.1365-2966.2010.16920.x}
\end{barticle}
\endbibitem

\bibitem[\protect\citeauthoryear{{Eke} et~al.}{2001}]{eke01}
\begin{barticle}
\bauthor{\binits{V.R.} \bsnm{{Eke}}},
\bauthor{\binits{J.F.} \bsnm{{Navarro}}},
\bauthor{\binits{M.} \bsnm{{Steinmetz}}},
\batitle{{The Power Spectrum Dependence of Dark Matter Halo Concentrations}}.
\bjtitle{\apj}
\bvolume{554},
\bfpage{114}--\blpage{125}
(\byear{2001}).
doi:\doiurl{10.1086/321345}
\end{barticle}
\endbibitem

\bibitem[\protect\citeauthoryear{{Ettori} et~al.}{2009}]{ett+al09}
\begin{barticle}
\bauthor{\binits{S.} \bsnm{{Ettori}}},
\bauthor{\binits{A.} \bsnm{{Morandi}}},
\bauthor{\binits{P.} \bsnm{{Tozzi}}},
\bauthor{\binits{I.} \bsnm{{Balestra}}},
\bauthor{\binits{S.} \bsnm{{Borgani}}},
\bauthor{\binits{P.} \bsnm{{Rosati}}},
\bauthor{\binits{L.} \bsnm{{Lovisari}}},
\bauthor{\binits{F.} \bsnm{{Terenziani}}},
\batitle{{The cluster gas mass fraction as a cosmological probe: a revised
  study}}.
\bjtitle{\aap}
\bvolume{501},
\bfpage{61}--\blpage{73}
(\byear{2009}).
doi:\doiurl{10.1051/0004-6361/200810878}
\end{barticle}
\endbibitem

\bibitem[\protect\citeauthoryear{{Evans} and {Bridle}}{2009}]{evans09}
\begin{barticle}
\bauthor{\binits{A.K.D.} \bsnm{{Evans}}},
\bauthor{\binits{S.} \bsnm{{Bridle}}},
\batitle{{A Detection of Dark Matter Halo Ellipticity using Galaxy Cluster
  Lensing in the SDSS}}.
\bjtitle{\apj}
\bvolume{695},
\bfpage{1446}--\blpage{1456}
(\byear{2009}).
doi:\doiurl{10.1088/0004-637X/695/2/1446}
\end{barticle}
\endbibitem

\bibitem[\protect\citeauthoryear{{Evrard} et~al.}{2002}]{evrard02}
\begin{barticle}
\bauthor{\binits{A.E.} \bsnm{{Evrard}}},
\bauthor{\binits{T.J.} \bsnm{{MacFarland}}},
\bauthor{\binits{H.M.P.} \bsnm{{Couchman}}},
\bauthor{\binits{J.M.} \bsnm{{Colberg}}},
\bauthor{\binits{N.} \bsnm{{Yoshida}}},
\bauthor{\binits{S.D.M.} \bsnm{{White}}},
\bauthor{\binits{A.} \bsnm{{Jenkins}}},
\bauthor{\binits{C.S.} \bsnm{{Frenk}}},
\bauthor{\binits{F.R.} \bsnm{{Pearce}}},
\bauthor{\binits{J.A.} \bsnm{{Peacock}}},
\bauthor{\binits{P.A.} \bsnm{{Thomas}}},
\batitle{{Galaxy Clusters in Hubble Volume Simulations: Cosmological
  Constraints from Sky Survey Populations}}.
\bjtitle{\apj}
\bvolume{573},
\bfpage{7}--\blpage{36}
(\byear{2002}).
doi:\doiurl{10.1086/340551}
\end{barticle}
\endbibitem

\bibitem[\protect\citeauthoryear{{Fabricant} et~al.}{1984}]{fabricant84}
\begin{barticle}
\bauthor{\binits{D.} \bsnm{{Fabricant}}},
\bauthor{\binits{G.} \bsnm{{Rybicki}}},
\bauthor{\binits{P.} \bsnm{{Gorenstein}}},
\batitle{{X-ray measurements of the nonspherical mass distribution in the
  cluster of galaxies A2256}}.
\bjtitle{\apj}
\bvolume{286},
\bfpage{186}--\blpage{195}
(\byear{1984}).
doi:\doiurl{10.1086/162586}
\end{barticle}
\endbibitem

\bibitem[\protect\citeauthoryear{{Fasano} and {Vio}}{1991}]{fa+vi91}
\begin{barticle}
\bauthor{\binits{G.} \bsnm{{Fasano}}},
\bauthor{\binits{R.} \bsnm{{Vio}}},
\batitle{{Apparent and true flattening distribution of elliptical galaxies}}.
\bjtitle{MNRAS}
\bvolume{249},
\bfpage{629}--\blpage{633}
(\byear{1991})
\end{barticle}
\endbibitem

\bibitem[\protect\citeauthoryear{{Fox} and {Pen}}{2002}]{foxpen}
\begin{barticle}
\bauthor{\binits{D.C.} \bsnm{{Fox}}},
\bauthor{\binits{U.-L.} \bsnm{{Pen}}},
\batitle{{The Distance to Clusters: Correcting for Asphericity}}.
\bjtitle{\apj}
\bvolume{574},
\bfpage{38}--\blpage{50}
(\byear{2002}).
doi:\doiurl{10.1086/340897}
\end{barticle}
\endbibitem

\bibitem[\protect\citeauthoryear{{Frenk} et~al.}{1988}]{frenk88}
\begin{barticle}
\bauthor{\binits{C.S.} \bsnm{{Frenk}}},
\bauthor{\binits{S.D.M.} \bsnm{{White}}},
\bauthor{\binits{M.} \bsnm{{Davis}}},
\bauthor{\binits{G.} \bsnm{{Efstathiou}}},
\batitle{{The formation of dark halos in a universe dominated by cold dark
  matter}}.
\bjtitle{\apj}
\bvolume{327},
\bfpage{507}--\blpage{525}
(\byear{1988}).
doi:\doiurl{10.1086/166213}
\end{barticle}
\endbibitem

\bibitem[\protect\citeauthoryear{{Gao} et~al.}{2008}]{gao+al08}
\begin{barticle}
\bauthor{\binits{L.} \bsnm{{Gao}}},
\bauthor{\binits{J.F.} \bsnm{{Navarro}}},
\bauthor{\binits{S.} \bsnm{{Cole}}},
\bauthor{\binits{C.S.} \bsnm{{Frenk}}},
\bauthor{\binits{S.D.M.} \bsnm{{White}}},
\bauthor{\binits{V.} \bsnm{{Springel}}},
\bauthor{\binits{A.} \bsnm{{Jenkins}}},
\bauthor{\binits{A.F.} \bsnm{{Neto}}},
\batitle{{The redshift dependence of the structure of massive {$\Lambda$} cold
  dark matter haloes}}.
%\bjtitle{\mnras}
\bvolume{387},
\bfpage{536}--\blpage{544}
(\byear{2008}).
doi:\doiurl{10.1111/j.1365-2966.2008.13277.x}
\end{barticle}
\endbibitem

\bibitem[\protect\citeauthoryear{{Gao} et~al.}{2012}]{phoenix}
\begin{botherref}
\oauthor{\binits{L.} \bsnm{{Gao}}},
\oauthor{\binits{J.F.} \bsnm{{Navarro}}},
\oauthor{\binits{C.S.} \bsnm{{Frenk}}},
\oauthor{\binits{A.} \bsnm{{Jenkins}}},
\oauthor{\binits{V.} \bsnm{{Springel}}},
\oauthor{\binits{S.D.M.} \bsnm{{White}}},
{The Phoenix Project: the Dark Side of Rich Galaxy Clusters}.
ArXiv e-prints
(2012)
\end{botherref}
\endbibitem

\bibitem[\protect\citeauthoryear{{Gavazzi}}{2005}]{gavazzi05}
\begin{barticle}
\bauthor{\binits{R.} \bsnm{{Gavazzi}}},
\batitle{{Projection effects in cluster mass estimates: the case of
  MS2137-23}}.
\bjtitle{\aap}
\bvolume{443},
\bfpage{793}--\blpage{804}
(\byear{2005}).
doi:\doiurl{10.1051/0004-6361:20053166}
\end{barticle}
\endbibitem

\bibitem[\protect\citeauthoryear{{Gehrels}}{1986}]{gehrels1986}
\begin{barticle}
\bauthor{\binits{N.} \bsnm{{Gehrels}}},
\batitle{{Confidence limits for small numbers of events in astrophysical
  data}}.
\bjtitle{\apj}
\bvolume{303},
\bfpage{336}--\blpage{346}
(\byear{1986}).
doi:\doiurl{10.1086/164079}
\end{barticle}
\endbibitem

\bibitem[\protect\citeauthoryear{{Gitti} et~al.}{2012}]{gitti2012}
\begin{botherref}
\oauthor{\binits{M.} \bsnm{{Gitti}}},
\oauthor{\binits{F.} \bsnm{{Brighenti}}},
\oauthor{\binits{B.R.} \bsnm{{McNamara}}},
{Evidence for AGN Feedback in Galaxy Clusters and Groups}.
Advances in Astronomy
\textbf{2012}
(2012).
doi:\doiurl{10.1155/2012/950641}
\end{botherref}
\endbibitem

\bibitem[\protect\citeauthoryear{{Gralla} et~al.}{2011}]{kerenSZ}
\begin{barticle}
\bauthor{\binits{M.B.} \bsnm{{Gralla}}},
\bauthor{\binits{K.} \bsnm{{Sharon}}},
\bauthor{\binits{M.D.} \bsnm{{Gladders}}},
\bauthor{\binits{D.P.} \bsnm{{Marrone}}},
\bauthor{\binits{L.F.} \bsnm{{Barrientos}}},
\bauthor{\binits{M.} \bsnm{{Bayliss}}},
\bauthor{\binits{M.} \bsnm{{Bonamente}}},
\bauthor{\binits{E.} \bsnm{{Bulbul}}},
\bauthor{\binits{J.E.} \bsnm{{Carlstrom}}},
\bauthor{\binits{T.} \bsnm{{Culverhouse}}},
\bauthor{\binits{D.G.} \bsnm{{Gilbank}}},
\bauthor{\binits{C.} \bsnm{{Greer}}},
\bauthor{\binits{N.} \bsnm{{Hasler}}},
\bauthor{\binits{D.} \bsnm{{Hawkins}}},
\bauthor{\binits{R.} \bsnm{{Hennessy}}},
\bauthor{\binits{M.} \bsnm{{Joy}}},
\bauthor{\binits{B.} \bsnm{{Koester}}},
\bauthor{\binits{J.} \bsnm{{Lamb}}},
\bauthor{\binits{E.} \bsnm{{Leitch}}},
\bauthor{\binits{A.} \bsnm{{Miller}}},
\bauthor{\binits{T.} \bsnm{{Mroczkowski}}},
\bauthor{\binits{S.} \bsnm{{Muchovej}}},
\bauthor{\binits{M.} \bsnm{{Oguri}}},
\bauthor{\binits{T.} \bsnm{{Plagge}}},
\bauthor{\binits{C.} \bsnm{{Pryke}}},
\bauthor{\binits{D.} \bsnm{{Woody}}},
\batitle{{Sunyaev-Zel'dovich Effect Observations of Strong Lensing Galaxy
  Clusters: Probing the Overconcentration Problem}}.
\bjtitle{\apj}
\bvolume{737},
\bfpage{74}
(\byear{2011}).
doi:\doiurl{10.1088/0004-637X/737/2/74}
\end{barticle}
\endbibitem

\bibitem[\protect\citeauthoryear{{Guennou} et~al.}{2010}]{guennou10}
\begin{barticle}
\bauthor{\binits{L.} \bsnm{{Guennou}}},
\bauthor{\binits{C.} \bsnm{{Adami}}},
\bauthor{\binits{M.P.} \bsnm{{Ulmer}}},
\bauthor{\binits{V.} \bsnm{{Lebrun}}},
\bauthor{\binits{F.} \bsnm{{Durret}}},
\bauthor{\binits{D.} \bsnm{{Johnston}}},
\bauthor{\binits{O.} \bsnm{{Ilbert}}},
\bauthor{\binits{D.} \bsnm{{Clowe}}},
\bauthor{\binits{R.} \bsnm{{Gavazzi}}},
\bauthor{\binits{K.} \bsnm{{Murphy}}},
\bauthor{\binits{T.} \bsnm{{Schrabback}}},
\bauthor{\binits{S.} \bsnm{{Allam}}},
\bauthor{\binits{J.} \bsnm{{Annis}}},
\bauthor{\binits{S.} \bsnm{{Basa}}},
\bauthor{\binits{C.} \bsnm{{Benoist}}},
\bauthor{\binits{A.} \bsnm{{Biviano}}},
\bauthor{\binits{A.} \bsnm{{Cappi}}},
\bauthor{\binits{J.M.} \bsnm{{Kubo}}},
\bauthor{\binits{P.} \bsnm{{Marshall}}},
\bauthor{\binits{A.} \bsnm{{Mazure}}},
\bauthor{\binits{F.} \bsnm{{Rostagni}}},
\bauthor{\binits{D.} \bsnm{{Russeil}}},
\bauthor{\binits{E.} \bsnm{{Slezak}}},
\batitle{{The DAFT/FADA survey. I. Photometric redshifts along lines of sight
  to clusters in the z = [0.4, 0.9] interval}}.
\bjtitle{\aap}
\bvolume{523},
\bfpage{21}
(\byear{2010}).
doi:\doiurl{10.1051/0004-6361/201015174}
\end{barticle}
\endbibitem

\bibitem[\protect\citeauthoryear{{Guennou} et~al.}{2012}]{guennou12}
\begin{barticle}
\bauthor{\binits{L.} \bsnm{{Guennou}}},
\bauthor{\binits{C.} \bsnm{{Adami}}},
\bauthor{\binits{C.} \bsnm{{Da Rocha}}},
\bauthor{\binits{F.} \bsnm{{Durret}}},
\bauthor{\binits{M.P.} \bsnm{{Ulmer}}},
\bauthor{\binits{S.} \bsnm{{Allam}}},
\bauthor{\binits{S.} \bsnm{{Basa}}},
\bauthor{\binits{C.} \bsnm{{Benoist}}},
\bauthor{\binits{A.} \bsnm{{Biviano}}},
\bauthor{\binits{D.} \bsnm{{Clowe}}},
\bauthor{\binits{R.} \bsnm{{Gavazzi}}},
\bauthor{\binits{C.} \bsnm{{Halliday}}},
\bauthor{\binits{O.} \bsnm{{Ilbert}}},
\bauthor{\binits{D.} \bsnm{{Johnston}}},
\bauthor{\binits{D.} \bsnm{{Just}}},
\bauthor{\binits{R.} \bsnm{{Kron}}},
\bauthor{\binits{J.M.} \bsnm{{Kubo}}},
\bauthor{\binits{V.} \bsnm{{Le Brun}}},
\bauthor{\binits{P.} \bsnm{{Marshall}}},
\bauthor{\binits{A.} \bsnm{{Mazure}}},
\bauthor{\binits{K.J.} \bsnm{{Murphy}}},
\bauthor{\binits{D.N.E.} \bsnm{{Pereira}}},
\bauthor{\binits{C.R.} \bsnm{{Raba{\c c}a}}},
\bauthor{\binits{F.} \bsnm{{Rostagni}}},
\bauthor{\binits{G.} \bsnm{{Rudnick}}},
\bauthor{\binits{D.} \bsnm{{Russeil}}},
\bauthor{\binits{T.} \bsnm{{Schrabback}}},
\bauthor{\binits{E.} \bsnm{{Slezak}}},
\bauthor{\binits{D.} \bsnm{{Tucker}}},
\bauthor{\binits{D.} \bsnm{{Zaritsky}}},
\batitle{{Intracluster light in clusters of galaxies at redshifts 0.4  z
   0.8}}.
\bjtitle{\aap}
\bvolume{537},
\bfpage{64}
(\byear{2012}).
doi:\doiurl{10.1051/0004-6361/201117482}
\end{barticle}
\endbibitem

\bibitem[\protect\citeauthoryear{{Halkola} et~al.}{2006}]{halkola}
\begin{barticle}
\bauthor{\binits{A.} \bsnm{{Halkola}}},
\bauthor{\binits{S.} \bsnm{{Seitz}}},
\bauthor{\binits{M.} \bsnm{{Pannella}}},
\batitle{{Parametric strong gravitational lensing analysis of Abell 1689}}.
%\bjtitle{\mnras}
\bvolume{372},
\bfpage{1425}--\blpage{1462}
(\byear{2006}).
doi:\doiurl{10.1111/j.1365-2966.2006.10948.x}
\end{barticle}
\endbibitem

\bibitem[\protect\citeauthoryear{{Hayashi} et~al.}{2007}]{hayashi07}
\begin{barticle}
\bauthor{\binits{E.} \bsnm{{Hayashi}}},
\bauthor{\binits{J.F.} \bsnm{{Navarro}}},
\bauthor{\binits{V.} \bsnm{{Springel}}},
\batitle{{The shape of the gravitational potential in cold dark matter
  haloes}}.
%\bjtitle{\mnras}
\bvolume{377},
\bfpage{50}--\blpage{62}
(\byear{2007}).
doi:\doiurl{10.1111/j.1365-2966.2007.11599.x}
\end{barticle}
\endbibitem

\bibitem[\protect\citeauthoryear{{Hennawi} et~al.}{2007}]{hennawi07}
\begin{barticle}
\bauthor{\binits{J.F.} \bsnm{{Hennawi}}},
\bauthor{\binits{N.} \bsnm{{Dalal}}},
\bauthor{\binits{P.} \bsnm{{Bode}}},
\bauthor{\binits{J.P.} \bsnm{{Ostriker}}},
\batitle{{Characterizing the Cluster Lens Population}}.
\bjtitle{\apj}
\bvolume{654},
\bfpage{714}--\blpage{730}
(\byear{2007}).
doi:\doiurl{10.1086/497362}
\end{barticle}
\endbibitem

\bibitem[\protect\citeauthoryear{{Hoekstra} et~al.}{2011}]{henk2011}
\begin{barticle}
\bauthor{\binits{H.} \bsnm{{Hoekstra}}},
\bauthor{\binits{J.} \bsnm{{Hartlap}}},
\bauthor{\binits{S.} \bsnm{{Hilbert}}},
\bauthor{\binits{E.} \bsnm{{van Uitert}}},
\batitle{{Effects of distant large-scale structure on the precision of weak
  lensing mass measurements}}.
%\bjtitle{\mnras}
\bvolume{412},
\bfpage{2095}--\blpage{2103}
(\byear{2011}).
doi:\doiurl{10.1111/j.1365-2966.2010.18053.x}
\end{barticle}
\endbibitem

\bibitem[\protect\citeauthoryear{{Hopkins} et~al.}{2005}]{hopkins05}
\begin{barticle}
\bauthor{\binits{P.F.} \bsnm{{Hopkins}}},
\bauthor{\binits{N.A.} \bsnm{{Bahcall}}},
\bauthor{\binits{P.} \bsnm{{Bode}}},
\batitle{{Cluster Alignments and Ellipticities in {$\Lambda$}CDM Cosmology}}.
\bjtitle{\apj}
\bvolume{618},
\bfpage{1}--\blpage{15}
(\byear{2005}).
doi:\doiurl{10.1086/425993}
\end{barticle}
\endbibitem

\bibitem[\protect\citeauthoryear{{Hubble}}{1926}]{hub26}
\begin{barticle}
\bauthor{\binits{E.P.} \bsnm{{Hubble}}},
\batitle{{Extragalactic nebulae.}}
\bjtitle{ApJ}
\bvolume{64},
\bfpage{321}--\blpage{369}
(\byear{1926})
\end{barticle}
\endbibitem

\bibitem[\protect\citeauthoryear{{Itoh} et~al.}{1998}]{itoh1998}
\begin{barticle}
\bauthor{\binits{N.} \bsnm{{Itoh}}},
\bauthor{\binits{Y.} \bsnm{{Kohyama}}},
\bauthor{\binits{S.} \bsnm{{Nozawa}}},
\batitle{{Relativistic Corrections to the Sunyaev-Zeldovich Effect for Clusters
  of Galaxies}}.
\bjtitle{\apj}
\bvolume{502},
\bfpage{7}
(\byear{1998}).
doi:\doiurl{10.1086/305876}
\end{barticle}
\endbibitem

\bibitem[\protect\citeauthoryear{{Jing} and {Suto}}{2002}]{jing2002}
\begin{barticle}
\bauthor{\binits{Y.P.} \bsnm{{Jing}}},
\bauthor{\binits{Y.} \bsnm{{Suto}}},
\batitle{{Triaxial Modeling of Halo Density Profiles with High-Resolution
  N-Body Simulations}}.
\bjtitle{\apj}
\bvolume{574},
\bfpage{538}--\blpage{553}
(\byear{2002}).
doi:\doiurl{10.1086/341065}
\end{barticle}
\endbibitem

\bibitem[\protect\citeauthoryear{{Jullo} et~al.}{2007}]{jullo07}
\begin{barticle}
\bauthor{\binits{E.} \bsnm{{Jullo}}},
\bauthor{\binits{J.-P.} \bsnm{{Kneib}}},
\bauthor{\binits{M.} \bsnm{{Limousin}}},
\bauthor{\binits{{\'A}.} \bsnm{{El{\'{\i}}asd{\'o}ttir}}},
\bauthor{\binits{P.J.} \bsnm{{Marshall}}},
\bauthor{\binits{T.} \bsnm{{Verdugo}}},
\batitle{{A Bayesian approach to strong lensing modelling of galaxy clusters}}.
\bjtitle{New Journal of Physics}
\bvolume{9},
\bfpage{447}
(\byear{2007}).
doi:\doiurl{10.1088/1367-2630/9/12/447}
\end{barticle}
\endbibitem

\bibitem[\protect\citeauthoryear{{Kartaltepe} et~al.}{2008}]{macslargescale}
\begin{barticle}
\bauthor{\binits{J.S.} \bsnm{{Kartaltepe}}},
\bauthor{\binits{H.} \bsnm{{Ebeling}}},
\bauthor{\binits{C.J.} \bsnm{{Ma}}},
\bauthor{\binits{D.} \bsnm{{Donovan}}},
\batitle{{Probing the large-scale structure around the most distant galaxy
  clusters from the massive cluster survey}}.
%\bjtitle{\mnras}
\bvolume{389},
\bfpage{1240}--\blpage{1248}
(\byear{2008}).
doi:\doiurl{10.1111/j.1365-2966.2008.13620.x}
\end{barticle}
\endbibitem

\bibitem[\protect\citeauthoryear{{Kasun} and {Evrard}}{2005}]{kasunevrard05}
\begin{barticle}
\bauthor{\binits{S.F.} \bsnm{{Kasun}}},
\bauthor{\binits{A.E.} \bsnm{{Evrard}}},
\batitle{{Shapes and Alignments of Galaxy Cluster Halos}}.
\bjtitle{\apj}
\bvolume{629},
\bfpage{781}--\blpage{790}
(\byear{2005}).
doi:\doiurl{10.1086/430811}
\end{barticle}
\endbibitem

\bibitem[\protect\citeauthoryear{{Kawahara}}{2010}]{kawahara}
\begin{barticle}
\bauthor{\binits{H.} \bsnm{{Kawahara}}},
\batitle{{The Axis Ratio Distribution of X-ray Clusters Observed by
  XMM-Newton}}.
\bjtitle{\apj}
\bvolume{719},
\bfpage{1926}--\blpage{1931}
(\byear{2010}).
doi:\doiurl{10.1088/0004-637X/719/2/1926}
\end{barticle}
\endbibitem

\bibitem[\protect\citeauthoryear{{Kawaharada} et~al.}{2010}]{suzaku1689}
\begin{barticle}
\bauthor{\binits{M.} \bsnm{{Kawaharada}}},
\bauthor{\binits{N.} \bsnm{{Okabe}}},
\bauthor{\binits{K.} \bsnm{{Umetsu}}},
\bauthor{\binits{M.} \bsnm{{Takizawa}}},
\bauthor{\binits{K.} \bsnm{{Matsushita}}},
\bauthor{\binits{Y.} \bsnm{{Fukazawa}}},
\bauthor{\binits{T.} \bsnm{{Hamana}}},
\bauthor{\binits{S.} \bsnm{{Miyazaki}}},
\bauthor{\binits{K.} \bsnm{{Nakazawa}}},
\bauthor{\binits{T.} \bsnm{{Ohashi}}},
\batitle{{Suzaku Observation of A1689: Anisotropic Temperature and Entropy
  Distributions Associated with the Large-scale Structure}}.
\bjtitle{\apj}
\bvolume{714},
\bfpage{423}--\blpage{441}
(\byear{2010}).
doi:\doiurl{10.1088/0004-637X/714/1/423}
\end{barticle}
\endbibitem

\bibitem[\protect\citeauthoryear{{Keeton}}{2001a}]{kee01b}
\begin{botherref}
\oauthor{\binits{C.R.} \bsnm{{Keeton}}},
{A Catalog of Mass Models for Gravitational Lensing}.

astro-ph/0102341(2001a)
\end{botherref}
\endbibitem

\bibitem[\protect\citeauthoryear{{Keeton}}{2001b}]{kee01a}
\begin{botherref}
\oauthor{\binits{C.R.} \bsnm{{Keeton}}},
{Computational Methods for Gravitational Lensing}.

astro-ph/0102340(2001b)
\end{botherref}
\endbibitem

\bibitem[\protect\citeauthoryear{{King} et~al.}{2002}]{king02a}
\begin{barticle}
\bauthor{\binits{L.J.} \bsnm{{King}}},
\bauthor{\binits{D.I.} \bsnm{{Clowe}}},
\bauthor{\binits{P.} \bsnm{{Schneider}}},
\batitle{{Parameterised models for the lensing cluster Abell 1689}}.
\bjtitle{\aap}
\bvolume{383},
\bfpage{118}--\blpage{124}
(\byear{2002}).
doi:\doiurl{10.1051/0004-6361:20011722}
\end{barticle}
\endbibitem

\bibitem[\protect\citeauthoryear{{Kitayama} et~al.}{2004}]{kitayama}
\begin{barticle}
\bauthor{\binits{T.} \bsnm{{Kitayama}}},
\bauthor{\binits{E.} \bsnm{{Komatsu}}},
\bauthor{\binits{N.} \bsnm{{Ota}}},
\bauthor{\binits{T.} \bsnm{{Kuwabara}}},
\bauthor{\binits{Y.} \bsnm{{Suto}}},
\bauthor{\binits{K.} \bsnm{{Yoshikawa}}},
\bauthor{\binits{M.} \bsnm{{Hattori}}},
\bauthor{\binits{H.} \bsnm{{Matsuo}}},
\batitle{{Exploring Cluster Physics with High-Resolution Sunyaev--Zel'dovich
  Effect Images and X-Ray Data: The Case of the Most X-Ray-Luminous Galaxy
  Cluster RX J1347-1145}}.
%\bjtitle{\pasj}
\bvolume{56},
\bfpage{17}--\blpage{28}
(\byear{2004})
\end{barticle}
\endbibitem

\bibitem[\protect\citeauthoryear{{Lau} et~al.}{2009}]{lau09}
\begin{barticle}
\bauthor{\binits{E.T.} \bsnm{{Lau}}},
\bauthor{\binits{A.V.} \bsnm{{Kravtsov}}},
\bauthor{\binits{D.} \bsnm{{Nagai}}},
\batitle{{Residual Gas Motions in the Intracluster Medium and Bias in
  Hydrostatic Measurements of Mass Profiles of Clusters}}.
\bjtitle{\apj}
\bvolume{705},
\bfpage{1129}--\blpage{1138}
(\byear{2009}).
doi:\doiurl{10.1088/0004-637X/705/2/1129}
\end{barticle}
\endbibitem

\bibitem[\protect\citeauthoryear{{Lau} et~al.}{2012}]{lau2012}
\begin{botherref}
\oauthor{\binits{E.T.} \bsnm{{Lau}}},
\oauthor{\binits{D.} \bsnm{{Nagai}}},
\oauthor{\binits{A.V.} \bsnm{{Kravtsov}}},
\oauthor{\binits{A.} \bsnm{{Vikhlinin}}},
\oauthor{\binits{A.R.} \bsnm{{Zentner}}},
{Constraining Cluster Physics with the Shape of X-ray Clusters: Comparison of
  Local X-ray Clusters versus LCDM Clusters}.
ArXiv e-prints
(2012)
\end{botherref}
\endbibitem

\bibitem[\protect\citeauthoryear{{Lee} and {Suto}}{2003}]{lee2003}
\begin{barticle}
\bauthor{\binits{J.} \bsnm{{Lee}}},
\bauthor{\binits{Y.} \bsnm{{Suto}}},
\batitle{{Modeling Intracluster Gas in Triaxial Dark Halos: An Analytic Ap
  proach}}.
\bjtitle{\apj}
\bvolume{585},
\bfpage{151}--\blpage{160}
(\byear{2003}).
doi:\doiurl{10.1086/345931}
\end{barticle}
\endbibitem

\bibitem[\protect\citeauthoryear{{Lee} and {Suto}}{2004}]{lee2004}
\begin{barticle}
\bauthor{\binits{J.} \bsnm{{Lee}}},
\bauthor{\binits{Y.} \bsnm{{Suto}}},
\batitle{{Reconstructing the Three-dimensional Structure of Underlying Triaxial
  Dark Halos from X-Ray and Sunyaev-Zel'dovich Effect Observations of Galaxy
  Clusters}}.
\bjtitle{\apj}
\bvolume{601},
\bfpage{599}--\blpage{609}
(\byear{2004}).
doi:\doiurl{10.1086/380506}
\end{barticle}
\endbibitem

\bibitem[\protect\citeauthoryear{{Lee} et~al.}{2005}]{lee05}
\begin{barticle}
\bauthor{\binits{J.} \bsnm{{Lee}}},
\bauthor{\binits{Y.P.} \bsnm{{Jing}}},
\bauthor{\binits{Y.} \bsnm{{Suto}}},
\batitle{{An Analytic Model for the Axis Ratio Distribution of Dark Matter
  Halos from the Primordial Gaussian Density Field}}.
\bjtitle{\apj}
\bvolume{632},
\bfpage{706}--\blpage{712}
(\byear{2005}).
doi:\doiurl{10.1086/444345}
\end{barticle}
\endbibitem

\bibitem[\protect\citeauthoryear{{Lemze} et~al.}{2008}]{lemze}
\begin{barticle}
\bauthor{\binits{D.} \bsnm{{Lemze}}},
\bauthor{\binits{R.} \bsnm{{Barkana}}},
\bauthor{\binits{T.J.} \bsnm{{Broadhurst}}},
\bauthor{\binits{Y.} \bsnm{{Rephaeli}}},
\batitle{{Mass and gas profiles in A1689: joint X-ray and lensing analysis}}.
%\bjtitle{\mnras}
\bvolume{386},
\bfpage{1092}--\blpage{1106}
(\byear{2008}).
doi:\doiurl{10.1111/j.1365-2966.2008.13116.x}
\end{barticle}
\endbibitem

\bibitem[\protect\citeauthoryear{{Limousin} et~al.}{2007}]{mypaperIII}
\begin{barticle}
\bauthor{\binits{M.} \bsnm{{Limousin}}},
\bauthor{\binits{J.} \bsnm{{Richard}}},
\bauthor{\binits{E.} \bsnm{{Jullo}}},
\bauthor{\binits{J.-P.} \bsnm{{Kneib}}},
\bauthor{\binits{B.} \bsnm{{Fort}}},
\bauthor{\binits{G.} \bsnm{{Soucail}}},
\bauthor{\binits{A.} \bsnm{{El{\'{\i}}asd{\'o}ttir}}},
\bauthor{\binits{P.} \bsnm{{Natarajan}}},
\bauthor{\binits{R.S.} \bsnm{{Ellis}}},
\bauthor{\binits{I.} \bsnm{{Smail}}},
\bauthor{\binits{O.} \bsnm{{Czoske}}},
\bauthor{\binits{G.P.} \bsnm{{Smith}}},
\bauthor{\binits{P.} \bsnm{{Hudelot}}},
\bauthor{\binits{S.} \bsnm{{Bardeau}}},
\bauthor{\binits{H.} \bsnm{{Ebeling}}},
\bauthor{\binits{E.} \bsnm{{Egami}}},
\bauthor{\binits{K.K.} \bsnm{{Knudsen}}},
\batitle{{Combining Strong and Weak Gravitational Lensing in Abell 1689}}.
\bjtitle{\apj}
\bvolume{668},
\bfpage{643}--\blpage{666}
(\byear{2007}).
doi:\doiurl{10.1186/383259}
\end{barticle}
\endbibitem

\bibitem[\protect\citeauthoryear{{Limousin} et~al.}{2008}]{mypaperV}
\begin{barticle}
\bauthor{\binits{M.} \bsnm{{Limousin}}},
\bauthor{\binits{J.} \bsnm{{Richard}}},
\bauthor{\binits{J.-P.} \bsnm{{Kneib}}},
\bauthor{\binits{H.} \bsnm{{Brink}}},
\bauthor{\binits{R.} \bsnm{{Pell{\'o}}}},
\bauthor{\binits{E.} \bsnm{{Jullo}}},
\bauthor{\binits{H.} \bsnm{{Tu}}},
\bauthor{\binits{J.} \bsnm{{Sommer-Larsen}}},
\bauthor{\binits{E.} \bsnm{{Egami}}},
\bauthor{\binits{M.J.} \bsnm{{Micha{\l}owski}}},
\bauthor{\binits{R.} \bsnm{{Cabanac}}},
\bauthor{\binits{D.P.} \bsnm{{Stark}}},
\batitle{{Strong lensing in Abell 1703: constraints on the slope of the inner
  daribution}}.
\bjtitle{\aap}
\bvolume{489},
\bfpage{23}--\blpage{35}
(\byear{2008}).
doi:\doiurl{10.1051/0004-6361:200809646}
\end{barticle}
\endbibitem

\bibitem[\protect\citeauthoryear{{Limousin} et~al.}{2012}]{my0717}
\begin{barticle}
\bauthor{\binits{M.} \bsnm{{Limousin}}},
\bauthor{\binits{H.} \bsnm{{Ebeling}}},
\bauthor{\binits{J.} \bsnm{{Richard}}},
\bauthor{\binits{A.M.} \bsnm{{Swinbank}}},
\bauthor{\binits{G.P.} \bsnm{{Smith}}},
\bauthor{\binits{M.} \bsnm{{Jauzac}}},
\bauthor{\binits{S.} \bsnm{{Rodionov}}},
\bauthor{\binits{C.-J.} \bsnm{{Ma}}},
\bauthor{\binits{I.} \bsnm{{Smail}}},
\bauthor{\binits{A.C.} \bsnm{{Edge}}},
\bauthor{\binits{E.} \bsnm{{Jullo}}},
\bauthor{\binits{J.-P.} \bsnm{{Kneib}}},
\batitle{{Strong lensing by a node of the cosmic web. The core of MACS
  J0717.5+3745 at z = 0.55}}.
\bjtitle{\aap}
\bvolume{544},
\bfpage{71}
(\byear{2012}).
doi:\doiurl{10.1051/0004-6361/201117921}
\end{barticle}
\endbibitem

\bibitem[\protect\citeauthoryear{{{\L}okas} et~al.}{2006}]{lokas}
\begin{barticle}
\bauthor{\binits{E.L.} \bsnm{{{\L}okas}}},
\bauthor{\binits{F.} \bsnm{{Prada}}},
\bauthor{\binits{R.} \bsnm{{Wojtak}}},
\bauthor{\binits{M.} \bsnm{{Moles}}},
\bauthor{\binits{S.} \bsnm{{Gottl{\"o}ber}}},
\batitle{{The complex velocity distribution of galaxies in Abell 1689:
  implications for mass modelling}}.
%\bjtitle{\mnras}
\bvolume{366},
\bfpage{26}--\blpage{30}
(\byear{2006}).
doi:\doiurl{10.1111/j.1745-3933.2005.00125.x}
\end{barticle}
\endbibitem

\bibitem[\protect\citeauthoryear{{Ludlow} et~al.}{2012}]{ludlow}
\begin{botherref}
\oauthor{\binits{A.D.} \bsnm{{Ludlow}}},
\oauthor{\binits{J.F.} \bsnm{{Navarro}}},
\oauthor{\binits{M.} \bsnm{{Li}}},
\oauthor{\binits{R.E.} \bsnm{{Angulo}}},
\oauthor{\binits{M.} \bsnm{{Boylan-Kolchin}}},
\oauthor{\binits{P.E.} \bsnm{{Bett}}},
{The Dynamical State and Mass-Concentration Relation of Galaxy Clusters}.
ArXiv e-prints
(2012)
\end{botherref}
\endbibitem

\bibitem[\protect\citeauthoryear{{Macci{\`o}} et~al.}{2008}]{mac+al08}
\begin{barticle}
\bauthor{\binits{A.V.} \bsnm{{Macci{\`o}}}},
\bauthor{\binits{A.A.} \bsnm{{Dutton}}},
\bauthor{\binits{F.C.} \bsnm{{van den Bosch}}},
\batitle{{Concentration, spin and shape of dark matter haloes as a function of
  the cosmological model: WMAP1, WMAP3 and WMAP5 results}}.
%\bjtitle{\mnras}
\bvolume{391},
\bfpage{1940}--\blpage{1954}
(\byear{2008}).
doi:\doiurl{10.1111/j.1365-2966.2008.14029.x}
\end{barticle}
\endbibitem

\bibitem[\protect\citeauthoryear{{Mahdavi} and {Chang}}{2011}]{mahdavi}
\begin{barticle}
\bauthor{\binits{A.} \bsnm{{Mahdavi}}},
\bauthor{\binits{W.} \bsnm{{Chang}}},
\batitle{{Model-independent Limits on the Line-of-sight Depth of Clusters of
  Galaxies Using X-Ray and Sunyaev-Zel'dovich data}}.
%\bjtitle{\apjl}
\bvolume{735},
\bfpage{4}
(\byear{2011}).
doi:\doiurl{10.1088/2041-8205/735/1/L4}
\end{barticle}
\endbibitem

\bibitem[\protect\citeauthoryear{{Mahdavi} et~al.}{2007}]{jaco}
\begin{barticle}
\bauthor{\binits{A.} \bsnm{{Mahdavi}}},
\bauthor{\binits{H.} \bsnm{{Hoekstra}}},
\bauthor{\binits{A.} \bsnm{{Babul}}},
\bauthor{\binits{J.} \bsnm{{Sievers}}},
\bauthor{\binits{S.T.} \bsnm{{Myers}}},
\bauthor{\binits{J.P.} \bsnm{{Henry}}},
\batitle{{Joint Analysis of Cluster Observations. I. Mass Profile of Abell 478
  from Combined X-Ray, Sunyaev-Zel'dovich, and Weak-Lensing Data}}.
\bjtitle{\apj}
\bvolume{664},
\bfpage{162}--\blpage{180}
(\byear{2007}).
doi:\doiurl{10.1086/517958}
\end{barticle}
\endbibitem

\bibitem[\protect\citeauthoryear{{Mahdavi} et~al.}{2008}]{mahdavi08}
\begin{barticle}
\bauthor{\binits{A.} \bsnm{{Mahdavi}}},
\bauthor{\binits{H.} \bsnm{{Hoekstra}}},
\bauthor{\binits{A.} \bsnm{{Babul}}},
\bauthor{\binits{J.P.} \bsnm{{Henry}}},
\batitle{{Evidence for non-hydrostatic gas from the cluster X-ray to lensing
  mass ratio}}.
%\bjtitle{\mnras}
\bvolume{384},
\bfpage{1567}--\blpage{1574}
(\byear{2008}).
doi:\doiurl{10.1111/j.1365-2966.2007.12796.x}
\end{barticle}
\endbibitem

\bibitem[\protect\citeauthoryear{{Mann} and {Ebeling}}{2011}]{mannebeling}
\begin{botherref}
\oauthor{\binits{A.W.} \bsnm{{Mann}}},
\oauthor{\binits{H.} \bsnm{{Ebeling}}},
{X-ray/optical classification of cluster mergers and the evolution of the
  cluster merger fraction}.
ArXiv e-prints
(2011)
\end{botherref}
\endbibitem

\bibitem[\protect\citeauthoryear{{Marshall} et~al.}{2003}]{phil2003}
\begin{barticle}
\bauthor{\binits{P.J.} \bsnm{{Marshall}}},
\bauthor{\binits{M.P.} \bsnm{{Hobson}}},
\bauthor{\binits{A.} \bsnm{{Slosar}}},
\batitle{{Bayesian joint analysis of cluster weak lensing and
  Sunyaev-Zel'dovich effect data}}.
%\bjtitle{\mnras}
\bvolume{346},
\bfpage{489}--\blpage{500}
(\byear{2003}).
doi:\doiurl{10.1046/j.1365-2966.2003.07111.x}
\end{barticle}
\endbibitem

\bibitem[\protect\citeauthoryear{{Mazzotta} et~al.}{2004}]{mazzotta2004}
\begin{barticle}
\bauthor{\binits{P.} \bsnm{{Mazzotta}}},
\bauthor{\binits{E.} \bsnm{{Rasia}}},
\bauthor{\binits{L.} \bsnm{{Moscardini}}},
\bauthor{\binits{G.} \bsnm{{Tormen}}},
\batitle{{Comparing the temperatures of galaxy clusters from hydrodynamical
  N-body simulations to Chandra and XMM-Newton observations}}.
%\bjtitle{\mnras}
\bvolume{354},
\bfpage{10}--\blpage{24}
(\byear{2004}).
doi:\doiurl{10.1111/j.1365-2966.2004.08167.x}
\end{barticle}
\endbibitem

\bibitem[\protect\citeauthoryear{{Medezinski} et~al.}{2007}]{elinor}
\begin{barticle}
\bauthor{\binits{E.} \bsnm{{Medezinski}}},
\bauthor{\binits{T.} \bsnm{{Broadhurst}}},
\bauthor{\binits{K.} \bsnm{{Umetsu}}},
\bauthor{\binits{D.} \bsnm{{Coe}}},
\bauthor{\binits{N.} \bsnm{{Ben{\'{\i}}tez}}},
\bauthor{\binits{H.} \bsnm{{Ford}}},
\bauthor{\binits{Y.} \bsnm{{Rephaeli}}},
\bauthor{\binits{N.} \bsnm{{Arimoto}}},
\bauthor{\binits{X.} \bsnm{{Kong}}},
\batitle{{Using Weak-Lensing Dilution to Improve Measurements of the Luminous
  and Dark Matter in A1689}}.
\bjtitle{\apj}
\bvolume{663},
\bfpage{717}--\blpage{733}
(\byear{2007}).
doi:\doiurl{10.1086/518638}
\end{barticle}
\endbibitem

\bibitem[\protect\citeauthoryear{{Meneghetti} et~al.}{2003}]{massimo03}
\begin{barticle}
\bauthor{\binits{M.} \bsnm{{Meneghetti}}},
\bauthor{\binits{M.} \bsnm{{Bartelmann}}},
\bauthor{\binits{L.} \bsnm{{Moscardini}}},
\batitle{{cD galaxy contribution to the strong lensing cross-sections of galaxy
  clusters}}.
%\bjtitle{\mnras}
\bvolume{346},
\bfpage{67}--\blpage{77}
(\byear{2003}).
doi:\doiurl{10.1046/j.1365-2966.2003.07068.x}
\end{barticle}
\endbibitem

\bibitem[\protect\citeauthoryear{{Meneghetti} et~al.}{2001}]{massimo01}
\begin{barticle}
\bauthor{\binits{M.} \bsnm{{Meneghetti}}},
\bauthor{\binits{N.} \bsnm{{Yoshida}}},
\bauthor{\binits{M.} \bsnm{{Bartelmann}}},
\bauthor{\binits{L.} \bsnm{{Moscardini}}},
\bauthor{\binits{V.} \bsnm{{Springel}}},
\bauthor{\binits{G.} \bsnm{{Tormen}}},
\bauthor{\binits{S.D.M.} \bsnm{{White}}},
\batitle{{Giant cluster arcs as a constraint on the scattering cross-section of
  dark matter}}.
%\bjtitle{\mnras}
\bvolume{325},
\bfpage{435}--\blpage{442}
(\byear{2001}).
doi:\doiurl{10.1046/j.1365-8711.2001.04477.x}
\end{barticle}
\endbibitem

\bibitem[\protect\citeauthoryear{{Meneghetti} et~al.}{2010a}]{massimo2010}
\begin{barticle}
\bauthor{\binits{M.} \bsnm{{Meneghetti}}},
\bauthor{\binits{C.} \bsnm{{Fedeli}}},
\bauthor{\binits{F.} \bsnm{{Pace}}},
\bauthor{\binits{S.} \bsnm{{Gottl{\"o}ber}}},
\bauthor{\binits{G.} \bsnm{{Yepes}}},
\batitle{{Strong lensing in the MARENOSTRUM UNIVERSE. I. Biases in the cluster
  lens population}}.
\bjtitle{\aap}
\bvolume{519},
\bfpage{90}
(\byear{2010}a).
doi:\doiurl{10.1051/0004-6361/201014098}
\end{barticle}
\endbibitem

\bibitem[\protect\citeauthoryear{{Meneghetti} et~al.}{2010b}]{massimo2010a}
\begin{barticle}
\bauthor{\binits{M.} \bsnm{{Meneghetti}}},
\bauthor{\binits{E.} \bsnm{{Rasia}}},
\bauthor{\binits{J.} \bsnm{{Merten}}},
\bauthor{\binits{F.} \bsnm{{Bellagamba}}},
\bauthor{\binits{S.} \bsnm{{Ettori}}},
\bauthor{\binits{P.} \bsnm{{Mazzotta}}},
\bauthor{\binits{K.} \bsnm{{Dolag}}},
\bauthor{\binits{S.} \bsnm{{Marri}}},
\batitle{{Weighing simulated galaxy clusters using lensing and X-ray}}.
\bjtitle{\aap}
\bvolume{514},
\bfpage{93}
(\byear{2010}b).
doi:\doiurl{10.1051/0004-6361/200913222}
\end{barticle}
\endbibitem

\bibitem[\protect\citeauthoryear{{Miralda-Escude}}{2000}]{miralda_escude2000}
\begin{botherref}
\oauthor{\binits{J.} \bsnm{{Miralda-Escude}}},
{A Test of the Collisional Dark Matter Hypothesis from Cluster Lensing}.
ArXiv Astrophysics e-prints
(2000)
\end{botherref}
\endbibitem

\bibitem[\protect\citeauthoryear{{Mohr} et~al.}{1995}]{moh+al95}
\begin{barticle}
\bauthor{\binits{J.J.} \bsnm{{Mohr}}},
\bauthor{\binits{A.E.} \bsnm{{Evrard}}},
\bauthor{\binits{D.G.} \bsnm{{Fabricant}}},
\bauthor{\binits{M.J.} \bsnm{{Geller}}},
\batitle{{Cosmological Constraints from Observed Cluster X-Ray Morphologies}}.
\bjtitle{ApJ}
\bvolume{447},
\bfpage{8}
(\byear{1995}).
doi:\doiurl{10.1086/175852}
\end{barticle}
\endbibitem

\bibitem[\protect\citeauthoryear{{Molnar} et~al.}{2010}]{molnar}
\begin{barticle}
\bauthor{\binits{S.M.} \bsnm{{Molnar}}},
\bauthor{\binits{I.-N.} \bsnm{{Chiu}}},
\bauthor{\binits{K.} \bsnm{{Umetsu}}},
\bauthor{\binits{P.} \bsnm{{Chen}}},
\bauthor{\binits{N.} \bsnm{{Hearn}}},
\bauthor{\binits{T.} \bsnm{{Broadhurst}}},
\bauthor{\binits{G.} \bsnm{{Bryan}}},
\bauthor{\binits{C.} \bsnm{{Shang}}},
\batitle{{Testing Strict Hydrostatic Equilibrium in Simulated Clusters of
  Galaxies: Implications for A1689}}.
%\bjtitle{\apjl}
\bvolume{724},
\bfpage{1}--\blpage{4}
(\byear{2010}).
doi:\doiurl{10.1088/2041-8205/724/1/L1}
\end{barticle}
\endbibitem

\bibitem[\protect\citeauthoryear{{Morandi} and
  {Ettori}}{2007}]{morandi_ettori07}
\begin{barticle}
\bauthor{\binits{A.} \bsnm{{Morandi}}},
\bauthor{\binits{S.} \bsnm{{Ettori}}},
\batitle{{Entropy profiles in X-ray luminous galaxy clusters at z 0.1}}.
%\bjtitle{\mnras}
\bvolume{380},
\bfpage{1521}--\blpage{1532}
(\byear{2007}).
doi:\doiurl{10.1111/j.1365-2966.2007.12158.x}
\end{barticle}
\endbibitem

\bibitem[\protect\citeauthoryear{{Morandi} and {Limousin}}{2012}]{morandi2012a}
\begin{barticle}
\bauthor{\binits{A.} \bsnm{{Morandi}}},
\bauthor{\binits{M.} \bsnm{{Limousin}}},
\batitle{{Triaxiality, principal axis orientation and non-thermal pressure in
  Abell 383}}.
%\bjtitle{\mnras}
\bvolume{421},
\bfpage{3147}--\blpage{3158}
(\byear{2012}).
doi:\doiurl{10.1111/j.1365-2966.2012.20537.x}
\end{barticle}
\endbibitem

\bibitem[\protect\citeauthoryear{{Morandi} et~al.}{2010}]{morandi2010a}
\begin{barticle}
\bauthor{\binits{A.} \bsnm{{Morandi}}},
\bauthor{\binits{K.} \bsnm{{Pedersen}}},
\bauthor{\binits{M.} \bsnm{{Limousin}}},
\batitle{{Unveiling the Three-dimensional Structure of Galaxy Clusters:
  Resolving the Discrepancy Between X-ray and Lensing Masses}}.
\bjtitle{\apj}
\bvolume{713},
\bfpage{491}--\blpage{502}
(\byear{2010}).
doi:\doiurl{10.1088/0004-637X/713/1/491}
\end{barticle}
\endbibitem

\bibitem[\protect\citeauthoryear{{Morandi} et~al.}{2011a}]{morandi2011a}
\begin{barticle}
\bauthor{\binits{A.} \bsnm{{Morandi}}},
\bauthor{\binits{K.} \bsnm{{Pedersen}}},
\bauthor{\binits{M.} \bsnm{{Limousin}}},
\batitle{{Reconstructing the Triaxiality of the Galaxy Cluster A1689: Solving
  the X-ray and Strong Lensing Mass Discrepancy}}.
\bjtitle{\apj}
\bvolume{729},
\bfpage{37}
(\byear{2011}a).
doi:\doiurl{10.1088/0004-637X/729/1/37}
\end{barticle}
\endbibitem

\bibitem[\protect\citeauthoryear{{Morandi} et~al.}{2012b}]{morandi2012b}
\begin{botherref}
\oauthor{\binits{A.} \bsnm{{Morandi}}},
\oauthor{\binits{M.} \bsnm{{Limousin}}},
\oauthor{\binits{J.} \bsnm{{Sayers}}},
\oauthor{\binits{S.R.} \bsnm{{Golwala}}},
\oauthor{\binits{N.G.} \bsnm{{Czakon}}},
\oauthor{\binits{E.} \bsnm{{Pierpaoli}}},
\oauthor{\binits{S.} \bsnm{{Ameglio}}},
{X-ray, lensing and Sunyaev Zel'dovich triaxial analysis of Abell 1835 out to
  R200}.

ArXiv e-prints(2011b)
\end{botherref}
\endbibitem

\bibitem[\protect\citeauthoryear{{Morandi} et~al.}{2011b}]{morandi2011b}
\begin{barticle}
\bauthor{\binits{A.} \bsnm{{Morandi}}},
\bauthor{\binits{M.} \bsnm{{Limousin}}},
\bauthor{\binits{Y.} \bsnm{{Rephaeli}}},
\bauthor{\binits{K.} \bsnm{{Umetsu }}},
\bauthor{\binits{R.} \bsnm{{Barkana}}},
\bauthor{\binits{T.} \bsnm{{Broadhurst}}},
\bauthor{\binits{H.} \bsnm{{Dahle}}},
\batitle{{Triaxiality and non-thermal gas pressure in Abell 1689}}.
%\bjtitle{\mnras}
\bvolume{416},

\bfpage{2567}--\blpage{2573}(\byear{2011}b).
doi:\doiurl{10.1111/j.1365-2966.2011.19175.x}
\end{barticle}
\endbibitem

\bibitem[\protect\citeauthoryear{{Mortonson} et~al.}{2011}]{mortonson}
\begin{barticle}
\bauthor{\binits{M.J.} \bsnm{{Mortonson}}},
\bauthor{\binits{W.} \bsnm{{Hu}}},
\bauthor{\binits{D.} \bsnm{{Huterer}}},
\batitle{{Simultaneous falsification of {$\Lambda$}CDM and quintessence with
  massive, distant clusters}}.
%\bjtitle{\prd}
\bvolume{83}(\bissue{2}),
\bfpage{023015}
(\byear{2011}).
doi:\doiurl{10.1103/PhysRevD.83.023015}
\end{barticle}
\endbibitem

\bibitem[\protect\citeauthoryear{{Mu{\~n}oz-Cuartas}
  et~al.}{2011}]{munozcuartas}
\begin{barticle}
\bauthor{\binits{J.C.} \bsnm{{Mu{\~n}oz-Cuartas}}},
\bauthor{\binits{A.V.} \bsnm{{Macci{\`o}}}},
\bauthor{\binits{S.} \bsnm{{Gottl{\"o}ber}}},
\bauthor{\binits{A.A.} \bsnm{{Dutton}}},
\batitle{{The redshift evolution of {$\Lambda$} cold dark matter halo
  parameters: concentration, spin and shape}}.
%\bjtitle{\mnras}
\bvolume{411},
\bfpage{584}--\blpage{594}
(\byear{2011}).
doi:\doiurl{10.1111/j.1365-2966.2010.17704.x}
\end{barticle}
\endbibitem

\bibitem[\protect\citeauthoryear{{Neto} et~al.}{2007}]{neto}
\begin{barticle}
\bauthor{\binits{A.F.} \bsnm{{Neto}}},
\bauthor{\binits{L.} \bsnm{{Gao}}},
\bauthor{\binits{P.} \bsnm{{Bett}}},
\bauthor{\binits{S.} \bsnm{{Cole}}},
\bauthor{\binits{J.F.} \bsnm{{Navarro}}},
\bauthor{\binits{C.S.} \bsnm{{Frenk}}},
\bauthor{\binits{S.D.M.} \bsnm{{White}}},
\bauthor{\binits{V.} \bsnm{{Springel}}},
\bauthor{\binits{A.} \bsnm{{Jenkins}}},
\batitle{{The statistics of {$\Lambda$} CDM halo concentrations}}.
%\bjtitle{\mnras}
\bvolume{381},
\bfpage{1450}--\blpage{1462}
(\byear{2007}).
doi:\doiurl{10.1111/j.1365-2966.2007.12381.x}
\end{barticle}
\endbibitem

\bibitem[\protect\citeauthoryear{{Newman} et~al.}{2011}]{newman}
\begin{barticle}
\bauthor{\binits{A.B.} \bsnm{{Newman}}},
\bauthor{\binits{T.} \bsnm{{Treu}}},
\bauthor{\binits{R.S.} \bsnm{{Ellis}}},
\bauthor{\binits{D.J.} \bsnm{{Sand}}},
\batitle{{The Dark Matter Distribution in A383: Evidence for a Shallow Density
  Cusd Lensing, Stellar Kinematic, and X-ray Data}}.
%\bjtitle{\apjl}
\bvolume{728},
\bfpage{39}
(\byear{2011})
\end{barticle}
\endbibitem

\bibitem[\protect\citeauthoryear{{Newman} et~al.}{2012a}]{newman2012a}
\begin{botherref}
\oauthor{\binits{A.B.} \bsnm{{Newman}}},
\oauthor{\binits{T.} \bsnm{{Treu}}},
\oauthor{\binits{R.S.} \bsnm{{Ellis}}},
\oauthor{\binits{D.J.} \bsnm{{Sand}}},
\oauthor{\binits{C.} \bsnm{{Nipoti}}},
\oauthor{\binits{J.} \bsnm{{Richard}}},
\oauthor{\binits{E.} \bsnm{{Jullo}}},
{The Density Profiles of Massive, Relaxed Galaxy Clusters: I. The Total Density
  Over 3 Decades in Radius}.

ArXiv e-prints(2012a)
\end{botherref}
\endbibitem

\bibitem[\protect\citeauthoryear{{Newman} et~al.}{2012b}]{newman2012b}
\begin{botherref}
\oauthor{\binits{A.B.} \bsnm{{Newman}}},
\oauthor{\binits{T.} \bsnm{{Treu}}},
\oauthor{\binits{R.S.} \bsnm{{Ellis}}},
\oauthor{\binits{D.J.} \bsnm{{Sand}}},
{The Density Profiles of Massive, Relaxed Galaxy Clusters: II. Separating
  Luminous and Dark Matter in Cluster Cores}.

ArXiv e-prints(2012b)
\end{botherref}
\endbibitem

\bibitem[\protect\citeauthoryear{{Noerdlinger}}{1979}]{noe79}
\begin{barticle}
\bauthor{\binits{P.D.} \bsnm{{Noerdlinger}}},
\batitle{{The intrinsic flattening of galaxies}}.
\bjtitle{ApJ}
\bvolume{234},
\bfpage{802}--\blpage{809}
(\byear{1979}).
doi:\doiurl{10.1086/157559}
\end{barticle}
\endbibitem

\bibitem[\protect\citeauthoryear{{Nord} et~al.}{2009}]{nord09}
\begin{barticle}
\bauthor{\binits{M.} \bsnm{{Nord}}},
\bauthor{\binits{K.} \bsnm{{Basu}}},
\bauthor{\binits{F.} \bsnm{{Pacaud}}},
\bauthor{\binits{P.A.R.} \bsnm{{Ade}}},
\bauthor{\binits{A.N.} \bsnm{{Bender}}},
\bauthor{\binits{B.A.} \bsnm{{Benson}}},
\bauthor{\binits{F.} \bsnm{{Bertoldi}}},
\bauthor{\binits{H.-M.} \bsnm{{Cho}}},
\bauthor{\binits{G.} \bsnm{{Chon}}},
\bauthor{\binits{J.} \bsnm{{Clarke}}},
\bauthor{\binits{M.} \bsnm{{Dobbs}}},
\bauthor{\binits{D.} \bsnm{{Ferrusca}}},
\bauthor{\binits{N.W.} \bsnm{{Halverson}}},
\bauthor{\binits{W.L.} \bsnm{{Holzapfel}}},
\bauthor{\binits{C.} \bsnm{{Horellou}}},
\bauthor{\binits{D.} \bsnm{{Johansson}}},
\bauthor{\binits{J.} \bsnm{{Kennedy}}},
\bauthor{\binits{Z.} \bsnm{{Kermish}}},
\bauthor{\binits{R.} \bsnm{{Kneissl}}},
\bauthor{\binits{T.} \bsnm{{Lanting}}},
\bauthor{\binits{A.T.} \bsnm{{Lee}}},
\bauthor{\binits{M.} \bsnm{{Lueker}}},
\bauthor{\binits{J.} \bsnm{{Mehl}}},
\bauthor{\binits{K.M.} \bsnm{{Menten}}},
\bauthor{\binits{T.} \bsnm{{Plagge}}},
\bauthor{\binits{C.L.} \bsnm{{Reichardt}}},
\bauthor{\binits{P.L.} \bsnm{{Richards}}},
\bauthor{\binits{R.} \bsnm{{Schaaf}}},
\bauthor{\binits{D.} \bsnm{{Schwan}}},
\bauthor{\binits{H.} \bsnm{{Spieler}}},
\bauthor{\binits{C.} \bsnm{{Tucker}}},
\bauthor{\binits{A.} \bsnm{{Weiss}}},
\bauthor{\binits{O.} \bsnm{{Zahn}}},
\batitle{{Multi-frequency imaging of the galaxy cluster Abell 2163 using the
  Sunyaev-Zel'dovich effect}}.
\bjtitle{\aap}
\bvolume{506},
\bfpage{623}--\blpage{636}
(\byear{2009}).
doi:\doiurl{10.1051/0004-6361/200911746}
\end{barticle}
\endbibitem

\bibitem[\protect\citeauthoryear{{Oguri} and {Blandford}}{2009}]{oguri09a}
\begin{barticle}
\bauthor{\binits{M.} \bsnm{{Oguri}}},
\bauthor{\binits{R.D.} \bsnm{{Blandford}}},
\batitle{{What is the largest Einstein radius in the universe?}}
%\bjtitle{\mnras}
\bvolume{392},
\bfpage{930}--\blpage{944}
(\byear{2009}).
doi:\doiurl{10.1111/j.1365-2966.2008.14154.x}
\end{barticle}
\endbibitem

\bibitem[\protect\citeauthoryear{{Oguri} et~al.}{2005}]{oguri}
\begin{barticle}
\bauthor{\binits{M.} \bsnm{{Oguri}}},
\bauthor{\binits{M.} \bsnm{{Takada}}},
\bauthor{\binits{K.} \bsnm{{Umetsu}}},
\bauthor{\binits{T.} \bsnm{{Broadhurst}}},
\batitle{{Can the Steep Mass Profile of A1689 Be Explained by a Triaxial Dark
  Halo?}}
\bjtitle{\apj}
\bvolume{632},
\bfpage{841}--\blpage{846}
(\byear{2005}).
doi:\doiurl{10.1086/452629}
\end{barticle}
\endbibitem

\bibitem[\protect\citeauthoryear{{Oguri} et~al.}{2009}]{oguri09b}
\begin{barticle}
\bauthor{\binits{M.} \bsnm{{Oguri}}},
\bauthor{\binits{J.F.} \bsnm{{Hennawi}}},
\bauthor{\binits{M.D.} \bsnm{{Gladders}}},
\bauthor{\binits{H.} \bsnm{{Dahle}}},
\bauthor{\binits{P.} \bsnm{{Natarajan}}},
\bauthor{\binits{N.} \bsnm{{Dalal}}},
\bauthor{\binits{B.P.} \bsnm{{Koester}}},
\bauthor{\binits{K.} \bsnm{{Sharon}}},
\bauthor{\binits{M.} \bsnm{{Bayliss}}},
\batitle{{Subaru Weak Lensing Measurements of Four Strong Lensing Clusters: Are
  Lensing Clusters Overconcentrated?}}
\bjtitle{\apj}
\bvolume{699},
\bfpage{1038}--\blpage{1052}
(\byear{2009}).
doi:\doiurl{10.1088/0004-637X/699/2/1038}
\end{barticle}
\endbibitem

\bibitem[\protect\citeauthoryear{{Oguri} et~al.}{2010}]{oguri10}
\begin{barticle}
\bauthor{\binits{M.} \bsnm{{Oguri}}},
\bauthor{\binits{M.} \bsnm{{Takada}}},
\bauthor{\binits{N.} \bsnm{{Okabe}}},
\bauthor{\binits{G.P.} \bsnm{{Smith}}},
\batitle{{Direct measurement of dark matter halo ellipticity from
  two-dimensional lensing shear maps of 25 massive clusters}}.
%\bjtitle{\mnras}
\bvolume{405},
\bfpage{2215}--\blpage{2230}
(\byear{2010}).
doi:\doiurl{10.1111/j.1365-2966.2010.16622.x}
\end{barticle}
\endbibitem

\bibitem[\protect\citeauthoryear{{Oguri} et~al.}{2012}]{oguri11}
\begin{barticle}
\bauthor{\binits{M.} \bsnm{{Oguri}}},
\bauthor{\binits{M.B.} \bsnm{{Bayliss}}},
\bauthor{\binits{H.} \bsnm{{Dahle}}},
\bauthor{\binits{K.} \bsnm{{Sharon}}},
\bauthor{\binits{M.D.} \bsnm{{Gladders}}},
\bauthor{\binits{P.} \bsnm{{Natarajan}}},
\bauthor{\binits{J.F.} \bsnm{{Hennawi}}},
\bauthor{\binits{B.P.} \bsnm{{Koester}}},
\batitle{{Combined strong and weak lensing analysis of 28 clusters from the
  Sloan Giant Arcs Survey}}.
%\bjtitle{\mnras}
\bvolume{420},
\bfpage{3213}--\blpage{3239}
(\byear{2012}).
doi:\doiurl{10.1111/j.1365-2966.2011.20248.x}
\end{barticle}
\endbibitem

\bibitem[\protect\citeauthoryear{{Okabe} et~al.}{2010}]{okabe10}
\begin{barticle}
\bauthor{\binits{N.} \bsnm{{Okabe}}},
\bauthor{\binits{M.} \bsnm{{Takada}}},
\bauthor{\binits{K.} \bsnm{{Umetsu}}},
\bauthor{\binits{T.} \bsnm{{Futamase}}},
\bauthor{\binits{G.P.} \bsnm{{Smith}}},
\batitle{{LoCuSS: Subaru Weak Lensing Study of 30 Galaxy Clusters}}.
%\bjtitle{\pasj}
\bvolume{62},
\bfpage{811}
(\byear{2010})
\end{barticle}
\endbibitem

\bibitem[\protect\citeauthoryear{{Patiri} et~al.}{2006}]{patiri06}
\begin{barticle}
\bauthor{\binits{S.G.} \bsnm{{Patiri}}},
\bauthor{\binits{A.J.} \bsnm{{Cuesta}}},
\bauthor{\binits{F.} \bsnm{{Prada}}},
\bauthor{\binits{J.} \bsnm{{Betancort-Rijo}}},
\bauthor{\binits{A.} \bsnm{{Klypin}}},
\batitle{{The Alignment of Dark Matter Halos with the Cosmic Web}}.
%\bjtitle{\apjl}
\bvolume{652},
\bfpage{75}--\blpage{78}
(\byear{2006}).
doi:\doiurl{10.1086/510330}
\end{barticle}
\endbibitem

\bibitem[\protect\citeauthoryear{{Paz} et~al.}{2006a}]{paz06}
\begin{barticle}
\bauthor{\binits{D.J.} \bsnm{{Paz}}},
\bauthor{\binits{D.G.} \bsnm{{Lambas}}},
\bauthor{\binits{N.} \bsnm{{Padilla}}},
\bauthor{\binits{M.} \bsnm{{Merch{\'a}n}}},
\batitle{{Shapes of clusters and groups of galaxies: comparison of model
  predictions with observations}}.
%\bjtitle{\mnras}
\bvolume{366},
\bfpage{1503}--\blpage{1510}
(\byear{2006}a).
doi:\doiurl{10.1111/j.1365-2966.2005.09934.x}
\end{barticle}
\endbibitem

\bibitem[\protect\citeauthoryear{{Paz} et~al.}{2006b}]{paz+al06}
\begin{barticle}
\bauthor{\binits{D.J.} \bsnm{{Paz}}},
\bauthor{\binits{D.G.} \bsnm{{Lambas}}},
\bauthor{\binits{N.} \bsnm{{Padilla}}},
\bauthor{\binits{M.} \bsnm{{Merch{\'a}n}}},
\batitle{{Shapes of clusters and groups of galaxies: comparison of model
  predictions with observations}}.
\bjtitle{MNRAS}
\bvolume{366},
\bfpage{1503}--\blpage{1510}
(\byear{2006}b).
doi:\doiurl{10.1111/j.1365-2966.2005.09934.x}
\end{barticle}
\endbibitem

\bibitem[\protect\citeauthoryear{{Peng} et~al.}{2009}]{peng}
\begin{barticle}
\bauthor{\binits{E.-H.} \bsnm{{Peng}}},
\bauthor{\binits{K.} \bsnm{{Andersson}}},
\bauthor{\binits{M.W.} \bsnm{{Bautz}}},
\bauthor{\binits{G.P.} \bsnm{{Garmire}}},
\batitle{{Discrepant Mass Estimates in the Cluster of Galaxies Abell 1689}}.
\bjtitle{\apj}
\bvolume{701},
\bfpage{1283}--\blpage{1299}
(\byear{2009}).
doi:\doiurl{10.1088/0004-637X/701/2/1283}
\end{barticle}
\endbibitem

\bibitem[\protect\citeauthoryear{{Peter} et~al.}{2012}]{peter12}
\begin{botherref}
\oauthor{\binits{A.H.G.} \bsnm{{Peter}}},
\oauthor{\binits{M.} \bsnm{{Rocha}}},
\oauthor{\binits{J.S.} \bsnm{{Bullock}}},
\oauthor{\binits{M.} \bsnm{{Kaplinghat}}},
{Cosmological Simulations with Self-Interacting Dark Matter II: Halo Shapes vs.
  Observations}.
ArXiv e-prints
(2012)
\end{botherref}
\endbibitem

\bibitem[\protect\citeauthoryear{{Piffaretti} and
  {Valdarnini}}{2008}]{piffaretti08}
\begin{barticle}
\bauthor{\binits{R.} \bsnm{{Piffaretti}}},
\bauthor{\binits{R.} \bsnm{{Valdarnini}}},
\batitle{{Total mass biases in X-ray galaxy clusters}}.
\bjtitle{\aap}
\bvolume{491},
\bfpage{71}--\blpage{87}
(\byear{2008}).
doi:\doiurl{10.1051/0004-6361:200809739}
\end{barticle}
\endbibitem

\bibitem[\protect\citeauthoryear{{Piffaretti} et~al.}{2003}]{roco03}
\begin{barticle}
\bauthor{\binits{R.} \bsnm{{Piffaretti}}},
\bauthor{\binits{P.} \bsnm{{Jetzer}}},
\bauthor{\binits{S.} \bsnm{{Schindler}}},
\batitle{{Aspherical galaxy clusters: Effects on cluster masses and g as mass
  fractions}}.
\bjtitle{\aap}
\bvolume{398},
\bfpage{41}--\blpage{48}
(\byear{2003}).
doi:\doiurl{10.1051/0004-6361:20021648}
\end{barticle}
\endbibitem

\bibitem[\protect\citeauthoryear{{Plionis} et~al.}{2004}]{pli+al04}
\begin{barticle}
\bauthor{\binits{M.} \bsnm{{Plionis}}},
\bauthor{\binits{S.} \bsnm{{Basilakos}}},
\bauthor{\binits{H.M.} \bsnm{{Tovmassian}}},
\batitle{{The shape of poor groups of galaxies}}.
\bjtitle{MNRAS}
\bvolume{352},
\bfpage{1323}--\blpage{1328}
(\byear{2004}).
doi:\doiurl{10.1111/j.1365-2966.2004.08023.x}
\end{barticle}
\endbibitem

\bibitem[\protect\citeauthoryear{{Postman} et~al.}{2012}]{clash}
\begin{barticle}
\bauthor{\binits{M.} \bsnm{{Postman}}},
\bauthor{\binits{D.} \bsnm{{Coe}}},
\bauthor{\binits{N.} \bsnm{{Ben{\'{\i}}tez}}},
\bauthor{\binits{L.} \bsnm{{Bradley}}},
\bauthor{\binits{T.} \bsnm{{Broadhurst}}},
\bauthor{\binits{M.} \bsnm{{Donahue}}},
\bauthor{\binits{H.} \bsnm{{Ford}}},
\bauthor{\binits{O.} \bsnm{{Graur}}},
\bauthor{\binits{G.} \bsnm{{Graves}}},
\bauthor{\binits{S.} \bsnm{{Jouvel}}},
\bauthor{\binits{A.} \bsnm{{Koekemoer}}},
\bauthor{\binits{D.} \bsnm{{Lemze}}},
\bauthor{\binits{E.} \bsnm{{Medezinski}}},
\bauthor{\binits{A.} \bsnm{{Molino}}},
\bauthor{\binits{L.} \bsnm{{Moustakas}}},
\bauthor{\binits{S.} \bsnm{{Ogaz}}},
\bauthor{\binits{A.} \bsnm{{Riess}}},
\bauthor{\binits{S.} \bsnm{{Rodney}}},
\bauthor{\binits{P.} \bsnm{{Rosati}}},
\bauthor{\binits{K.} \bsnm{{Umetsu}}},
\bauthor{\binits{W.} \bsnm{{Zheng}}},
\bauthor{\binits{A.} \bsnm{{Zitrin}}},
\bauthor{\binits{M.} \bsnm{{Bartelmann}}},
\bauthor{\binits{R.} \bsnm{{Bouwens}}},
\bauthor{\binits{N.} \bsnm{{Czakon}}},
\bauthor{\binits{S.} \bsnm{{Golwala}}},
\bauthor{\binits{O.} \bsnm{{Host}}},
\bauthor{\binits{L.} \bsnm{{Infante}}},
\bauthor{\binits{S.} \bsnm{{Jha}}},
\bauthor{\binits{Y.} \bsnm{{Jimenez-Teja}}},
\bauthor{\binits{D.} \bsnm{{Kelson}}},
\bauthor{\binits{O.} \bsnm{{Lahav}}},
\bauthor{\binits{R.} \bsnm{{Lazkoz}}},
\bauthor{\binits{D.} \bsnm{{Maoz}}},
\bauthor{\binits{C.} \bsnm{{McCully}}},
\bauthor{\binits{P.} \bsnm{{Melchior}}},
\bauthor{\binits{M.} \bsnm{{Meneghetti}}},
\bauthor{\binits{J.} \bsnm{{Merten}}},
\bauthor{\binits{J.} \bsnm{{Moustakas}}},
\bauthor{\binits{M.} \bsnm{{Nonino}}},
\bauthor{\binits{B.} \bsnm{{Patel}}},
\bauthor{\binits{E.} \bsnm{{Reg{\"o}s}}},
\bauthor{\binits{J.} \bsnm{{Sayers}}},
\bauthor{\binits{S.} \bsnm{{Seitz}}},
\bauthor{\binits{A.} \bsnm{{Van der Wel}}},
\batitle{{The Cluster Lensing and Supernova Survey with Hubble: An Overview}}.
%\bjtitle{\apjs}
\bvolume{199},
\bfpage{25}
(\byear{2012}).
doi:\doiurl{10.1088/0067-0049/199/2/25}
\end{barticle}
\endbibitem

\bibitem[\protect\citeauthoryear{{Prada} et~al.}{2012}]{prada11}
\begin{barticle}
\bauthor{\binits{F.} \bsnm{{Prada}}},
\bauthor{\binits{A.A.} \bsnm{{Klypin}}},
\bauthor{\binits{A.J.} \bsnm{{Cuesta}}},
\bauthor{\binits{J.E.} \bsnm{{Betancort-Rijo}}},
\bauthor{\binits{J.} \bsnm{{Primack}}},
\batitle{{Halo concentrations in the standard {$\Lambda$} cold dark matter
  cosmology}}.
%\bjtitle{\mnras}
\bvolume{423},
\bfpage{3018}--\blpage{3030}
(\byear{2012}).
doi:\doiurl{10.1111/j.1365-2966.2012.21007.x}
\end{barticle}
\endbibitem

\bibitem[\protect\citeauthoryear{{Puchwein} and {Bartelmann}}{2006}]{puchwein}
\begin{barticle}
\bauthor{\binits{E.} \bsnm{{Puchwein}}},
\bauthor{\binits{M.} \bsnm{{Bartelmann}}},
\batitle{{Three-dimensional reconstruction of the intra-cluster medium}}.
\bjtitle{\aap}
\bvolume{455},
\bfpage{791}--\blpage{801}
(\byear{2006}).
doi:\doiurl{10.1051/0004-6361:20054717}
\end{barticle}
\endbibitem

\bibitem[\protect\citeauthoryear{{Rasia} et~al.}{2006}]{rasia06}
\begin{barticle}
\bauthor{\binits{E.} \bsnm{{Rasia}}},
\bauthor{\binits{S.} \bsnm{{Ettori}}},
\bauthor{\binits{L.} \bsnm{{Moscardini}}},
\bauthor{\binits{P.} \bsnm{{Mazzotta}}},
\bauthor{\binits{S.} \bsnm{{Borgani}}},
\bauthor{\binits{K.} \bsnm{{Dolag}}},
\bauthor{\binits{G.} \bsnm{{Tormen}}},
\bauthor{\binits{L.M.} \bsnm{{Cheng}}},
\bauthor{\binits{A.} \bsnm{{Diaferio}}},
\batitle{{Systematics in the X-ray cluster mass estimators}}.
%\bjtitle{\mnras}
\bvolume{369},
\bfpage{2013}--\blpage{2024}
(\byear{2006}).
doi:\doiurl{10.1111/j.1365-2966.2006.10466.x}
\end{barticle}
\endbibitem

\bibitem[\protect\citeauthoryear{{Reblinsky}}{2000}]{reblinsky}
\begin{barticle}
\bauthor{\binits{K.} \bsnm{{Reblinsky}}},
\batitle{{Cluster deprojection combining multiple observable data sets}}.
\bjtitle{\aap}
\bvolume{364},
\bfpage{377}--\blpage{390}
(\byear{2000})
\end{barticle}
\endbibitem

\bibitem[\protect\citeauthoryear{{Redlich} et~al.}{2012}]{redlich12}
\begin{botherref}
\oauthor{\binits{M.} \bsnm{{Redlich}}},
\oauthor{\binits{M.} \bsnm{{Bartelmann}}},
\oauthor{\binits{J.-C.} \bsnm{{Waizmann}}},
\oauthor{\binits{C.} \bsnm{{Fedeli}}},
{The strongest gravitational lenses: I. The statistical impact of cluster
  mergers}.
ArXiv e-prints
(2012)
\end{botherref}
\endbibitem

\bibitem[\protect\citeauthoryear{{Riemer-S{\o}rensen} et~al.}{2009}]{signe}
\begin{barticle}
\bauthor{\binits{S.} \bsnm{{Riemer-S{\o}rensen}}},
\bauthor{\binits{D.} \bsnm{{Paraficz}}},
\bauthor{\binits{D.D.M.} \bsnm{{Ferreira}}},
\bauthor{\binits{K.} \bsnm{{Pedersen}}},
\bauthor{\binits{M.} \bsnm{{Limousin}}},
\bauthor{\binits{H.} \bsnm{{Dahle}}},
\batitle{{Resolving the Discrepancy Between Lensing and X-Ray Mass Estimates of
  the Complex Galaxy Cluster Abell 1689}}.
\bjtitle{\apj}
\bvolume{693},
\bfpage{1570}--\blpage{1578}
(\byear{2009}).
doi:\doiurl{10.1088/0004-637X/693/2/1570}
\end{barticle}
\endbibitem

\bibitem[\protect\citeauthoryear{{Rocha} et~al.}{2012}]{rocha12}
\begin{botherref}
\oauthor{\binits{M.} \bsnm{{Rocha}}},
\oauthor{\binits{A.H.G.} \bsnm{{Peter}}},
\oauthor{\binits{J.S.} \bsnm{{Bullock}}},
\oauthor{\binits{M.} \bsnm{{Kaplinghat}}},
\oauthor{\binits{S.} \bsnm{{Garrison-Kimmel}}},
\oauthor{\binits{J.} \bsnm{{Onorbe}}},
\oauthor{\binits{L.A.} \bsnm{{Moustakas}}},
{Cosmological Simulations with Self-Interacting Dark Matter I: Constant Density
  Cores and Substructure}.
ArXiv e-prints
(2012)
\end{botherref}
\endbibitem

\bibitem[\protect\citeauthoryear{{Rossi} et~al.}{2011}]{rossi11}
\begin{barticle}
\bauthor{\binits{G.} \bsnm{{Rossi}}},
\bauthor{\binits{R.K.} \bsnm{{Sheth}}},
\bauthor{\binits{G.} \bsnm{{Tormen}}},
\batitle{{Modelling the shapes of the largest gravitationally bound objects}}.
%\bjtitle{\mnras}
\bvolume{416},
\bfpage{248}--\blpage{261}
(\byear{2011}).
doi:\doiurl{10.1111/j.1365-2966.2011.19028.x}
\end{barticle}
\endbibitem

\bibitem[\protect\citeauthoryear{{Ryden}}{1996}]{ryd96}
\begin{barticle}
\bauthor{\binits{B.S.} \bsnm{{Ryden}}},
\batitle{{The Intrinsic Shapes of Stellar Systems}}.
\bjtitle{\apj}
\bvolume{461},
\bfpage{146}
(\byear{1996}).
doi:\doiurl{10.1086/177043}
\end{barticle}
\endbibitem

\bibitem[\protect\citeauthoryear{{Samsing} et~al.}{2012}]{samsing}
\begin{barticle}
\bauthor{\binits{J.} \bsnm{{Samsing}}},
\bauthor{\binits{A.} \bsnm{{Skielboe}}},
\bauthor{\binits{S.H.} \bsnm{{Hansen}}},
\batitle{{Measuring the Three-dimensional Shape of X-Ray Clusters}}.
\bjtitle{\apj}
\bvolume{748},
\bfpage{21}
(\byear{2012}).
doi:\doiurl{10.1088/0004-637X/748/1/21}
\end{barticle}
\endbibitem

\bibitem[\protect\citeauthoryear{{Sand} et~al.}{2002}]{sand02}
\begin{barticle}
\bauthor{\binits{D.J.} \bsnm{{Sand}}},
\bauthor{\binits{T.} \bsnm{{Treu}}},
\bauthor{\binits{R.S.} \bsnm{{Ellis}}},
\batitle{{The Dark Matter Density Profile of the Lensing Cluster MS 2137-23: A
  Test of the Cold Dark Matter Paradigm}}.
%\bjtitle{\apjl}
\bvolume{574},
\bfpage{129}--\blpage{133}
(\byear{2002}).
doi:\doiurl{10.1086/342530}
\end{barticle}
\endbibitem

\bibitem[\protect\citeauthoryear{{Sand} et~al.}{2004}]{sand04}
\begin{barticle}
\bauthor{\binits{D.J.} \bsnm{{Sand}}},
\bauthor{\binits{T.} \bsnm{{Treu}}},
\bauthor{\binits{G.P.} \bsnm{{Smith}}},
\bauthor{\binits{R.S.} \bsnm{{Ellis}}},
\batitle{{The Dark Matter Distribution in the Central Regions of Galaxy
  Clusters: Implications for Cold Dark Matter}}.
\bjtitle{\apj}
\bvolume{604},
\bfpage{88}--\blpage{107}
(\byear{2004}).
doi:\doiurl{10.1086/382146}
\end{barticle}
\endbibitem

\bibitem[\protect\citeauthoryear{{Sand} et~al.}{2008}]{sand07}
\begin{barticle}
\bauthor{\binits{D.J.} \bsnm{{Sand}}},
\bauthor{\binits{T.} \bsnm{{Treu}}},
\bauthor{\binits{R.S.} \bsnm{{Ellis}}},
\bauthor{\binits{G.P.} \bsnm{{Smith}}},
\bauthor{\binits{J.-P.} \bsnm{{Kneib}}},
\batitle{{Separating Baryons and Dark Matter in Cluster Cores: A Full
  Two-dimensional Lensing and Dynamic Analysis of Abell 383 and MS 2137-23}}.
\bjtitle{\apj}
\bvolume{674},
\bfpage{711}--\blpage{727}
(\byear{2008}).
doi:\doiurl{10.1086/524652}
\end{barticle}
\endbibitem

\bibitem[\protect\citeauthoryear{{Sayers} et~al.}{2011a}]{sayers2011a}
\begin{barticle}
\bauthor{\binits{J.} \bsnm{{Sayers}}},
\bauthor{\binits{S.R.} \bsnm{{Golwala}}},
\bauthor{\binits{S.} \bsnm{{Ameglio}}},
\bauthor{\binits{E.} \bsnm{{Pierpa oli}}},
\batitle{{Cluster Morphologies and Model-independent Y $_{SZ}$ Estimates from
  Bolocam Sunyaev-Zel'dovich Images}}.
\bjtitle{\apj}
\bvolume{728},

\bfpage{39}(\byear{2011}a).
doi:\doiurl{10.1088/0004-637X/728/1/39}
\end{barticle}
\endbibitem

\bibitem[\protect\citeauthoryear{{Sereno}}{2007}]{ser07}
\begin{barticle}
\bauthor{\binits{M.} \bsnm{{Sereno}}},
\batitle{{On the deprojection of clusters of galaxies combining X-ray,
  Sunyaev-Zeldovich temperature decrement and gravitational lensing maps}}.
%\bjtitle{\mnras}
\bvolume{380},
\bfpage{1207}--\blpage{1218}
(\byear{2007}).
doi:\doiurl{10.1111/j.1365-2966.2007.12171.x}
\end{barticle}
\endbibitem

\bibitem[\protect\citeauthoryear{{Sereno} and {Umetsu}}{2011}]{sereno_umetsu}
\begin{barticle}
\bauthor{\binits{M.} \bsnm{{Sereno}}},
\bauthor{\binits{K.} \bsnm{{Umetsu}}},
\batitle{{Weak- and strong-lensing analyses of the triaxial matter distribution
  of Abell 1689}}.
%\bjtitle{\mnras}
\bvolume{416},
\bfpage{3187}--\blpage{3200}
(\byear{2011}).
doi:\doiurl{10.1111/j.1365-2966.2011.19274.x}
\end{barticle}
\endbibitem

\bibitem[\protect\citeauthoryear{{Sereno} and {Zitrin}}{2012}]{sereno_zitrin}
\begin{barticle}
\bauthor{\binits{M.} \bsnm{{Sereno}}},
\bauthor{\binits{A.} \bsnm{{Zitrin}}},
\batitle{{Triaxial strong-lensing analysis of the z  0.5 MACS clusters:
  the mass-concentration relation}}.
%\bjtitle{\mnras}
\bvolume{419},
\bfpage{3280}--\blpage{3291}
(\byear{2012}).
doi:\doiurl{10.1111/j.1365-2966.2011.19968.x}
\end{barticle}
\endbibitem

\bibitem[\protect\citeauthoryear{{Sereno} et~al.}{2012}]{sereno_ettori}
\begin{barticle}
\bauthor{\binits{M.} \bsnm{{Sereno}}},
\bauthor{\binits{S.} \bsnm{{Ettori}}},
\bauthor{\binits{A.} \bsnm{{Baldi}}},
\batitle{{Shape and orientation of the gas distribution in A1689}}.
%\bjtitle{\mnras}
\bvolume{419},
\bfpage{2646}--\blpage{2656}
(\byear{2012}).
doi:\doiurl{10.1111/j.1365-2966.2011.19914.x}
\end{barticle}
\endbibitem

\bibitem[\protect\citeauthoryear{{Sereno}
  et~al.}{2010a}]{sereno_overconcentrated}
\begin{barticle}
\bauthor{\binits{M.} \bsnm{{Sereno}}},
\bauthor{\binits{P.} \bsnm{{Jetzer}}},
\bauthor{\binits{M.} \bsnm{{Lubini}}},
\batitle{{On the overconcentration problem of strong lensing clusters}}.
%\bjtitle{\mnras}
\bvolume{403},
\bfpage{2077}--\blpage{2087}
(\byear{2010}a).
doi:\doiurl{10.1111/j.1365-2966.2010.16248.x}
\end{barticle}
\endbibitem

\bibitem[\protect\citeauthoryear{{Sereno} et~al.}{2010b}]{ser+al10b}
\begin{barticle}
\bauthor{\binits{M.} \bsnm{{Sereno}}},
\bauthor{\binits{M.} \bsnm{{Lubini}}},
\bauthor{\binits{P.} \bsnm{{Jetzer}}},
\batitle{{A multiwavelength strong lensing analysis of baryons and dark matter
  in the dynamically active cluster AC 114}}.
\bjtitle{\aap}
\bvolume{518},
\bfpage{55}
(\byear{2010}b).
doi:\doiurl{10.1051/0004-6361/200913843}
\end{barticle}
\endbibitem

\bibitem[\protect\citeauthoryear{{Sereno} et~al.}{2006}]{filippis06}
\begin{barticle}
\bauthor{\binits{M.} \bsnm{{Sereno}}},
\bauthor{\binits{E.} \bsnm{{De Filippis}}},
\bauthor{\binits{G.} \bsnm{{Longo}}},
\bauthor{\binits{M.W.} \bsnm{{Bautz}}},
\batitle{{Measuring the Three-dimensional Structure of Galaxy Clusters. II. Are
  Clusters of Galaxies Oblate or Prolate?}}
\bjtitle{\apj}
\bvolume{645},
\bfpage{170}--\blpage{178}
(\byear{2006}).
doi:\doiurl{10.1086/503198}
\end{barticle}
\endbibitem

\bibitem[\protect\citeauthoryear{{Shaw} et~al.}{2006}]{shaw06}
\begin{barticle}
\bauthor{\binits{L.D.} \bsnm{{Shaw}}},
\bauthor{\binits{J.} \bsnm{{Weller}}},
\bauthor{\binits{J.P.} \bsnm{{Ostriker}}},
\bauthor{\binits{P.} \bsnm{{Bode}}},
\batitle{{Statistics of Physical Properties of Dark Matter Clusters}}.
\bjtitle{\apj}
\bvolume{646},
\bfpage{815}--\blpage{833}
(\byear{2006}).
doi:\doiurl{10.1086/505016}
\end{barticle}
\endbibitem

\bibitem[\protect\citeauthoryear{{Shaw} et~al.}{2010}]{shaw2010}
\begin{barticle}
\bauthor{\binits{L.D.} \bsnm{{Shaw}}},
\bauthor{\binits{D.} \bsnm{{Nagai}}},
\bauthor{\binits{S.} \bsnm{{Bhattacharya}}},
\bauthor{\binits{E.T.} \bsnm{{Lau}}},
\batitle{{Impact of Cluster Physics on the Sunyaev-Zel'dovich Power Spec
  trum}}.
\bjtitle{\apj}
\bvolume{725},
\bfpage{1452}--\blpage{1465}
(\byear{2010}).
doi:\doiurl{10.1088/0004-637X/725/2/1452}
\end{barticle}
\endbibitem

\bibitem[\protect\citeauthoryear{{Silk} and {White}}{1978}]{silk_white}
\begin{barticle}
\bauthor{\binits{J.} \bsnm{{Silk}}},
\bauthor{\binits{S.D.M.} \bsnm{{White}}},
\batitle{{The determination of Q$_{0}$ using X-ray and microwave observations
  of galaxy clusters}}.
%\bjtitle{\apjl}
\bvolume{226},
\bfpage{103}--\blpage{106}
(\byear{1978}).
doi:\doiurl{10.1086/182841}
\end{barticle}
\endbibitem

\bibitem[\protect\citeauthoryear{{Skielboe} et~al.}{2012}]{skielboe}
\begin{botherref}
\oauthor{\binits{A.} \bsnm{{Skielboe}}},
\oauthor{\binits{R.} \bsnm{{Wojtak}}},
\oauthor{\binits{K.} \bsnm{{Pedersen}}},
\oauthor{\binits{E.} \bsnm{{Rozo}}},
\oauthor{\binits{E.S.} \bsnm{{Rykoff}}},
{Spatial anisotropy of galaxy kinematics in SDSS galaxy clusters}.
ArXiv e-prints
(2012)
\end{botherref}
\endbibitem

\bibitem[\protect\citeauthoryear{{Smith} et~al.}{2009}]{1149}
\begin{barticle}
\bauthor{\binits{G.P.} \bsnm{{Smith}}},
\bauthor{\binits{H.} \bsnm{{Ebeling}}},
\bauthor{\binits{M.} \bsnm{{Limousin}}},
\bauthor{\binits{J.-P.} \bsnm{{Kneib}}},
\bauthor{\binits{A.M.} \bsnm{{Swinbank}}},
\bauthor{\binits{C.-J.} \bsnm{{Ma}}},
\bauthor{\binits{M.} \bsnm{{Jauzac}}},
\bauthor{\binits{J.} \bsnm{{Richard}}},
\bauthor{\binits{E.} \bsnm{{Jullo}}},
\bauthor{\binits{D.J.} \bsnm{{Sand}}},
\bauthor{\binits{A.C.} \bsnm{{Edge}}},
\bauthor{\binits{I.} \bsnm{{Smail}}},
\batitle{{Hubble Space Telescope Observations of a Spectacular New
  Strong-Lensing Galaxy Cluster: MACS J1149.5+2223 at z = 0.544}}.
%\bjtitle{\apjl}
\bvolume{707},
\bfpage{163}--\blpage{168}
(\byear{2009}).
doi:\doiurl{10.1088/0004-637X/707/2/L163}
\end{barticle}
\endbibitem

\bibitem[\protect\citeauthoryear{{Smith} et~al.}{2010}]{locussbcg}
\begin{barticle}
\bauthor{\binits{G.P.} \bsnm{{Smith}}},
\bauthor{\binits{H.G.} \bsnm{{Khosroshahi}}},
\bauthor{\binits{A.} \bsnm{{Dariush}}},
\bauthor{\binits{A.J.R.} \bsnm{{Sanderson}}},
\bauthor{\binits{T.J.} \bsnm{{Ponman}}},
\bauthor{\binits{J.P.} \bsnm{{Stott}}},
\bauthor{\binits{C.P.} \bsnm{{Haines}}},
\bauthor{\binits{E.} \bsnm{{Egami}}},
\bauthor{\binits{D.P.} \bsnm{{Stark}}},
\batitle{{LoCuSS: connecting the dominance and shape of brightest cluster
  galaxies with the assembly history of massive clusters}}.
%\bjtitle{\mnras}
\bvolume{409},
\bfpage{169}--\blpage{183}
(\byear{2010}).
doi:\doiurl{10.1111/j.1365-2966.2010.17311.x}
\end{barticle}
\endbibitem

\bibitem[\protect\citeauthoryear{{Sommer-Larsen} and {Limousin}}{2010}]{jesper}
\begin{barticle}
\bauthor{\binits{J.} \bsnm{{Sommer-Larsen}}},
\bauthor{\binits{M.} \bsnm{{Limousin}}},
\batitle{{Moderate steepening of galaxy cluster dark matter profiles by
  baryonic pinching}}.
%\bjtitle{\mnras}
\bvolume{408},
\bfpage{1998}--\blpage{2007}
(\byear{2010}).
doi:\doiurl{10.1111/j.1365-2966.2010.17260.x}
\end{barticle}
\endbibitem

\bibitem[\protect\citeauthoryear{{Soucail} et~al.}{1987}]{soucail87}
\begin{barticle}
\bauthor{\binits{G.} \bsnm{{Soucail}}},
\bauthor{\binits{B.} \bsnm{{Fort}}},
\bauthor{\binits{Y.} \bsnm{{Mellier}}},
\bauthor{\binits{J.P.} \bsnm{{Picat}}},
\batitle{{A blue ring-like structure, in the center of the A 370 cluster of
  galaxies}}.
\bjtitle{\aap}
\bvolume{172},
\bfpage{14}--\blpage{16}
(\byear{1987})
\end{barticle}
\endbibitem

\bibitem[\protect\citeauthoryear{{Springel} et~al.}{2008}]{simu_aquarius}
\begin{barticle}
\bauthor{\binits{V.} \bsnm{{Springel}}},
\bauthor{\binits{J.} \bsnm{{Wang}}},
\bauthor{\binits{M.} \bsnm{{Vogelsberger}}},
\bauthor{\binits{A.} \bsnm{{Ludlow}}},
\bauthor{\binits{A.} \bsnm{{Jenkins}}},
\bauthor{\binits{A.} \bsnm{{Helmi}}},
\bauthor{\binits{J.F.} \bsnm{{Navarro}}},
\bauthor{\binits{C.S.} \bsnm{{Frenk}}},
\bauthor{\binits{S.D.M.} \bsnm{{White}}},
\batitle{{The Aquarius Project: the subhaloes of galactic haloes}}.
%\bjtitle{\mnras}
\bvolume{391},
\bfpage{1685}--\blpage{1711}
(\byear{2008}).
doi:\doiurl{10.1111/j.1365-2966.2008.14066.x}
\end{barticle}
\endbibitem

\bibitem[\protect\citeauthoryear{{Stark}}{1977}]{sta77}
\begin{barticle}
\bauthor{\binits{A.A.} \bsnm{{Stark}}},
\batitle{{Triaxial Models of the Bulge of M31}}.
\bjtitle{\apj}
\bvolume{213},
\bfpage{368}--\blpage{373}
(\byear{1977}).
doi:\doiurl{10.1086/155164}
\end{barticle}
\endbibitem

\bibitem[\protect\citeauthoryear{{Thakur} and {Chakraborty}}{2001}]{th+ch01}
\begin{barticle}
\bauthor{\binits{P.} \bsnm{{Thakur}}},
\bauthor{\binits{D.K.} \bsnm{{Chakraborty}}},
\batitle{{Correlated projected properties of some triaxial mass models:
  implications for their intrinsic shapes}}.
\bjtitle{MNRAS}
\bvolume{328},
\bfpage{330}--\blpage{338}
(\byear{2001}).
doi:\doiurl{10.1046/j.1365-8711.2001.04794.x}
\end{barticle}
\endbibitem

\bibitem[\protect\citeauthoryear{{Umetsu} and {Broadhurst}}{2008}]{um+br08}
\begin{barticle}
\bauthor{\binits{K.} \bsnm{{Umetsu}}},
\bauthor{\binits{T.} \bsnm{{Broadhurst}}},
\batitle{{Combining Lens Distortion and Depletion to Map the Mass Distribution
  of A1689}}.
\bjtitle{\apj}
\bvolume{684},
\bfpage{177}--\blpage{203}
(\byear{2008}).
doi:\doiurl{10.1086/589683}
\end{barticle}
\endbibitem

\bibitem[\protect\citeauthoryear{{Umetsu} et~al.}{2009}]{umetsu09}
\begin{barticle}
\bauthor{\binits{K.} \bsnm{{Umetsu}}},
\bauthor{\binits{M.} \bsnm{{Birkinshaw}}},
\bauthor{\binits{G.-C.} \bsnm{{Liu}}},
\bauthor{\binits{J.-H.P.} \bsnm{{Wu}}},
\bauthor{\binits{E.} \bsnm{{Medezinski}}},
\bauthor{\binits{T.} \bsnm{{Broadhurst}}},
\bauthor{\binits{D.} \bsnm{{Lemze}}},
\bauthor{\binits{A.} \bsnm{{Zitrin}}},
\bauthor{\binits{P.T.P.} \bsnm{{Ho}}},
\bauthor{\binits{C.-W.L.} \bsnm{{Huang}}},
\bauthor{\binits{P.M.} \bsnm{{Koch}}},
\bauthor{\binits{Y.-W.} \bsnm{{Liao}}},
\bauthor{\binits{K.-Y.} \bsnm{{Lin}}},
\bauthor{\binits{S.M.} \bsnm{{Molnar}}},
\bauthor{\binits{H.} \bsnm{{Nishioka}}},
\bauthor{\binits{F.-C.} \bsnm{{Wang}}},
\bauthor{\binits{P.} \bsnm{{Altamirano}}},
\bauthor{\binits{C.-H.} \bsnm{{Chang}}},
\bauthor{\binits{S.-H.} \bsnm{{Chang}}},
\bauthor{\binits{S.-W.} \bsnm{{Chang}}},
\bauthor{\binits{M.-T.} \bsnm{{Chen}}},
\bauthor{\binits{C.-C.} \bsnm{{Han}}},
\bauthor{\binits{Y.-D.} \bsnm{{Huang}}},
\bauthor{\binits{Y.-J.} \bsnm{{Hwang}}},
\bauthor{\binits{H.} \bsnm{{Jiang}}},
\bauthor{\binits{M.} \bsnm{{Kesteven}}},
\bauthor{\binits{D.Y.} \bsnm{{Kubo}}},
\bauthor{\binits{C.-T.} \bsnm{{Li}}},
\bauthor{\binits{P.} \bsnm{{Martin-Cocher}}},
\bauthor{\binits{P.} \bsnm{{Oshiro}}},
\bauthor{\binits{P.} \bsnm{{Raffin}}},
\bauthor{\binits{T.} \bsnm{{Wei}}},
\bauthor{\binits{W.} \bsnm{{Wilson}}},
\batitle{{Mass and Hot Baryons in Massive Galaxy Clusters from Subaru
  Weak-Lensing and AMiBA Sunyaev-Zel'Dovich Effect Observations}}.
\bjtitle{\apj}
\bvolume{694},
\bfpage{1643}--\blpage{1663}
(\byear{2009}).
doi:\doiurl{10.1088/0004-637X/694/2/1643}
\end{barticle}
\endbibitem

\bibitem[\protect\citeauthoryear{{Umetsu} et~al.}{2011}]{umetsu11}
\begin{barticle}
\bauthor{\binits{K.} \bsnm{{Umetsu}}},
\bauthor{\binits{T.} \bsnm{{Broadhurst}}},
\bauthor{\binits{A.} \bsnm{{Zitrin}}},
\bauthor{\binits{E.} \bsnm{{Medezinski}}},
\bauthor{\binits{L.-Y.} \bsnm{{Hsu}}},
\batitle{{Cluster Mass Profiles from a Bayesian Analysis of Weak-lensing
  Distortion and Magnification Measurements: Applications to Subaru Data}}.
\bjtitle{\apj}
\bvolume{729},
\bfpage{127}
(\byear{2011}).
doi:\doiurl{10.1088/0004-637X/729/2/127}
\end{barticle}
\endbibitem

\bibitem[\protect\citeauthoryear{{Vera-Ciro} et~al.}{2011}]{aquarius}
\begin{barticle}
\bauthor{\binits{C.A.} \bsnm{{Vera-Ciro}}},
\bauthor{\binits{L.V.} \bsnm{{Sales}}},
\bauthor{\binits{A.} \bsnm{{Helmi}}},
\bauthor{\binits{C.S.} \bsnm{{Frenk}}},
\bauthor{\binits{J.F.} \bsnm{{Navarro}}},
\bauthor{\binits{V.} \bsnm{{Springel}}},
\bauthor{\binits{M.} \bsnm{{Vogelsberger}}},
\bauthor{\binits{S.D.M.} \bsnm{{White}}},
\batitle{{The shape of dark matter haloes in the Aquarius simulations:
  evolution and memory}}.
%\bjtitle{\mnras}
\bvolume{416},
\bfpage{1377}--\blpage{1391}
(\byear{2011}).
doi:\doiurl{10.1111/j.1365-2966.2011.19134.x}
\end{barticle}
\endbibitem

\bibitem[\protect\citeauthoryear{{Vikhlinin} et~al.}{2006a}]{vikhlinin06}
\begin{barticle}
\bauthor{\binits{A.} \bsnm{{Vikhlinin}}},
\bauthor{\binits{A.} \bsnm{{Kravtsov}}},
\bauthor{\binits{W.} \bsnm{{Forman}}},
\bauthor{\binits{C.} \bsnm{{Jones}}},
\bauthor{\binits{M.} \bsnm{{Markevitch}}},
\bauthor{\binits{S.S.} \bsnm{{Murray}}},
\bauthor{\binits{L.} \bsnm{{Van Speybroeck}}},
\batitle{{Chandra Sample of Nearby Relaxed Galaxy Clusters: Mass, Gas Fraction,
  and Mass-Temperature Relation}}.
\bjtitle{\apj}
\bvolume{640},
\bfpage{691}--\blpage{709}
(\byear{2006}a).
doi:\doiurl{10.1086/500288}
\end{barticle}
\endbibitem

\bibitem[\protect\citeauthoryear{{Vikhlinin} et~al.}{2006b}]{vik+al06}
\begin{barticle}
\bauthor{\binits{A.} \bsnm{{Vikhlinin}}},
\bauthor{\binits{A.} \bsnm{{Kravtsov}}},
\bauthor{\binits{W.} \bsnm{{Forman}}},
\bauthor{\binits{C.} \bsnm{{Jones}}},
\bauthor{\binits{M.} \bsnm{{Markevitch}}},
\bauthor{\binits{S.S.} \bsnm{{Murray}}},
\bauthor{\binits{L.} \bsnm{{Van Speybroeck}}},
\batitle{{Chandra Sample of Nearby Relaxed Galaxy Clusters: Mass, Gas Fraction,
  and Mass-Temperature Relation}}.
\bjtitle{\apj}
\bvolume{640},
\bfpage{691}--\blpage{709}
(\byear{2006}b).
doi:\doiurl{10.1086/500288}
\end{barticle}
\endbibitem

\bibitem[\protect\citeauthoryear{{Waizmann} et~al.}{2012}]{waizmann12}
\begin{botherref}
\oauthor{\binits{J.-C.} \bsnm{{Waizmann}}},
\oauthor{\binits{M.} \bsnm{{Redlich}}},
\oauthor{\binits{M.} \bsnm{{Bartelmann}}},
{The strongest gravitational lenses: II. Is the large Einstein radius of MACS
  J0717.5+3745 in conflict with LCDM?}
ArXiv e-prints
(2012)
\end{botherref}
\endbibitem

\bibitem[\protect\citeauthoryear{{Warren} et~al.}{1992}]{warren92}
\begin{barticle}
\bauthor{\binits{M.S.} \bsnm{{Warren}}},
\bauthor{\binits{P.J.} \bsnm{{Quinn}}},
\bauthor{\binits{J.K.} \bsnm{{Salmon}}},
\bauthor{\binits{W.H.} \bsnm{{Zurek}}},
\batitle{{Dark halos formed via dissipationless collapse. I - Shapes and
  alignment of angular momentum}}.
\bjtitle{\apj}
\bvolume{399},
\bfpage{405}--\blpage{425}
(\byear{1992}).
doi:\doiurl{10.1086/171937}
\end{barticle}
\endbibitem

\bibitem[\protect\citeauthoryear{{Wechsler} et~al.}{2002}]{wechsler02}
\begin{barticle}
\bauthor{\binits{R.H.} \bsnm{{Wechsler}}},
\bauthor{\binits{J.S.} \bsnm{{Bullock}}},
\bauthor{\binits{J.R.} \bsnm{{Primack}}},
\bauthor{\binits{A.V.} \bsnm{{Kravtsov}}},
\bauthor{\binits{A.} \bsnm{{Dekel}}},
\batitle{{Concentrations of Dark Halos from Their Assembly Histories}}.
\bjtitle{\apj}
\bvolume{568},
\bfpage{52}--\blpage{70}
(\byear{2002}).
doi:\doiurl{10.1086/338765}
\end{barticle}
\endbibitem

\bibitem[\protect\citeauthoryear{{Wyithe} et~al.}{2001}]{wyithe2001}
\begin{barticle}
\bauthor{\binits{J.S.B.} \bsnm{{Wyithe}}},
\bauthor{\binits{E.L.} \bsnm{{Turner}}},
\bauthor{\binits{D.N.} \bsnm{{Spergel}}},
\batitle{{Gravitational Lens Statistics for Generalized NFW Profiles: Pa
  rameter Degeneracy and Implications for Self-Interacting Cold Dark Matter}}.
\bjtitle{\apj}
\bvolume{555},
\bfpage{504}--\blpage{523}
(\byear{2001}).
doi:\doiurl{10.1086/321437}
\end{barticle}
\endbibitem

\bibitem[\protect\citeauthoryear{{Yoshida} et~al.}{2000}]{yoshida2000}
\begin{barticle}
\bauthor{\binits{N.} \bsnm{{Yoshida}}},
\bauthor{\binits{V.} \bsnm{{Springel}}},
\bauthor{\binits{S.D.M.} \bsnm{{White}}},
\bauthor{\binits{G.} \bsnm{{Tormen}}},
\batitle{{Collisional Dark Matter and the Structure of Dark Halos}}.
%\bjtitle{\apjl}
\bvolume{535},
\bfpage{103}--\blpage{106}
(\byear{2000}).
doi:\doiurl{10.1086/312707}
\end{barticle}
\endbibitem

\bibitem[\protect\citeauthoryear{{Yoshikawa} and {Suto}}{1999}]{yoshikawa}
\begin{barticle}
\bauthor{\binits{K.} \bsnm{{Yoshikawa}}},
\bauthor{\binits{Y.} \bsnm{{Suto}}},
\batitle{{Reconstructing the Radial Profiles of Gas Density and Temperature in
  Clusters of Galaxies from High-Resolution X-Ray and Radio Observations}}.
\bjtitle{\apj}
\bvolume{513},
\bfpage{549}--\blpage{554}
(\byear{1999}).
doi:\doiurl{10.1086/306908}
\end{barticle}
\endbibitem

\bibitem[\protect\citeauthoryear{{Yuan} et~al.}{2008}]{yuan}
\begin{barticle}
\bauthor{\binits{Q.} \bsnm{{Yuan}}},
\bauthor{\binits{T.-J.} \bsnm{{Zhang}}},
\bauthor{\binits{B.-Q.} \bsnm{{Wang}}},
\batitle{{Reconstruction of Gas Temperature and Density Profiles of the Galaxy
  Cluster RX J1347.5 1145}}.
%\bjtitle{\cjaa}
\bvolume{8},
\bfpage{671}--\blpage{676}
(\byear{2008}).
doi:\doiurl{10.1088/1009-9271/8/6/05}
\end{barticle}
\endbibitem

\bibitem[\protect\citeauthoryear{{Zaroubi} et~al.}{1998}]{zaroubi98}
\begin{barticle}
\bauthor{\binits{S.} \bsnm{{Zaroubi}}},
\bauthor{\binits{G.} \bsnm{{Squires}}},
\bauthor{\binits{Y.} \bsnm{{Hoffman}}},
\bauthor{\binits{J.} \bsnm{{Silk}}},
\batitle{{Deprojection of Rich Cluster Images}}.
%\bjtitle{\apjl}
\bvolume{500},
\bfpage{87}
(\byear{1998}).
doi:\doiurl{10.1086/311421}
\end{barticle}
\endbibitem

\bibitem[\protect\citeauthoryear{{Zaroubi} et~al.}{2001}]{zaroubi2001}
\begin{barticle}
\bauthor{\binits{S.} \bsnm{{Zaroubi}}},
\bauthor{\binits{G.} \bsnm{{Squires}}},
\bauthor{\binits{G.} \bsnm{{de Gasperis}}},
\bauthor{\binits{A.E.} \bsnm{{Evrard}}},
\bauthor{\binits{Y.} \bsnm{{Hoffman}}},
\bauthor{\binits{J.} \bsnm{{Silk}}},
\batitle{{Deprojection of Galaxy Cluster X-Ray, Sunyaev-Zeldovich Temperature
  Decrement, and Weak-Lensing Mass Maps}}.
\bjtitle{\apj}
\bvolume{561},
\bfpage{600}--\blpage{620}
(\byear{2001}).
doi:\doiurl{10.1086/323359}
\end{barticle}
\endbibitem

\bibitem[\protect\citeauthoryear{{Zhang} et~al.}{2010}]{zhang2010}
\begin{barticle}
\bauthor{\binits{Y.-Y.} \bsnm{{Zhang}}},
\bauthor{\binits{N.} \bsnm{{Okabe}}},
\bauthor{\binits{A.} \bsnm{{Finoguenov}}},
\bauthor{\binits{G.P.} \bsnm{{Smith}}},
\bauthor{\binits{R.} \bsnm{{Piffaretti}}},
\bauthor{\binits{R.} \bsnm{{Valdarnini}}},
\bauthor{\binits{A.} \bsnm{{Babul}}},
\bauthor{\binits{A.E.} \bsnm{{Evrard}}},
\bauthor{\binits{P.} \bsnm{{Mazzotta}}},
\bauthor{\binits{A.J.R.} \bsnm{{Sanderson}}},
\bauthor{\binits{D.P.} \bsnm{{Marrone}}},
\batitle{{LoCuSS: A Comparison of Cluster Mass Measurements from XMM-Newton and
  Subaru: Testing Deviation from Hydrostatic Equilibrium and Non-thermal
  Pressure Support}}.
\bjtitle{\apj}
\bvolume{711},
\bfpage{1033}--\blpage{1043}
(\byear{2010}).
doi:\doiurl{10.1088/0004-637X/711/2/1033}
\end{barticle}
\endbibitem

\bibitem[\protect\citeauthoryear{{Zhao} et~al.}{2003}]{zhao03}
\begin{barticle}
\bauthor{\binits{D.H.} \bsnm{{Zhao}}},
\bauthor{\binits{Y.P.} \bsnm{{Jing}}},
\bauthor{\binits{H.J.} \bsnm{{Mo}}},
\bauthor{\binits{G.} \bsnm{{B{\"o}rner}}},
\batitle{{Mass and Redshift Dependence of Dark Halo Structure}}.
%\bjtitle{\apjl}
\bvolume{597},
\bfpage{9}--\blpage{12}
(\byear{2003}).
doi:\doiurl{10.1086/379734}
\end{barticle}
\endbibitem

\bibitem[\protect\citeauthoryear{{Zitrin} and {Broadhurst}}{2009}]{zitrin1149}
\begin{barticle}
\bauthor{\binits{A.} \bsnm{{Zitrin}}},
\bauthor{\binits{T.} \bsnm{{Broadhurst}}},
\batitle{{Discovery of the Largest Known Lensed Images Formed by a Critically
  Convergent Lensing Cluster}}.
%\bjtitle{\apjl}
\bvolume{703},
\bfpage{132}--\blpage{136}
(\byear{2009}).
doi:\doiurl{10.1088/0004-637X/703/2/L132}
\end{barticle}
\endbibitem

\bibitem[\protect\citeauthoryear{{Zitrin} et~al.}{2009}]{adi0717}
\begin{barticle}
\bauthor{\binits{A.} \bsnm{{Zitrin}}},
\bauthor{\binits{T.} \bsnm{{Broadhurst}}},
\bauthor{\binits{Y.} \bsnm{{Rephaeli}}},
\bauthor{\binits{S.} \bsnm{{Sadeh}}},
\batitle{{The Largest Gravitational Lens: MACS J0717.5+3745 (z = 0 .546)}}.
%\bjtitle{\apjl}
\bvolume{707},
\bfpage{102}--\blpage{106}
(\byear{2009}).
doi:\doiurl{10.1088/0004-637X/707/1/L102}
\end{barticle}
\endbibitem

\bibitem[\protect\citeauthoryear{{Zitrin} et~al.}{2010}]{adisample}
\begin{botherref}
\oauthor{\binits{A.} \bsnm{{Zitrin}}},
\oauthor{\binits{T.} \bsnm{{Broadhurst}}},
\oauthor{\binits{R.} \bsnm{{Barkana}}},
\oauthor{\binits{Y.} \bsnm{{Rephaeli}}},
\oauthor{\binits{N.} \bsnm{{Benitez}}},
{Strong-Lensing Analysis of a Complete Sample of 12 MACS Clusters at z 0.5:
  Mass Models and Einstein Radii}.
ArXiv e-prints
(2010)
\end{botherref}
\endbibitem

\bibitem[\protect\citeauthoryear{{Zitrin} et~al.}{2012}]{zitrin12}
\begin{barticle}
\bauthor{\binits{A.} \bsnm{{Zitrin}}},
\bauthor{\binits{T.} \bsnm{{Broadhurst}}},
\bauthor{\binits{M.} \bsnm{{Bartelmann}}},
\bauthor{\binits{Y.} \bsnm{{Rephaeli}}},
\bauthor{\binits{M.} \bsnm{{Oguri}}},
\bauthor{\binits{N.} \bsnm{{Ben{\'{\i}}tez}}},
\bauthor{\binits{J.} \bsnm{{Hao}}},
\bauthor{\binits{K.} \bsnm{{Umetsu}}},
\batitle{{The universal Einstein radius distribution from 10 000 SDSS
  clusters}}.
%\bjtitle{\mnras}
\bvolume{423},
\bfpage{2308}--\blpage{2324}
(\byear{2012}).
doi:\doiurl{10.1111/j.1365-2966.2012.21041.x}
\end{barticle}
\endbibitem

\bibitem[\protect\citeauthoryear{}{{Rasia} et~al.}{2012}]{rasia2012}
\begin{botherref}
{Lensing and x-ray mass estimates of clusters (simulations)}.
New Journal of Physics
\textbf{14}(5),
055018
(2012).
doi:\doiurl{10.1088/1367-2630/14/5/055018}
\end{botherref}
\endbibitem

\end{thebibliography}
\nocite{*}

\end{document}